\documentclass[journal,twoside,web]{ieeecolor}

\usepackage{generic}
\usepackage[utf8]{inputenc}
\usepackage{cite}
\usepackage{qtree}
\usepackage{tikz}
\usepackage{amsmath,amssymb,amsfonts}
\usepackage{eufrak}
\usepackage{dsfont}
\usepackage{steinmetz}
\usepackage{algorithmicx}
\usepackage{algorithm}
\usepackage{algpseudocode}
\usepackage{graphicx}
\usepackage{amsmath,amssymb}
\usepackage{textcomp}
\usepackage{algpseudocode} 
\usepackage{mathtools}
\usepackage{nicefrac}
\usepackage{txfonts}
\usepackage{tikz}
\usepackage{dsfont}
\usepackage{graphicx}
\usepackage{amsmath,amssymb}
\usepackage{xcolor}
\usepackage{mathtools}
\usepackage{nicefrac}
\usepackage{txfonts}
\usepackage{todonotes}
\usepackage{cleveref}
\usepackage{xcolor}
\usepackage{subfig}
\usepackage{balance}
\usepackage[font=small]{caption}

\newcommand{\R}{\mathbb{R}}
\newcommand{\bN}{\mathbb{N}}
\newcommand{\bbP}{\mathbb{P}}

\newcommand{\U}{\mathcal{U}}
\newcommand{\E}{\mathbb{E}}

\newcommand{\T}{^{\mbox\tiny\mathsf{T}}}
\newcommand{\tr}{{\rm{tr}}}

\newcommand{\Q}{\mathcal{Q}}

\newcommand{\A}{\mathcal{A}}

\newcommand{\B}{\mathcal{B}}

\renewcommand{\baselinestretch}{0.997}

\DeclareMathOperator{\argmin}{\arg\!\min}

\def\mf{\mathbf}
\def\mb{\mathbb}
\def\mc{\mathcal}
\def\beq{\begin{equation*}}
\def\eeq{\end{equation*}}
\def\bql{\begin{equation}}
\def\eql{\end{equation}}
\def\bqn{\begin{eqnarray*}}
\def\eqn{\end{eqnarray*}}
\def\bnl{\begin{eqnarray}}
\def\enl{\end{eqnarray}}
\def\bna{\bql\begin{array}{rcl}}
\def\ena{\end{array}\eql}
\def\bnn{\beq\begin{array}{rcl}}
\def\enn{\end{array}\eeq}
\def\bma{\begin{bmatrix}}
\def\ema{\end{bmatrix}}
\def\bmx{\begin{matrix}}
\def\emx{\end{matrix}}
\def\ben{\begin{enumerate}}
\def\een{\end{enumerate}}
\def\bit{\begin{itemize}}
\def\eit{\end{itemize}}
\def\bei{\begin{itemize}}
\def\eei{\end{itemize}}
\def\bet{\begin{tabular}}
\def\eet{\end{tabular}}

\newtheorem{thm}{Theorem}
\newtheorem{lm}{ Lemma}

\newtheorem{rem}{ Remark}

\newtheorem{corr}{ Corollary}

\newtheorem{asm}{Assumption}

\title{Koopman Meets Limited Bandwidth: Effect of Quantization on Data-Driven Linear Prediction and Control of Nonlinear Systems}
\author{Shahab Ataei$^1$, Dipankar Maity$^2$, and Debdipta Goswami$^3$
\thanks{The work of S. Ataei and D. Goswami was supported by the OSU College of Engineering Strategic Research Initiative Grant.
The work of D. Maity was supported by the U.S. Army Small Business Innovation Research Program Office
and the Army Research Office under Contract No. W911NF-24-P-0002.}
\thanks{$^1$S. Ataei is with the Department of Electrical and Computer Engineering, The Ohio State University, Columbus,
OH, 43210, USA. Email:{ \tt ataei.3@osu.edu }}
\thanks{$^2$D. Maity is with the Department of Electrical and Computer Engineering and an affiliated faculty of the North Carolina Battery Complexity, Autonomous Vehicle, and Electrification Research Center (BATT CAVE), University of North Carolina at Charlotte,  NC, 28223, USA.
Email: {\tt {dmaity@charlotte.edu}}
}
\thanks{$^3$D. Goswami is with the Department of Mechanical and Aerospace Engineering and Electrical and Computer Engineering, The Ohio State University, Columbus,
OH, 43210, USA. Email:{ \tt goswami.78@osu.edu }}
}

\begin{document}

\maketitle
\thispagestyle{empty}

\begin{abstract}
Koopman-based lifted linear identification have been widely used for data-driven prediction and model predictive control (MPC) of nonlinear systems. It has found applications in flow-control, soft robotics, and unmanned aerial vehicles (UAV). For autonomous systems, this system identification method works by embedding the nonlinear system in a higher-dimensional linear space and computing a finite-dimensional approximation of the corresponding Koopman operator with the Extended Dynamic Mode Decomposition (EDMD) algorithm. EDMD is a data-driven algorithm that estimates an approximate linear system by lifting the state data-snapshots via nonlinear dictionary functions. For control systems, EDMD is further modified to utilize both state and control data-snapshots to estimate a lifted linear predictor with control input. 
    This article investigates how the estimation process is affected when the data is quantized.
    Specifically, we examine the fundamental connection between estimates of the linear predictor matrices obtained from unquantized data and those from quantized data via modified EDMD.
    Furthermore, using the law of large numbers, we demonstrate that, under a \textit{large data regime}, the quantized estimate can be considered a regularized version of the unquantized estimate. 
    %
    %
    We also explore the relationship between the two estimates in the \textit{finite data regime}.
    We further analyze the effect of nonlinear lifting functions on this regularization due to quantization.
    The theory is validated through repeated numerical experiments conducted on several control systems. The effect of quantization on the MPC performance is also demonstrated.
\end{abstract}


\section{Introduction} 

\IEEEPARstart{S}{ystem} identification is an essential component in controls and dynamical systems applications involving unknown or partially known dynamics.
In recent years, Koopman operator theory \cite{Koopman1931} based system identification methods have been widely used in various fields ranging from fluid mechanics \cite{Rowley2009} and plasma dynamics \cite{nayak2021koopman} to control of unmanned aircraft systems \cite{narayanan2023}, traffic prediction \cite{Avila2020}, and machine learning tasks for training deep neural networks \cite{Dogra2020}.
The reasons for the wide adoption of the Koopman operator based identification methods are manifold. The linearity of the Koopman operator provides linear surrogate dynamics and facilitates the development of efficient data-driven algorithms such as Extended Dynamic Mode Decomposition (EDMD) \cite{Williams2015}. The identified linear model is also amenable for data-driven control application via model predictive control (MPC) as demonstrated in \cite{korda2018linear}. It is well-understood that the quality of the estimated linear model improves/degrades with an increase/decrease in the amount of data, as expected \cite{Arbabi2017, Hirsh2020, Lee2024}. On the other hand, it is not clear how the quality of the data affects the estimation process, especially when the data undergoes a quantization process.

Existing work in EDMD and data-driven system identification typically assumes that these algorithms are implemented on systems with ample resources to handle large datasets generated from snapshots of the dynamical system. 
However, applying these data-intensive algorithms to resource-limited systems, such as low-powered, lightweight robotic applications \cite{Folkestad2022, Cleary2020}, may require quantization to meet hardware and other resource constraints. 
In fact, quantization naturally arises under communication and computation constraints, making it a common practice in networked control systems, multi-agent systems, and cyber-physical systems in general.

Quantization can significantly impact control systems, potentially causing a stabilizable system to become unstable if the quantization word-length drops below a critical threshold \cite{nair2004stabilizability}. 
Since system identification is typically the initial step in controlling unknown systems, the influence of quantization on the identification process subsequently affects controllers, state estimators, and ultimately the overall system performance. 
Moreover, the selection of an appropriate quantizer plays a crucial role, as it can directly influence the system's performance \cite{maity2021optimal, maity2023optimal}.

In this paper, we study the effects of \textit{dither quantization} \cite{gray1993dithered}---a highly effective and commonly used quantization method in controls, communications, and signal processing--- on Koopman-based linear predictor identification and MPC by extending our prior works \cite{maity2024effect, maity2024EDMD}. 
To the best of our knowledge, \cite{maity2024effect} and \cite{maity2024EDMD} were the first efforts to investigate the effect of quantization on Dynamic Mode Decomposition (DMD) and EDMD and develops and analyzes the \textit{Dither Quantized (E)DMD} method.
DMD and EDMD are used to identify autonomous nonlinear systems from data by estimating a finite-dimensional approximation of the corresponding Koopman operator. In contrast, identification of a nonlinear control system requires careful modification of these algorithms to incorporate control input data, thereby giving rise to an estimation problem with an objective to identify a linear time-invariant (LTI) predictor for the original nonlinear system \cite{korda2018linear}. Furthermore, the modified EDMD involves lifting the state-data via a dictionary of observable functions \cite{Williams2015}, and consequently,  those lifting functions (e.g., radial basis functions, Legendre polynomials) may further amplify the effects of quantization. This paper discusses the impact of quantization on the predictor identification and examines the role of lifting functions in influencing the resulting effects. It further demonstrates the effect of quantization on MPC performance when the predictive model is identified from quantized data.
%

The main contributions of this work are as follows: (i) We address the key question of whether and how the original solution for the linear predictor obtained from unquantized data can be recovered from quantized data. Using the law of large numbers, we prove in \Cref{thm:equivalence} that estimation with quantized data is equivalent to regularized estimation with unquantized data when a large number of data snapshots is available, revealing the link between the regularization parameter and quantization resolution. (ii) We extend this analysis to the small data regime, demonstrating analytically how the estimation difference relates to the quantization resolution. (iii) Our theory is validated through extensive experiments on four different systems, testing predictor and MPC performance with various quantization resolutions.

The rest of the paper is organized as follows: \Cref{sec:background} provides the necessary background materials on Koopman Operator theory, lifted linear predictor, and dither quantization. 
We define our problem statement in \Cref{sec:ProblemStatement} and analyze the dither quantized linear predictor estimation in \Cref{sec:QuantizedID}, demonstrating the connection between the solution obtained from quantized and unquantized data.
We discuss our observations from implementing the Koopman-based linear predictor and subsequent MPC on four dynamical systems in \Cref{sec:simulation}. We further analyze a special case of the large data-regime result from \Cref{sec:QuantizedID} where the Koopman observables themselves are quantized instead of the data-snapshots in \Cref{sec:SpecialCase}. Lastly, we provide some conclusions in \Cref{sec:conclusions}.

\textit{Notations:} Set of non-negative integers are denoted by $\mb{N}_0$. The space of real and complex numbers are denoted by $\mb{R}$ and $\mb{C}$, respectively.
$(\cdot)^{\dagger}$ and $(\cdot)^\top$ denote the Moore--Penrose inverse and transpose of a matrix, respectively. 
$\|\cdot\|$ denotes a norm, where we use Euclidean norm for vectors and Frobenius norms for matrices. 
The Big-O notation is denoted by $O(\cdot)$.
The composition of two functions are denoted by $\circ$. \\

\section{Background}\label{sec:background}
\subsection{Koopman operator theory}
Consider a discrete-time dynamical system on an $n$-dimensional compact manifold $\mathcal{M} \subset \mb{R}^n$, evolving according to the flow-map ${f}:\mc{M}\mapsto \mc{M}$ as follows: 
\begin{equation} \label{Eq: Dynamics}
    {x}_{t+1} = {f}({x}_{t}),\quad {x}_t\in\mc{M},\quad t\in\mb{N}_0.
\end{equation}
Let $\mc{F}$ be a Banach space of complex-valued observables  $\varphi:\mc{M}\rightarrow \mb{C}$. The discrete-time \emph{Koopman operator} $\mc{K}:\mc{F}\rightarrow \mc{F}$ is defined as
\begin{equation}
    \mc{K}\circ\varphi(\cdot) = \varphi \circ {f}(\cdot),\quad \text{with}~~\varphi({x}_{t+1})=\mc{K}\varphi({x}_{t}),
\end{equation}
 where $\mc{K}$ is infinite-dimensional, and linear over its argument. The scalar observables $\varphi$ are referred to as the \textit{Koopman observables}.

A \textit{Koopman eigenfunction} $\phi_i$ is a special  observable that satisfies $(\mc{K}\phi_i)(\cdot)=\lambda_i \phi_i(\cdot)$, for some eigenvalue $\lambda_i \in \mb{C}$. 
Considering the Koopman eigenfunctions (i.e., $\{\phi_i\}_{i \in \mb{N}}$) span the Koopman observables,  any {vector valued observable ${g}=[\varphi_1,~\varphi_2,~\ldots,~\varphi_p]^\top \in\mc{F}^p $} can be expressed as a sum of Koopman eigenfunctions ${g}(\cdot)=\sum_{i=1}^{\infty}\phi_i(\cdot){v}^{{g}}_i$, where ${v}^{{g}}_i\in\mb{R}^p$, for $i \in \mb{N}$ are called the \emph{Koopman modes} of the observable $g(\cdot)$. This modal decomposition provides the growth/decay rate $|\lambda_i|$ and frequency $\angle{\lambda_i}$ of different Koopman modes via its time evolution:
\begin{equation}\label{eq:koop_decomp}
    {g}({x}_t) = \sum\nolimits_{i=1}^{\infty}\lambda_i^t\phi_i({x}_0){v}^{{g}}_i.
\end{equation}
\noindent The Koopman eigenvalues ($\lambda_i$) and eigenfunctions ($\phi_i$) are properties of the dynamics only, whereas the Koopman modes ($v^i_g$) depend on the observable ($g$). 

Several methods have also been developed to compute the Koopman modal decomposition, e.g., DMD and EDMD \cite{schmid2010, Williams2015}, Ulam-Galerkin methods, and deep neural networks \cite{otto2019linearly, Yeung2019}. 

Koopman methods have been extended to nonlinear control systems by employing lifted linear \cite{korda2018linear} or bilinear predictors \cite{Goswami2021}. 
In this article, we utilize the extension defined in \cite{korda2018linear}. Consider a discrete-time nonlinear control system evolving on a compact manifold $\mc{M}\subset \mb{R}^n$ with a set of admissible control $\mc{U} \subset \mb{R}^m $,
\begin{equation} \label{Eq: Control_Dynamics}
    {x}_{t+1} = {f}({x}_{t}, u_t),\quad {x}_t\in\mc{M},\quad u_t \in \mc{U}, \quad t\in\mb{N}_0,
\end{equation}
with a flow map $f:\mc{M}\times \mc{U} \rightarrow \mc{M}$. Since, the Koopman operator is defined only for autonomous dynamical system \eqref{Eq: Dynamics}, we redefine \eqref{Eq: Control_Dynamics} as follows. 
First, let $l(\mc{U})$ denote the space of all possible sequences $\mf{u} = \{u_i\}_{i=0}^{\infty}$ with $u_i\in \mc{U}$ and denote $\mc{S}:l(\mc{U})\rightarrow l(\mc{U})$ to be the right shift operator, i.e., $\mc{S}^t\mf{u} = \{u_i\}_{i=t}^\infty$, for all $t \in \mb{N}_0$.
Then, define the new state ${\chi}_t = [x_t^\top,\,\mc{S}^t\mf{u}]^\top\in\mc{M} \times l(\mc{U})$, for all $t\in\mb{N}_0$. 
Consequently, the new flow-map $F:\mc{M} \times l(\mc{U}) \rightarrow \mc{M} \times l(\mc{U})$ is as follows:
\begin{equation} \label{Eq: Control_New}
    {\chi}_{t+1} = {F}({\chi}_{t}) = \begin{bmatrix}
        f(x_t, u_t) \\ \mc{S}\circ \mc{S}^t\mf{u}
    \end{bmatrix},\qquad \forall~ t \in \mb{N}_0.
\end{equation}
Now, modifying the Banach space $\mc{F}$ of observables $\psi:\mc{M}\times l(\mc{U}) \rightarrow \mb{C}$, the Koopman operator is redefined as 
\begin{equation}
    \mc{K}\circ\psi(\cdot) = \psi  \circ {F}(\cdot),\quad \text{with}~~\psi({x}_{t+1},\mc{S}^{t+1}\mf{u})=\mc{K}\psi({x}_{t},\mc{S}^t\mf{u}).
\end{equation}
The Koopman operator $\mc{K}$, such defined for a control system, is also infinite dimensional and linear in its arguments and paves the way to build a linear predictor for nonlinear system as described next. 

\subsection{Approximation of the lifted linear dynamics via EDMD}
Extended dynamic mode decomposition is a data-driven method for approximating the Koopman operator and dominant Koopman modes from a sequence of time-series data using a set of observable functions and matrix factorization. It was developed \cite{Williams2015} as a nonlinear extension of dynamic mode decomposition to extract spatio-temporal structures from intricate flows. This paper adopts a special version of EDMD to approximate the Koopman operator $\mc{K}$ for the controlled system from a series of data and control snapshots \cite{korda2018linear}. We assume a set of observables of the form 
\begin{equation} \label{eq:observables}
    \psi(x,\mf{u}) = \begin{bmatrix}
        \varphi(x) \\ \mf{u}(0)
    \end{bmatrix},
\end{equation}
where $\varphi(\cdot) = [\varphi^1(\cdot),\ldots,\varphi^N(\cdot)]^\top:\mc{M}\rightarrow \mb{R}^N$ and $\mf{u}(0)$ denotes the current value of the control sequence. 
That is, corresponding to the state at time $t$, $\mf{u}(0) = u_t$. 

Since we are interested only in the part of the Koopman operator $\mc{K}$ that maps the observables $\varphi^i(\cdot)$ from themselves and the current control input $u_t$ to the future value of $\varphi^i(\cdot)$ observables, we define a pair of snapshot matrices ${\Psi}$ and ${\Psi}^{+}$:
\begin{align} \label{eq:dataMatrix}
\begin{split} 
   {\Psi} &= \begin{bmatrix}
         \begin{bmatrix}\varphi(x_0)\\u_0\end{bmatrix} ~  \begin{bmatrix}\varphi(x_1)\\u_1\end{bmatrix} ~ \hdots ~ \begin{bmatrix}\varphi(x_{T-1})\\u_{T-1}\end{bmatrix}
    \end{bmatrix},  \\
    {\Psi}^{+} &= \begin{bmatrix}
         \begin{bmatrix}\varphi(x_1)\\u_1\end{bmatrix} ~  \begin{bmatrix}\varphi(x_2)\\u_2\end{bmatrix} ~ \hdots ~ \begin{bmatrix}\varphi(x_{T})\\u_{T}\end{bmatrix} 
    \end{bmatrix} .
\end{split}    
\end{align} 
The modified EDMD algorithm \cite{korda2018linear} aims to find the best linear approximator that relates the matrices $\mf{\Psi}$ and $\mf{\Psi}^{+}$ in the following manner: 
\begin{equation}
    \Phi^{+} \approx G {\Psi} = A {\Phi} + B {U},
\end{equation}
where 
\begin{align}
\label{eq:dataMatrix2}
\begin{split}
    {\Phi} &= [\varphi(x_0) ~ \hdots ~ \varphi(x_{T-1})],\\
    \Phi^{+} &= [\varphi(x_1) ~ \hdots ~ \varphi(x_{T})], \text{ and}\\
    {U} &= [u_0~\hdots~u_{T-1}].
\end{split}
\end{align}
The predictor matrices $G = [A, \, B]$ are determined by solving the least-square optimization problem
\begin{align}\label{Eq: optimization}
\begin{split}
    [A,\,B]=& \underset{\mc{A} {\in \R^{N \times N}},\,\mc{B} \in {\R^{N \times m}}}{\argmin} \dfrac{1}{T}\|\Phi^{+} - \mc{A} {\Phi} - \mc{B} {U}\|^2 \\
    = & \underset{\mc{G} \in {\R^{N \times (N+m)}}}{\argmin} \dfrac{1}{T}\|\Phi^{+} - \mc{G}\Psi\|^2.
    \end{split}
\end{align}
Once the $A$ and $B$ matrices are computed, we may predict the observable at time $t$ for any initial state $x_0$ and a control sequence $\{u_i\}_{i=0}^{t-1}$ as
\begin{align}\label{eq:lin_pred}
    \varphi(x_t) = A^t \varphi(x_0) + \sum_{k=0}^{t-1} A^{t-1-k} B u_k.
\end{align}
If the state $x \in \operatorname{span}\{\varphi^1(\cdot),\ldots,\varphi^N(\cdot)\}$, there exists a decoding matrix $C\in\R^{n\times N}$ such that $x = C \varphi(x)$. This decoding matrix can be identified as 
\begin{align*}
    C = \underset{\mc{C}}{\argmin} \dfrac{1}{T}\|X - \mc{C}{\Phi}\|^2.
\end{align*}
However, in practice, $\varphi(x)$ contains $x$, and hence, $C$ can be identified trivially. 

\begin{rem}
    The optimization problem \eqref{Eq: optimization} employs Koopman theory to approximate the nonlinear system \eqref{Eq: Dynamics} by a lifted linear time-invariant (LTI) system. However, a more accurate approximation is possible by employing a lifted bilinear control system \cite{Otto2020, Goswami2021}. However, this manuscript focuses on the lifted LTI system identification.
\end{rem}

\subsection{Dither quantization}
\textit{Dither quantization} refers to a specific quantization method where noise is added to the raw data before quantizing it \cite{schuchman1964dither, lipshitz1992quantization}. 
Then, the same noise is subtracted from the quantized data to reconstruct the original signal at the decoder. 
Although quite counter-intuitive, the addition of the noise actually helps in accurately reconstructing the signal compared to `standard' quantization methods that do not add noise \cite{roberts1962picture}. 
Albeit the noise must possess some essential properties to ensure optimal operation, as we will discuss later in this section. 

Dither quantitation has certain properties that significantly simplify theoretical analysis which would otherwise lead to an analytically intractable problem, making it an appropriate candidate for quantization in control theoretic analysis; see \cite{zames1977structural, mossaheb1983application, iannelli2003dither, silva2010framework, morita2011performance, kashima2014stationary, kawano2021effects} for a wide range of applications of dither quantization in controls.  

To discuss \textit{dither} quantization, let us start with the formal description of a quantizer. 
Let $q : (x_{\min}, x_{\max}) \subseteq \R \to \{0,\ldots, (2^b-1)\}$ denote a quantizer of word-length $b$.
For instance, a uniform quantizer is described by 
\begin{align}
    q(x)=\left\lfloor\frac{x-x_{\min}}{\epsilon} \right\rfloor,
\end{align}
where the quantization resolution $\epsilon$ is given by
\begin{align} \label{eq:quantizationResolution}
    \epsilon=\frac{x_{\max}-x_{\min}}{2^b}.
\end{align}
For all $x\in (x_{\min},x_{\max})$, $q(x)$ can be represented by a $b$-bit binary word, making the word-length of the quantizer to be $b$. 
The interval $(x_{\min},x_{\max})$ is known as the \textit{range} of the quantizer. 
One may extend the \textit{range} to the entire $\R$ by redefining $q$ as: 
\begin{align*}
    \bar{q}(x)=\begin{cases}
    q(x),\quad &x\in (x_{\min},x_{\max}),\\
    0, & x\le x_{\min},\\
    2^b-1, & x\ge  x_{\max}.
    \end{cases}
\end{align*}
In this case the region $(-\infty, x_{\min}) \cup (x_{\max}, \infty)$ is  called the \textit{saturation region} of the quantizer.

A decoder $\Q:\{0,\ldots, (2^b-1)\} \to \R$ decodes the $b$-bit word represented by $q$ (or, $\bar{q}$) and reconstructs the original signal value $x$.
For instance, the decoding of mid-point uniform quantizer is performed by
\begin{align*}
    \Q(x) =\epsilon q(x) + x_{\min} + \frac{\epsilon}{2}.
\end{align*}
The quantization error is defined to be
\begin{align*}
    e(x) \triangleq \mathcal{Q}(x) - x. 
\end{align*}
Consequently, $|e(x)| \le \frac{\epsilon}{2} $, for all $x\in (x_{\min}, x_{\max})$. 

The performance of a quantizer is optimal (i.e., minimum loss of statistical data due to quantization) when $e(x)$ is statistically independent of $x$ \cite{widrow1961statistical}. 
To achieve such statistical independence \textit{dither} quantization is used. 
Under \textit{dither} quantization, the encoded data is $x+w$, where $w$ is the added noise, and the decoded data is given by
\begin{align}\label{eq:decoding_dither}
    \tilde{x} = \Q(x+w) - w.
\end{align}
Accordingly, the quantization error is defined as $e \triangleq \tilde{x} - x = \Q(x+w) - x - w$. 
Note that we purposefully dropped the argument $x$ from $e$ as we will next discuss the statistical independence between $x$ and $e$. 

Schuchman's necessary and sufficient conditions \cite{schuchman1964dither} for the statistical independence (i.e., $\bbP(e|x)) = \bbP(e)$) is given in the following lemma.
\begin{lm}[\!\!\cite{gray1993dithered}] \label{lem:dither}
 $\bbP(e|x) = \bbP(e)$ holds if and only only if    
 \begin{align*}
     \mathcal{W}_w \left( \frac{k}{\Delta} \right) = 0,\qquad \forall ~k = \pm 1, \pm 2, \ldots,
 \end{align*}
 where $\mathcal{W}_w(s) = \E[e^{-jsw}]$ is the characteristic equation of the random variable $w$.
\end{lm}

There exists many noise distributions that satisfy the condition in \Cref{lem:dither}, see \cite{schuchman1964dither}. 
Among them, the uniform distribution $w \sim \mathcal{U}([-\frac{\Delta}{2}, \frac{\Delta}{2}])$ is the most popular one, perhaps due its simplicity. 
We will also use the uniform distribution for this work. 

\section{Problem Statement}\label{sec:ProblemStatement}
This manuscript aims to understand and quantify the effects of the dither quantization on the Koopman-based lifted linear predictor identified via the least-square optimization \eqref{Eq: optimization}. We assume that observables $\varphi(\cdot)$ are computed using decoded quantized data $\tilde{x}$ as defined in \eqref{eq:decoding_dither}.
That is, the data pertaining to the $i^{\text{th}}$ observable at time $t$ is $\varphi^i (\tilde{x}_t)$, whereas, in the unquantized case, that data is $\varphi^i(x_t)$. 
The quantization error for state data pertaining to time $t$ is denoted as $e_{t}^x \triangleq \tilde{x}_t - x_t$, where $\tilde{x}_t = \Q(x_t + w^x_t) - w^x_t$ is defined in \eqref{eq:decoding_dither} with $w^x_t$ being the dither noise. 
Furthermore, we assume that the control input snapshot at time $t$, i.e., $u_t$, is also in its decoded quantized form $\tilde{u}_t \triangleq \Q(u_t + w^u_t) - w^u_t$, where $w^u_t$ is the dither noise added for quantizing the input $u_t$.
The quantization error for control input data pertaining to time $t$ is denoted as $e^u_t \triangleq \tilde u_t - u_t$. Consequently, at time $t$, the available data is $(\bar \varphi (x_t), \tilde{u}_t)$, where 
\begin{align} \label{eq:tildeObservables}
    \bar \varphi (x_t) \triangleq \begin{bmatrix}
        \varphi^1(\tilde x_t), \cdots,
        \varphi^N(\tilde x_t)
    \end{bmatrix}^\top.
\end{align}

Let $\tilde{A},\,\tilde{B}$ denote the linear predictor matrices identified from the quantized data. That is, 
\begin{align} \label{eq:EDMD_quantized}
 \begin{split}     [\tilde{A},\,\tilde{B}]& \triangleq \underset{\mc{A},\,\mc{B}}{\argmin} \dfrac{1}{T}\|\bar{\Phi}^{+} - \mc{A} \bar{{\Phi}} - \mc{B} \bar{{U}}\|^2\\
     &= \underset{\mc{G}}{\argmin} \dfrac{1}{T}\|\bar{\Phi}^{+} - \mc{G}\bar{\Psi}\|^2,
\end{split}
\end{align}
where
\begin{align*}
    \bar{\Psi} &= \begin{bmatrix}
         \begin{bmatrix}\bar\varphi(x_0)\\\tilde u_0\end{bmatrix} ~  \begin{bmatrix}\bar\varphi(x_1)\\\tilde u_1\end{bmatrix} ~ \hdots ~ \begin{bmatrix}\bar\varphi(x_{T-1})\\\tilde u_{T-1}\end{bmatrix}
    \end{bmatrix},  \\
    \bar \Phi &= \begin{bmatrix}
        \bar \varphi(x_0) ~ \bar \varphi(x_1) ~ \hdots ~ \bar \varphi(x_{T-1}) 
    \end{bmatrix},~\\
    \bar \Phi^{+} &= \begin{bmatrix}
        \bar \varphi(x_1) ~ \bar \varphi(x_2) ~ \hdots ~ \bar \varphi(x_{T}) 
    \end{bmatrix} ,~\\
    \bar U &= \begin{bmatrix}
        \tilde u_0 ~ \tilde u_1 ~ \hdots ~ \tilde u_{T-1} 
    \end{bmatrix} ,
\end{align*}
and where $\bar{\varphi}(\cdot)$ is defined in \eqref{eq:tildeObservables}.
On the other hand, the predictor obtained from the unquantized data is
\begin{align} \label{eq:uqzEDMD}
\begin{split}
    [A,\,B] &= \underset{\mc{A},\,\mc{B}}{\argmin} \dfrac{1}{T}\|\Phi^{+} - \mc{A} {\Phi} - \mc{B} {U}\|^2 \\
    &= \underset{\mc{G}}{\argmin} \dfrac{1}{T}\|\Phi^{+} - \mc{G}\Psi\|^2,
    \end{split}
\end{align}
where $ \Phi,  \Phi^{+} \in \R^{N \times T}$ and $\Psi \in \R^{(N+m) \times T}$ are the data matrices defined in \eqref{eq:dataMatrix} and \eqref{eq:dataMatrix2}.
Note that $\bar\Psi = \begin{bmatrix}
    \bar \Phi \\ \bar U
\end{bmatrix}$ and $\Psi = \begin{bmatrix}
     \Phi \\  U
\end{bmatrix}$ are the quantized and unquantized augmented data matrices respectively.

\begin{rem}
It is noteworthy that the modified EDMD problem for lifted linear predictor identification under quantization resembles EDMD under noisy measurements (see, e.g.,\cite{dawson2016characterizing, haseli2019approximating}\footnote{These works concerns autonomous dynamical systems $x_{t+1} = f(x_t)$ and not controlled dynamical systems $x_{t+1} = f(x_t, u_t)$. To the best of our knowledge there are no works on investigating noisy measurements for Koopman linear predictor identification for controlled systems.}) with $e^x_t$ playing the role of the measurement noise. 
However, these works do not characterize how $\frac{\| A - \tilde{A}\|}{\|A\|}$ and $\frac{\| B - \tilde{B}\|}{\|B\|}$ change with the noise intensity, which is necessary for our work to understand how the `noise intensity' $\epsilon$ (equivalently, the quantization word-length $b$) affects the Koopman linear prediction. 
Furthermore, those works focus on analyzing the effect of noise using heuristic methods, whereas we adopt a principled approach using the Kolmogorov's law-of-large-number in this paper. 
Finally, our analysis provides insights on how the choice of the lifting functions (i.e., ${\Phi}$) amplifies/attenuates the quantization noise, which is missing in the existing literature.   
\end{rem}

\section{Linear Predictor Identification with Quantized Data}\label{sec:QuantizedID}
In this section, we study the optimization problem for linear predictor identification under both large (i.e, $T\to \infty$) and finite (i.e., $T<\infty$)  data regimes. 
In the large data regime we show that $[\tilde{A},\,\tilde{B}]$ and $[A,\, B]$ are connected via a regularized optimization problem.
In the finite data regime, we show that the difference between $[\tilde{A},\,\tilde{B}]$ and $[A,\, B]$ is $O(\epsilon)$, with $\epsilon$ being the quantization resolution. This insight helps us in two main ways: (i) It shows a direct link between the quantization resolution $\epsilon$ (equivalently, the word-length $b$) and the amount of distortion in the Koopman estimate, thus providing the necessary communication/computation resources for maintaining a prescribed level of distortion, and (ii) It shows how the amount of data (e.g., large-data vs. finite-data) can compensate for quantized measurement.

\begin{asm} \label{assm:BoundedPhi}
The observables are bounded functions and control input bounded.
That is, for all $i$ there exists $\ell_i < {h}_i$ such that  $\ell_i \le \varphi^i(x) \le {h}_i$ for all $x\in \R^n$.  Moreover, there exists $u_{\ell}$ and $u_{{h}}$ such that $\|u_t\| \in [u_{\text{min}}$,\, $u_{\text{max}}]$ for all $t$. 
\end{asm}

A direct consequence of this assumption\footnote{
For practical purposes, we only need the data matrices $\Phi$, $\Phi^{+}$, and $U$ to be bounded, since the EDMD algorithm deals only with the data and not the functions. 
Therefore, the observables do not need to be bounded functions as long as the measured data is bounded. 
} is that we may assume $\ell_{\min} \le \varphi^i(x) \le h_{\max}$ for all $i$.

\subsection{Large Data regime: $T\to \infty$}
Define the one-step least-square residual  $r:\mc{M}\times\mc{M}\times\mc{U}\rightarrow \R_+$ such that $r(x_{t+1},x_t, u_t)\triangleq \|\varphi(x_{t+1})- \mc{A}\varphi(x_t) -\mc{B}u_t\|^2$. This residual depends on the choice of the linear approximator matrices $\mc{A}$ and $\mc{B}$.
Therefore, we may write 
\begin{align*}
     \frac{1}{T} \|\Phi^{+} - \mc{A}\Phi - \mc{B} U \|^2 =  \sum\nolimits_{t=0}^{T-1} r(x_{t+1},x_t, u_t)
\end{align*}
For our analysis, we make the following assumptions. 
\begin{asm} \label{assm:boundedR}
    There exists $\mathbb{A} \subseteq \R^{N\times N}$, $\mb{B} \subseteq
    \R^{N\times m}$, and $c_r > 0$ such that $r(x_{t+1}, x_t, u_t)< c_r$ for all $t$ when $\mc{A}\in \mathbb{A}$ and $\mc{B}\in \mb{B}$.
\end{asm}

This assumption is necessary and sufficient to ensure that the unquantized EDMD (i.e., \eqref{eq:uqzEDMD}) under the large data regime (i.e., $T\to \infty$) is a well-posed problem.
Notice that the sets $\mb{A}$ and $\mb{B}$ in \Cref{assm:boundedR} are equivalent to the set 
\[\mb{A} \times \mb{B} = \left\{(\A,\B) : \lim_{T\to \infty} \frac{1}{T} \|\Phi^{+} - \A\Phi - \B U\|^2 <+\infty \right\}.\]
\begin{asm} \label{assm:AbsConvTaylor}
     $r(\cdot,\cdot, \cdot)$ has an absolutely convergent Taylor series for all $\mc{A}\in\mb{A}$ and $\mc{B}\in\mb{B}$, on the region of $\mc{M}\times\mc{M}\times\mc{U}$ from which $x_t,\, u_t$ data are collected, with a radius of convergence greater than $\epsilon/2$. 
\end{asm}
\begin{asm} \label{assm:BoundedPhi_derivative}
    There exists $c_\varphi > 0$  such that $\|\nabla \varphi^i(x) \| \le c_\varphi$ for all $x \in \R^n$ and $i=\{1,\ldots, N\}$.
\end{asm}

\Cref{assm:AbsConvTaylor} is used in \Cref{thm:equivalence} and \Cref{assm:BoundedPhi_derivative} is used in both Theorems~\ref{thm:equivalence} and \ref{thm:K_epsilon}.

\begin{thm}[Large data regime result]
\label{thm:equivalence}
    As $T \to \infty$, $[\tilde{A}, \tilde{B}]$ converges almost surely to the solution of the following regularized least-square optimization   \vspace{ -1mm}
    \begin{align} \label{eq:equivalence}
    \begin{split}
        \underset{\substack{\mathcal{A} \in \R^{N \times N}\\ \mathcal{B}\in \R^{N\times m}}}{\min} \limsup_{T\to \infty}\frac{1}{T}  \| \Phi^{+} - \mathcal{A} \Phi - \mathcal{B}U\|^2 + \tr(\mathcal{G}\beta(\epsilon)) + \tr(\mathcal{G}^\top \mathcal{G} \Gamma(\epsilon)),
        \end{split}\vspace{-2mm}
    \end{align} 
    where $\mathcal{G} = [\mathcal{A},\,\mathcal{B}]$, $\beta(\epsilon)$ and $\Gamma(\epsilon)$ are $O(\epsilon^2)$ functions. 
\end{thm}

\begin{proof}
    A proof is presented in \Cref{AP:thm:equivalence}.
\end{proof}

Theorem~\ref{thm:equivalence} implies that  $[\tilde{A},\,\tilde{B}]$ can be interpreted as a solution to a regularized least-squares problem, where the regularization parameter depends on the quantization resolution  $\epsilon$ and some matrices $\beta$ and $\Gamma$. These regularization matrices, in turn, depend on the Taylor series coefficients of the residual $r(\cdot,\cdot,\cdot)$. 
A consequence of \Cref{thm:equivalence} is that the solution $[\tilde{A},\,\tilde{B}]$ converges to $[A,\, B]$ almost surely as $\epsilon$ approaches to $0$. 
The quantization resolution $\epsilon$ is coupled with the quantization word-length $b$. 
Due to \eqref{eq:quantizationResolution}, $[\tilde{A},\,\tilde{B}] \rightarrow [A,\, B]$ almost surely at an \textit{exponential} rate with $b$.

\begin{rem}
   \Cref{thm:equivalence} establishes a fundamental relationship between $[A,\, B]$ and $[\tilde{A},\,\tilde{B}]$. It is important to note that the equivalence described in \eqref{eq:equivalence} relies on the quantization noises being i.i.d., a property ensured by the use of dither quantization. This conclusion may not necessarily extend to other forms of quantization.
\end{rem}

\begin{rem}
    The selection of the lifting functions $\varphi^i$ plays a critical role in amplifying or attenuating the effects of quantization through the terms $\beta(\epsilon)$ and $\Gamma(\epsilon)$, which are directly influenced by the derivatives of the lifting functions. Consequently, some types of lifting functions may be more advantageous than others in the context of quantization. A comprehensive analysis of this aspect lies beyond the scope of this paper but represents an intriguing direction for future research.
\end{rem}

\Cref{thm:equivalence} not only helps in identifying the relationship between $[A,\, B]$ and $[\tilde{A},\,\tilde{B}]$, but also provides a convenient framework to potentially recover $[A,\, B]$ from the quantized data, as discussed next. 

\subsection{Regularized Least-Square for Quantized Data}

\Cref{thm:equivalence} demonstrates that $\frac{1}{T}  \| \bar \Phi^{+} - \mc{A} \bar \Phi - \mc{B} \bar U\|^2 $ almost surely converges to $\frac{1}{T}  \| \Phi^{+} - \mc{A} \Phi - \mc{B} U\|^2 + \tr(\mc{G}\beta(\epsilon)) + \tr(\mc{G}\T \mc{G} \Gamma(\epsilon)) + \text{constant}$, as ${T\to \infty}$. 
Alternatively, one may state that $\frac{1}{T}  \| \bar \Phi^{+} - \mc{A} \bar \Phi - \mc{B} \bar U\|^2 -  \tr(\mc{G}\beta(\epsilon)) - \tr(\mc{G}\T \mc{G} \Gamma(\epsilon))$ almost surely converges to {\small$\frac{1}{T}  \| \Phi^{+} - \mc{A} \Phi - \mc{B} U\|^2$} + \text{constant}. 
Therefore, one may further claim that 
\begin{align} \label{eq:regularized_EDMD}
    \underset{\mc{A},\mc{B}}{\argmin} \limsup_{T\to \infty} \frac{1}{T}  \| \bar \Phi^{+} - \mc{A} \bar \Phi -\mc{B}\bar U\|^2 -  \tr(\mc{G}\beta(\epsilon)) - \tr(\mc{G}\T \mc{G} \Gamma(\epsilon)) \nonumber \\
   = \underset{\mc{A},\mc{B}}{\argmin} \limsup_{T\to \infty} \frac{1}{T}  \|  \Phi^{+} - \mc{A} \Phi - \mc{B} U\|^2 = [A,\, B].
\end{align}
In other words, $[A,\, B]$ can be recovered from quantized data by solving the regularized least-square problem defined in \eqref{eq:regularized_EDMD}, where $\beta(\epsilon), \Gamma(\epsilon)$ are the regularization parameters. 
The challenge in recovering $[A,\,B]$ from \eqref{eq:regularized_EDMD} is that the exact expressions of the regularization parameters are not easy to obtain. 
One potential approach would be to approximate these quantities by $\hat{\beta}(\epsilon)$ and $\hat{\Gamma}(\epsilon)$. 
Such approximation is beyond the scope of this paper, as the primary focus of this work is to \textit{analyze} the effect of quantization and the mitigation of such effects will be addressed in subsequent future works.



\subsection{Finite Data Regime}
\begin{thm}[Finite data regime result] \label{thm:K_epsilon}
    Let $\Phi$, $\bar \Phi$, $\Psi$, and $\bar\Psi$ be of full row rank. Then, $\exists \ G_\epsilon$ such that $\|G_\epsilon\| = O(\epsilon)$ and 
    \begin{align}
    \bar G = [\bar A,\ \bar B] = G + G_\epsilon = [A,\ B] + G_\epsilon .
    \end{align}
\end{thm}

\begin{proof}
    The closed form solution to the least-square problem in \eqref{eq:EDMD_quantized} with quantized data  is 
\begin{align} \label{eq:KDt_solution}
    \bar G = \bar\Phi^{+} \bar \Psi^\top \big( \bar \Psi \bar \Psi^\top\big)^{-1}, 
\end{align}
whereas that for the unquantized case is $G = \Phi^{+}\Psi^\top(\Psi\Psi^\top)^{-1}$.

Due to the mean-value theorem, we may write 
\begin{align} \label{eq:mean-value-theorem}
     \varphi^i(\tilde{x}_t) & = \varphi^i(x_t) + \underbrace{(e_t^x)^\top \nabla \varphi^i(x_t + \alpha^i_t e_t^x)}_{ \triangleq \delta^i_{t}} 
\end{align}
for some $\alpha^i_t \in [0,1]$. 
Consequently, $\bar \Phi = \Phi + \Phi_\epsilon$, where the $ij$-th element of $\Phi_\epsilon$ is the {\small$\delta^i_j$} defined in \eqref{eq:mean-value-theorem}. 
Notice that $|\delta^i_t| \le \frac{nc_\varphi}{2}\epsilon$  since $\| \nabla \varphi^i (x)\| \le c_\varphi$ for all $x$ due to \Cref{assm:BoundedPhi_derivative}, and $\|e_t^x\| \le \frac{\sqrt{n}}{2}\epsilon$ due to the quantization process.
Thus, $\| \Phi_\epsilon\| = O(\epsilon)$. Now 
\begin{align*}
    \bar\Psi = \begin{bmatrix}
        \bar\Phi \\ \bar U
    \end{bmatrix}
    = \begin{bmatrix}
        \Phi + \Phi_\epsilon \\  U + U_\epsilon
    \end{bmatrix}
    =\Psi + \Psi_\epsilon\ ,
\end{align*}
where $j$-th column of $U_\epsilon$ is $e_j^u$ and $\Psi_\epsilon = \begin{bmatrix}
    \Phi_\epsilon \\ U_\epsilon 
\end{bmatrix}$. 
Since each component of $e_j^u$ is uniformly distributed between $\left [ -\dfrac{\epsilon}{2},\ \dfrac{\epsilon}{2}\right]$, we conclude $\|U_\epsilon\| = O(\epsilon)$.
Combining $\| \Phi_\epsilon\| = O(\epsilon)$ and $\|U_\epsilon\| = O(\epsilon)$, we get $\|\Psi_\epsilon\| = O(\epsilon)$. Similarly, we can show $\bar\Phi^{+} = \Phi^{+} + \Phi^{+}_\epsilon$ with $\|\bar\Phi^{+}_\epsilon\| =  O(\epsilon)$ as well.

Substituting $\bar \Psi = \Psi + \Psi_\epsilon$ and $\bar \Phi^{+} = \Phi^{+} + \Phi^{+}_\epsilon$  in \eqref{eq:KDt_solution} followed by some simplifications yields
\begin{align*}
    \bar G & = G - G\big(  \Psi  \Psi^\top\Xi_\epsilon^{-1} + I  \big)^{-1} + \Pi_\epsilon \big( \bar \Psi \bar \Psi^\top\big)^{-1} ,
\end{align*}
where $\Xi_\epsilon = \Psi_\epsilon \Psi^\top + \Psi \Psi_\epsilon^\top + \Psi_\epsilon \Psi_\epsilon^\top$  and $\Pi_\epsilon = \Phi^{+}_\epsilon \Psi^\top + \Phi^{+} \Psi_\epsilon^\top + \Phi^{+}_\epsilon \Psi_\epsilon^\top$. 
Therefore, we may write 
\begin{align*}
    \bar G = [\bar A,\ \bar B] =  G + G_\epsilon = [A,\ B] + G_\epsilon,
\end{align*}
where $ G_\epsilon =  \Pi_\epsilon \big( \bar \Psi \bar \Psi^\top\big)^{-1}  - G\big(  \Psi  \Psi^\top\Xi_\epsilon^{-1} + I  \big)^{-1} $. 
The theorem is proven once we show that $\|\bar G_\epsilon\| = O(\epsilon)$. 
To that end, let us note that $\|\Psi_\epsilon\| = O(\epsilon)$ and $\|\Phi^{+}_\epsilon\| = O(\epsilon)$ implies $\|\Xi_\epsilon\| = O(\epsilon)$ and $\|\Pi_\epsilon\| = O(\epsilon)$, and therefore, $\|G_\epsilon \| = O(\epsilon)$. 
This concludes the proof.
\end{proof}


\section{Case Studies: Model Predictive Control} \label{sec:simulation}
In this section, we study the effect of quantization on EDMD-based system identification, prediction, and data-driven control from the identified surrogate model. As a control objective, we aim to track reference trajectories of the system states. The proposed model predictive controller (MPC) solves the following optimization problem for a predictive horizon $T_h$ and applies piecewise constant control for current time $t=0$:
\begin{align}\label{eq:NMPC}
        \begin{split}
        \underset{u_t,x_t}{\text{  minimize   }} J (\{u_t\},\{x_t\})   = &\sum_{t=0}^{T_h} ||x_t-x_{t,
 \text{ref}}||_{\boldsymbol{Q}}^2 + ||u_t||_{\boldsymbol{R}}^2\\
        \text{subject to: } 
        & x_{t+1} = f(x_t,\ u_t),\ t=0,\ldots,T_h -1,\\
        & x_t\in \mc{M}, u_t\in\mc{U},\\
        & x_{t=0} = x_0,
    \end{split}
\end{align}
where $\mf{Q}$ and $\mf{R}$ are the weighting matrices for state and control costs respectively. The optimization repeats at each time-step. We convert this problem into a linear MPC by employing the linear predictive model \eqref{eq:lin_pred} as follows:
\begin{align}\label{eq:LMPC}
        \begin{split}
        \underset{u_t,z_t}{\text{  minimize   }} J (\{u_t\},\{z_t\})   = &\sum_{t=0}^{T_h} (Cz_t-x_{t,
 \text{ref}})^{\top}Q(Cz_t-x_{t,
 \text{ref}}) + ||u_t||_{\boldsymbol{R}}^2\\
        \text{subject to: } 
        & z_{t+1} = Az_t + Bu_t,\ t=0,\ldots,T_h -1,\\
        & Cz_t\in \mc{M}, u_t\in\mc{U},\\
        & z_{t=0} = \varphi(x_0),
    \end{split}
\end{align}
where $A,\ B,$ and $C$ are linear predictor matrices obtained from \eqref{Eq: optimization} and $\varphi(\cdot)$ is the dictionary function.

The effect of dither quantization is demonstrated on Koopman-based linear system identification and subsequent MPC for a variety of systems: A pendulum with negative damping, the Van der Pol oscillator, a bilinear model of a field-controlled DC motor, and the Korteweg-de Vries (KdV) nonlinear partial differential equation (PDE). We used 50 independent Monte-Carlo realizations of the dither signal for data snapshots used in system identification and subsequent MPC. 

\begin{figure*}[t]
\centering 
\subfloat[]{\includegraphics[trim=1.9cm 7cm 2cm 7cm, clip=true, width=0.25\textwidth]{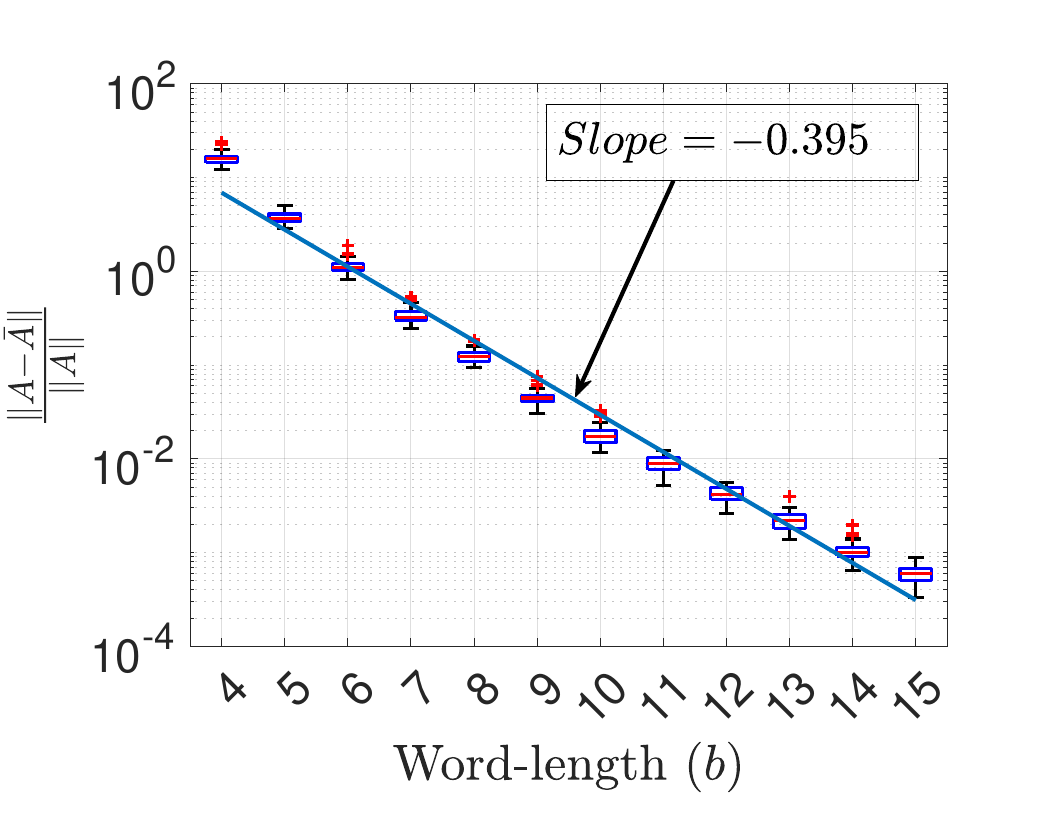}}
\subfloat[]{\includegraphics[trim=1.9cm 7cm 2cm 7cm, clip=true, width=0.25\textwidth]{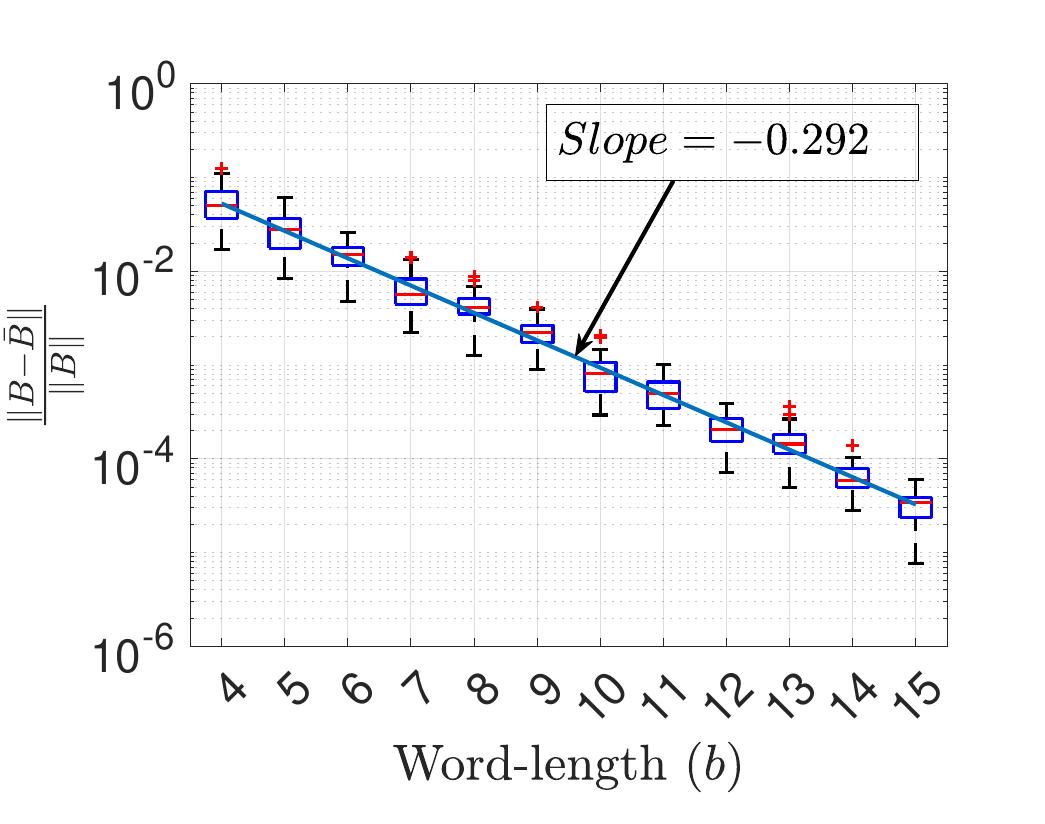}}
\subfloat[]{\includegraphics[trim=2cm 7cm 2cm 7cm, clip=true, width=0.25\textwidth]{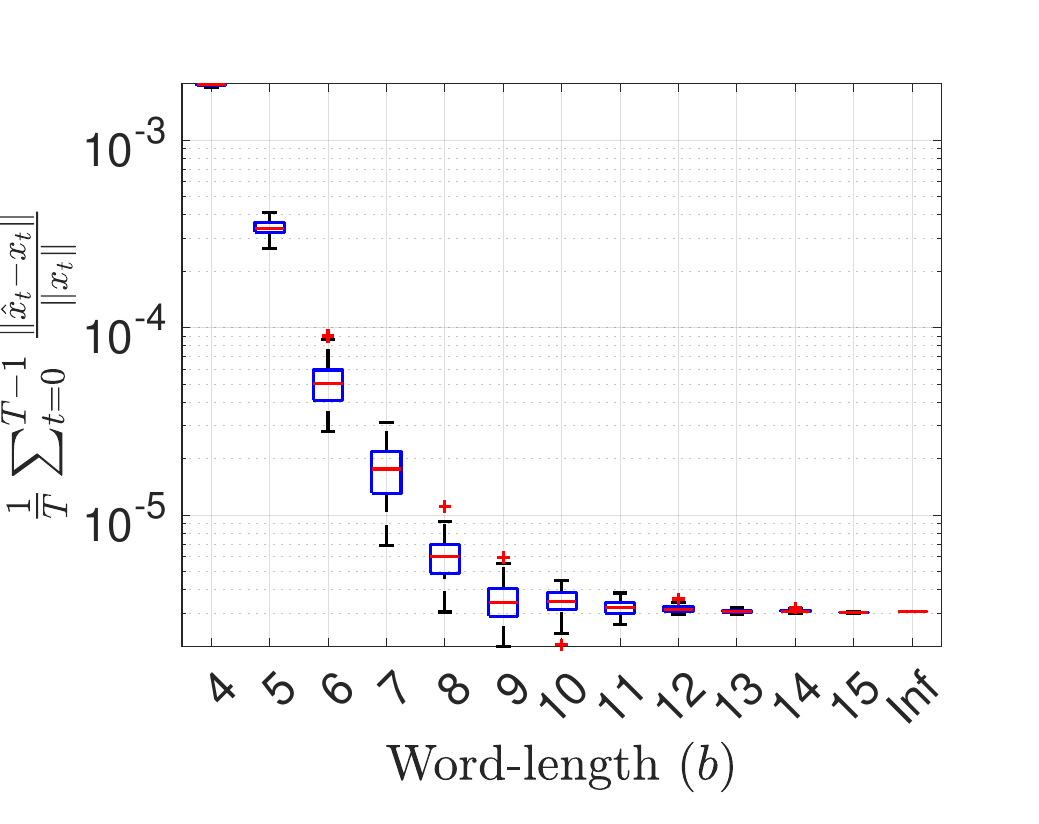}}
\subfloat[]{\includegraphics[trim=2cm 7cm 2cm 7cm, clip=true, width=0.25\textwidth]{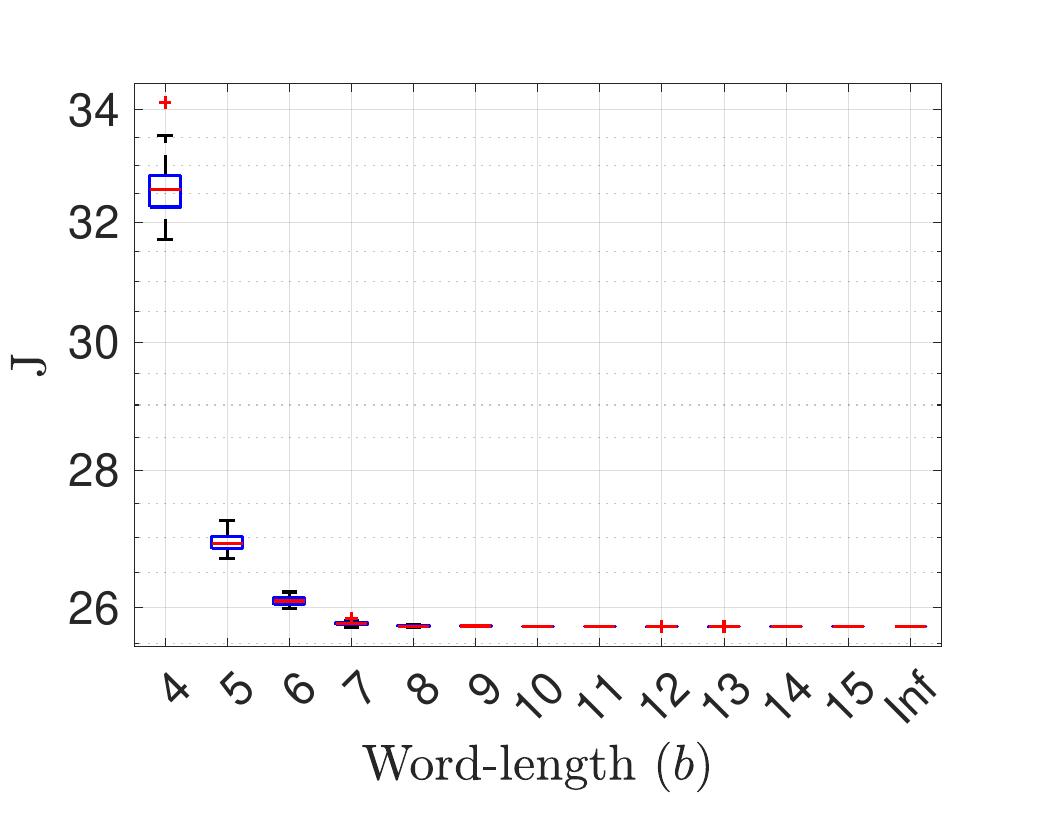}}\\
    \subfloat[]{\includegraphics[trim=2cm 7cm 2cm 7cm, clip=true, width=0.25\textwidth]{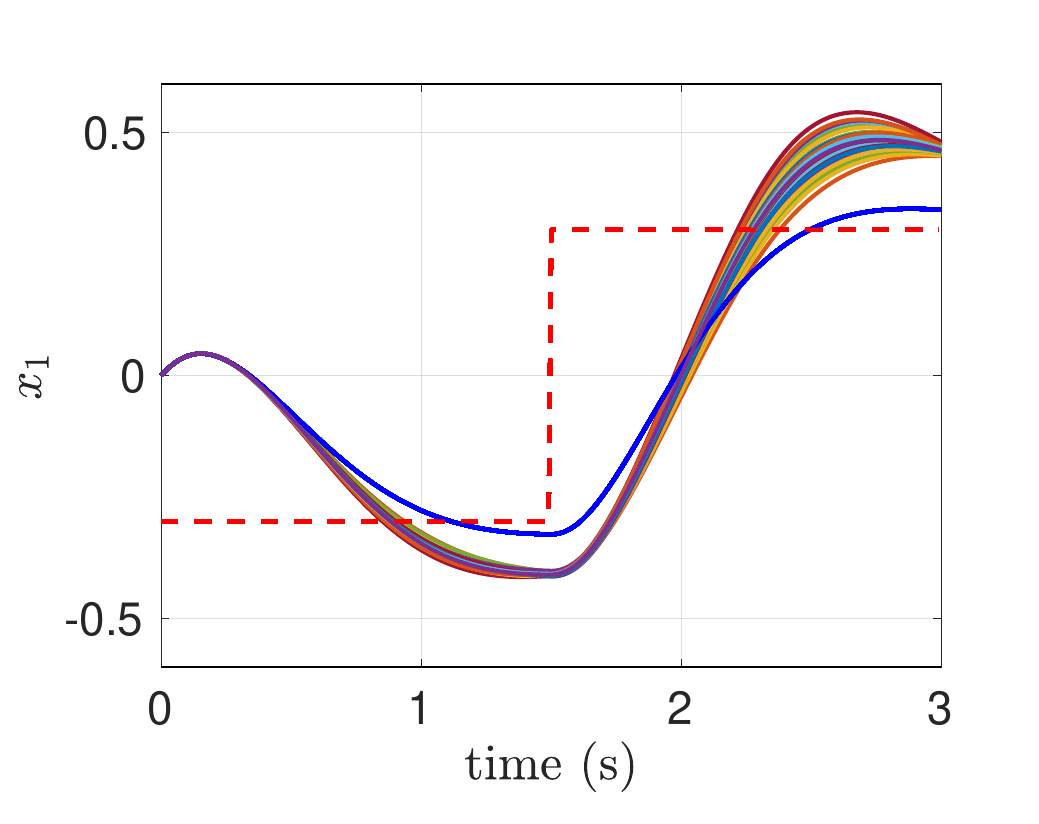}}
    \subfloat[]{\includegraphics[trim=2cm 7cm 2cm 7cm, clip=true, width=0.25\textwidth]{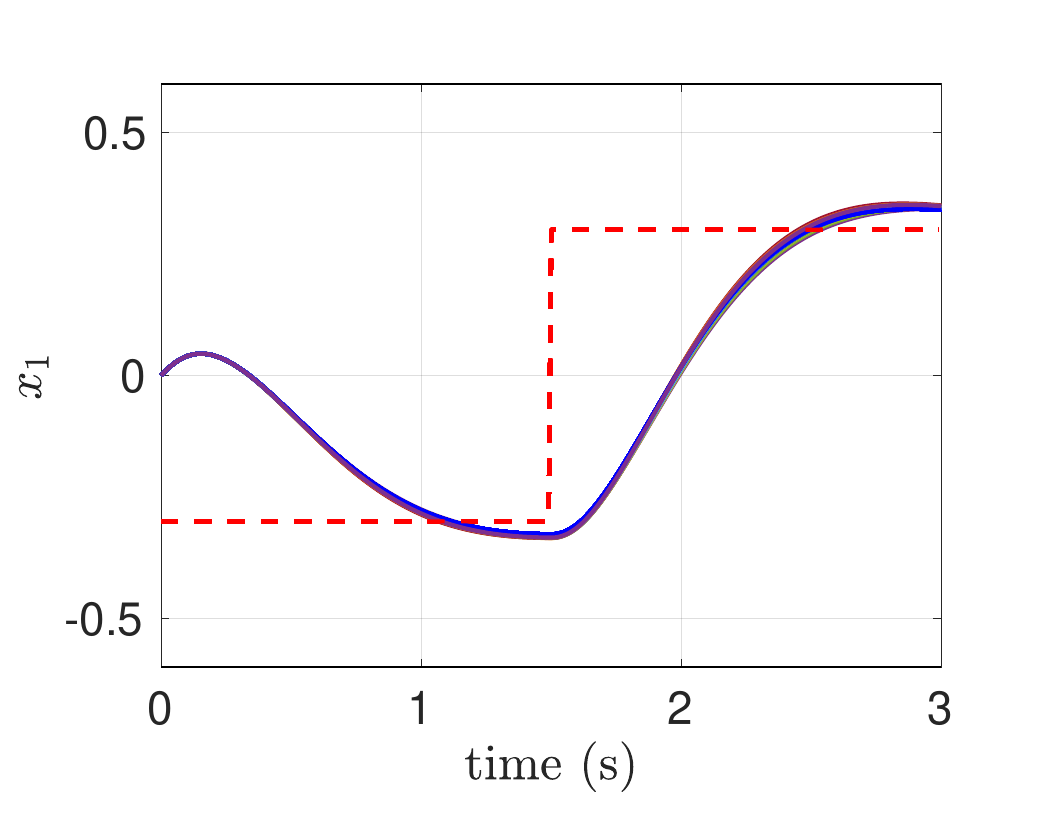}}
\subfloat[]{\includegraphics[trim=2cm 7cm 2cm 7cm, clip=true, width=0.25\textwidth]{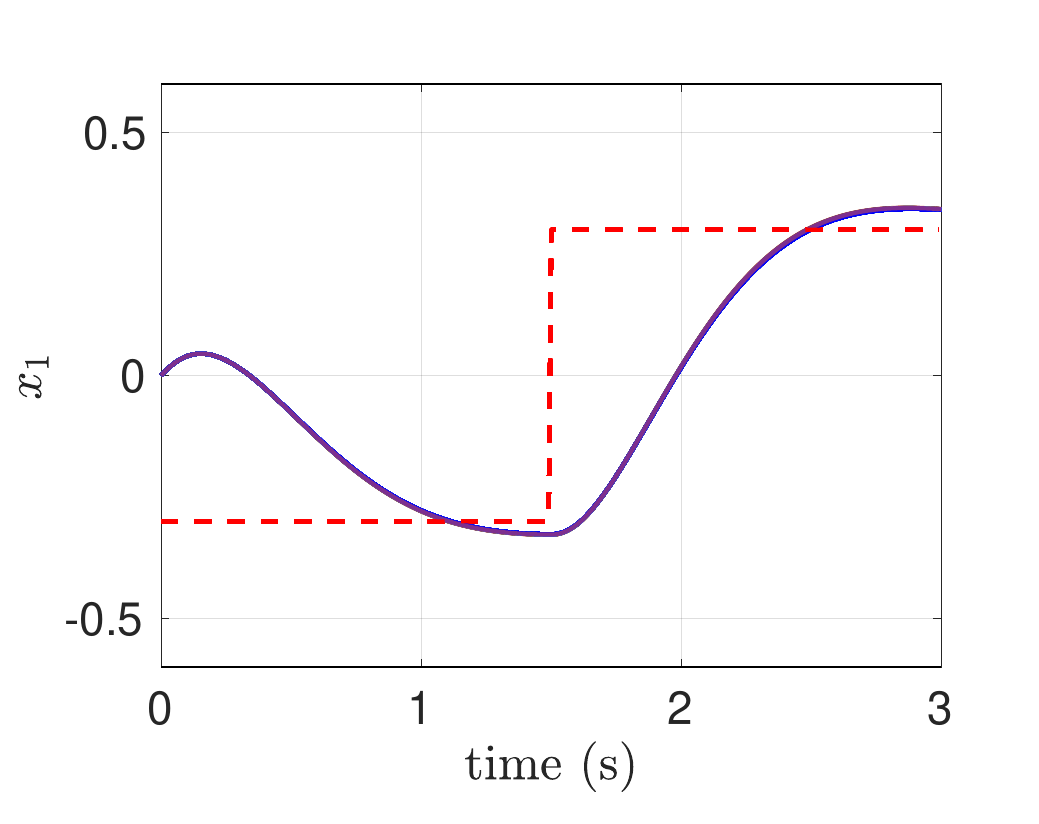}}
\subfloat[]{\includegraphics[trim=2cm 7cm 2cm 7cm, clip=true, width=0.25\textwidth]{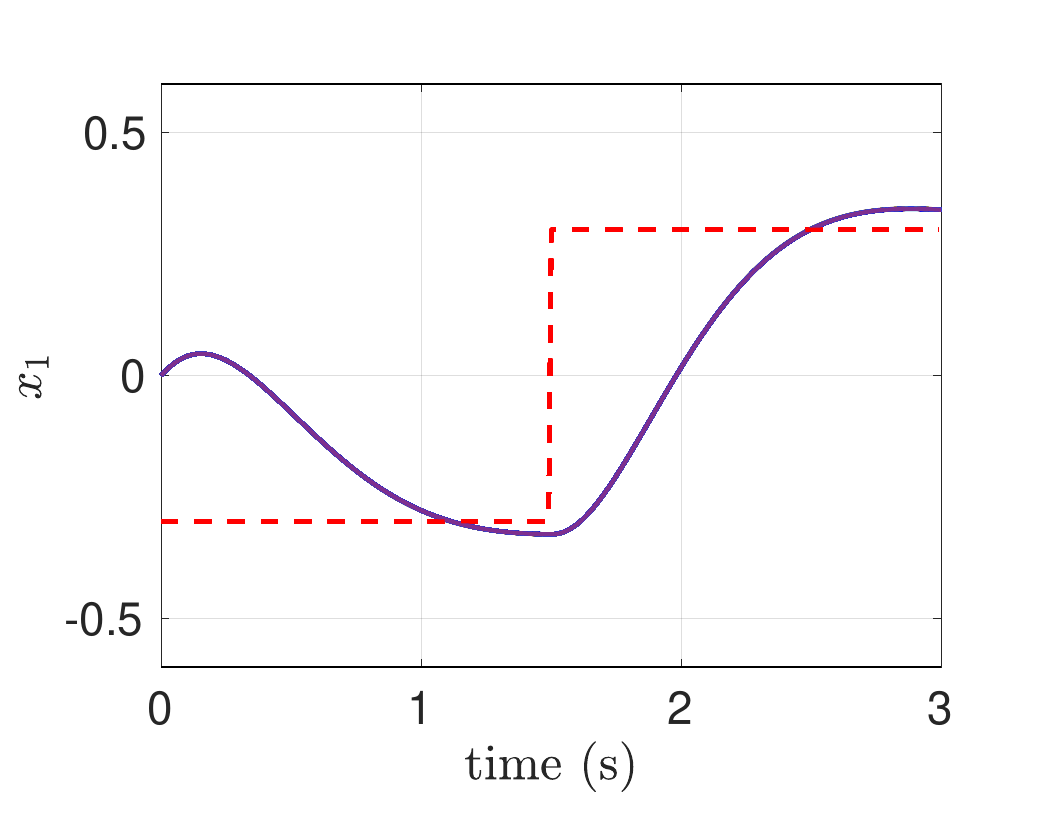}}
\caption{Error and prediction profile for a negatively-damped pendulum \eqref{Eq: Pendulum}: (a) relative error in matrix $A$; (b) relative error in matrix $B$; (c) time-averaged relative prediction error; (d) optimal cost achieved by the LMPC \eqref{eq:LMPC}; (e)--(h) LMPC tracking performance (with model identified from data snapshots quantized by 50 independent dither signal realization) for word lengths $b=4,\ 6, \ 8,\ 10$ respectively; dashed red line is the reference signal for LMPC.} \vspace{-1 em}
\label{Fig: Pendulum}
\end{figure*}

\subsection{Pendulum with negative damping}
A two-dimensional oscillatory system with slight instability is considered as a first example.
The dynamics of a simple pendulum with a destabilizing term is described by:
\bnl \label{Eq: Pendulum}
\dot{x}_1 &=& x_2\nonumber\\
\dot{x}_2&=&0.01x_2-\sin x_1 + u.
\enl
The dynamics is discretized using the fourth-order Runge-Kutta method with time-step $\Delta t = 0.01s$.
 The initial conditions are generated randomly with uniform distribution in the unit box $[-1, 1]^2$. The lifting functions $\varphi^i$ are chosen to be the state itself (i.e., $\varphi^1=x_1$, $\varphi^2=x_2$) and 100 thin plate spline radial basis functions with centers selected randomly with uniform distribution on the unit box\footnote{Thin plate spline radial basis function with center at $x_0$ is defined by $\psi(x) = \|x - x_0\|^2\log(\|x - x_0\|)$.}, leading to a lifted state space of dimension $N = 102$. The control input for each trajectory is chosen to be a uniformly distributed random signal on $[-1, 1]$. The system is simulated for 200 trajectories over 1000 sampling periods (i.e., 10 seconds per trajectory).

Relative 2-norm error $\frac{\| A - \tilde{A}\|}{\|A\|}$ and $\frac{\| B - \tilde{B}\|}{\|B\|}$ for linear predictor matrices, and time-average relative two norm error $\frac{1}{T}\sum_{t=0}^{T-1} \frac{\|\hat{x}_t - x_t \|}{\|x_t\|}$ of predictions using $[\tilde{A},\ \tilde{B}]$ for different word-length are shown in Fig.~\ref{Fig: Pendulum}(a)--(c). We notice that the logarithmic errors in $A$ and $B$ matrices decrease linearly with the word-length $b$ with a slope of $-0.395$ and $-0.292$. Note that, for a finite-data regime, the error should be $O(\epsilon) \approx k\epsilon \overset{\eqref{eq:quantizationResolution}}{=} k(u_{\max}-u_{\min})/2^b$ for some constant $k$ as $\epsilon \rightarrow 0$, i.e., the logarithm of the error should decrease linearly with $b$ at a slope of $-\log 2 = -0.301$. Moreover, the prediction error in Fig.~\ref{Fig: Pendulum}(c) decreases with the quantization word-length as expected. 

The identified $[\tilde{A},\ \tilde{B}]$ is then used to track a reference position $x_{1, ref}$ switching between two constant levels by solving the LMPC problem \eqref{eq:LMPC} with $Q = \operatorname{diag}([1\ 0])$, $R=0.01$, and $T_h = 100$, i.e., $1$s. The physical constraints on the control input and position $x_1$ are $u \in [-4, 4]$ and $x_1 \in [-0.6, 0.6]$ respectively. The minimum LMPC cost that is achieved for different word-lengths is demonstrated in Fig.~\ref{Fig: Pendulum} (d). As one would expect, achieved optimal cost decreases with higher quantization resolution, i.e., increased word-length, and asymptotically approaches the optimal LMPC cost achieved with the predictor identified from unquantized data. Fig.~\ref{Fig: Pendulum}(e)--(h) shows the controlled state-trajectories for $x_1$ with LMPC for different word-lengths. The tracking performance improves with higher quantization resolution, validating Fig.~\ref{Fig: Pendulum}(d).


\begin{figure*}[t]
\centering 
\subfloat[]{\includegraphics[trim=1.9cm 7cm 2cm 7cm, clip=true, width=0.25\textwidth]{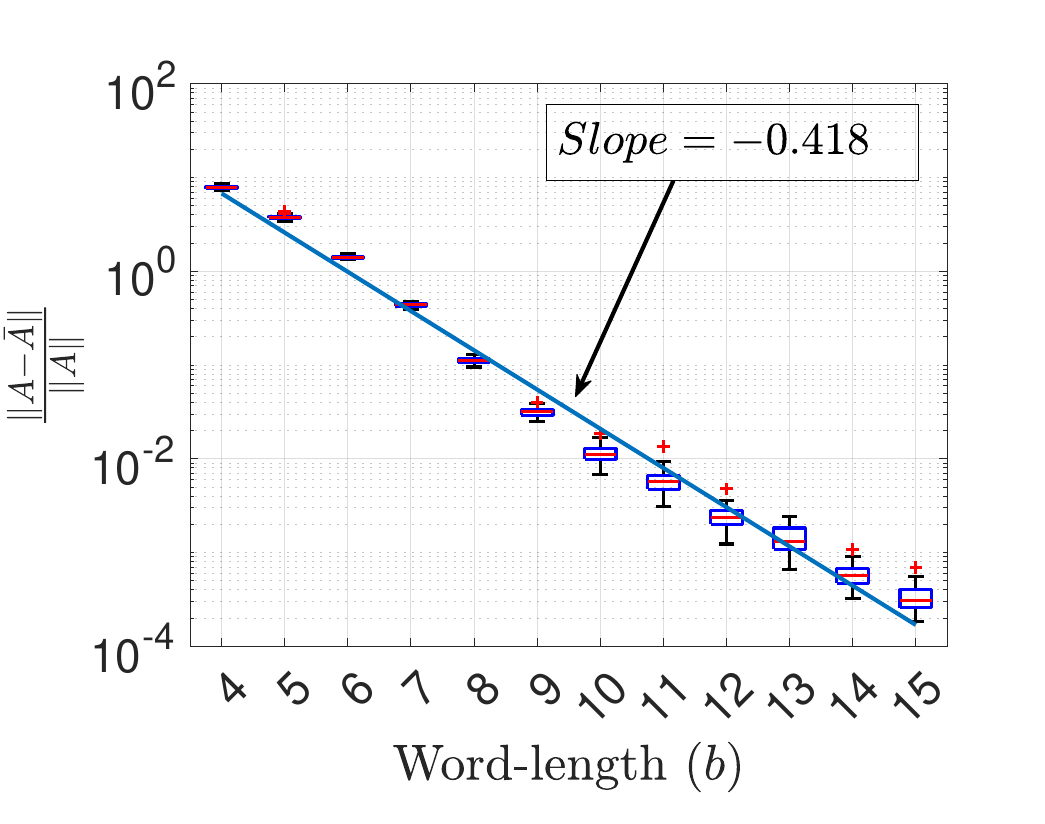}}
\subfloat[]{\includegraphics[trim=1.9cm 7cm 2cm 7cm, clip=true, width=0.25\textwidth]{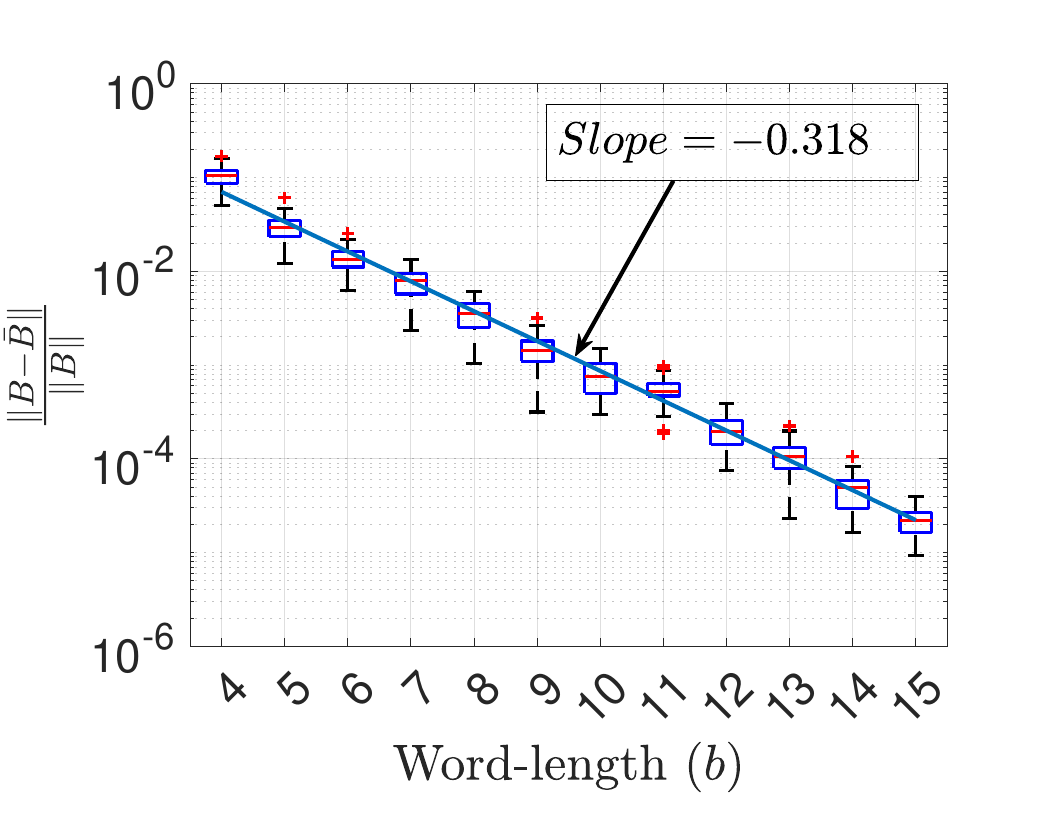}}
\subfloat[]{\includegraphics[trim=2cm 7cm 2cm 7cm, clip=true, width=0.25\textwidth]{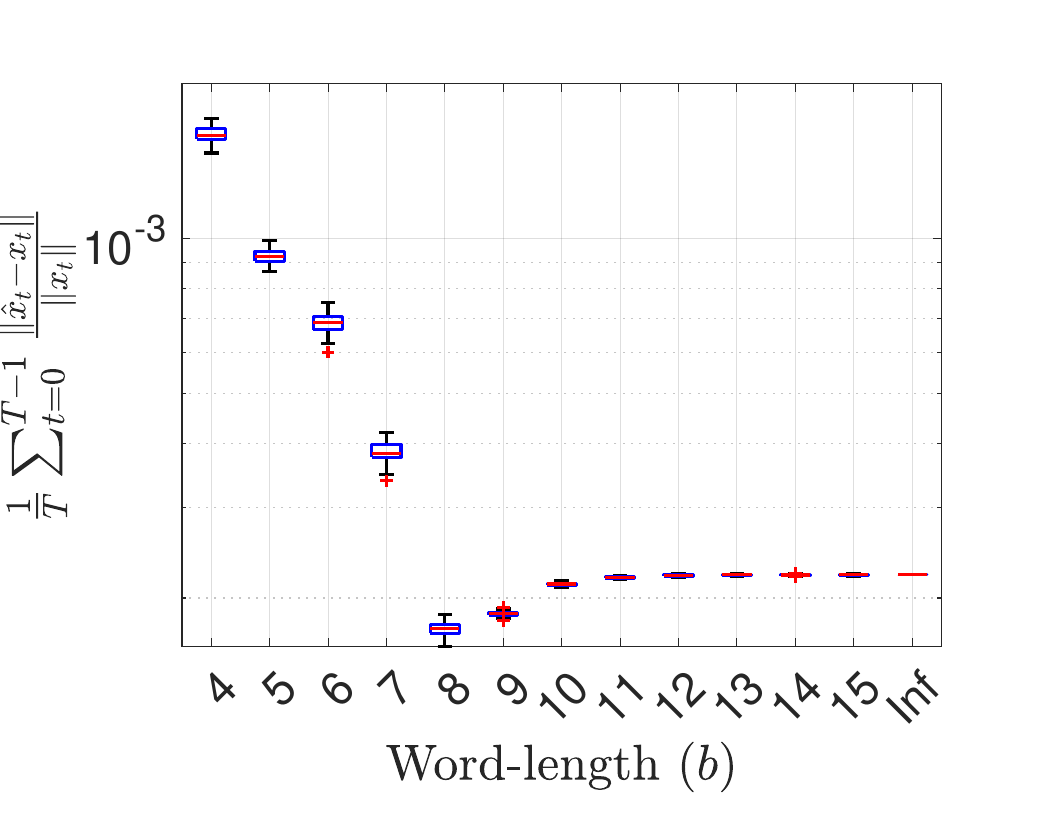}}
\subfloat[]{\includegraphics[trim=2cm 7cm 2cm 7cm, clip=true, width=0.25\textwidth]{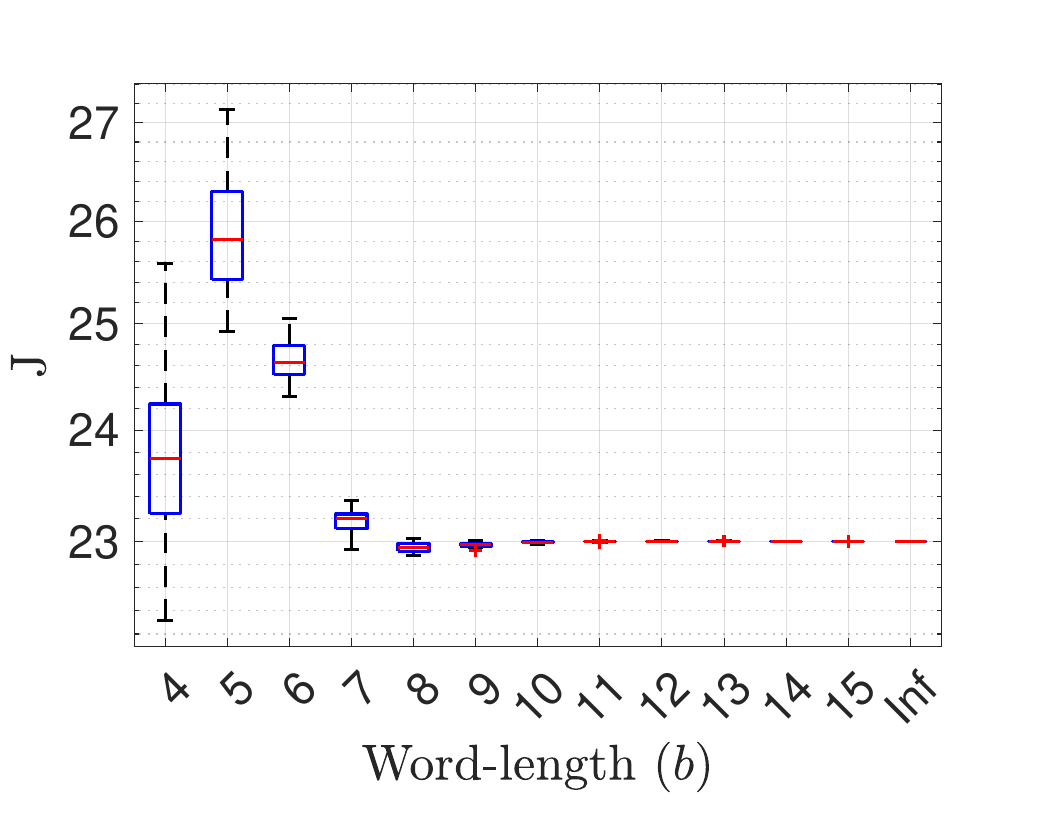}}\\
    \subfloat[]{\includegraphics[trim=2cm 7cm 2cm 7cm, clip=true, width=0.25\textwidth]{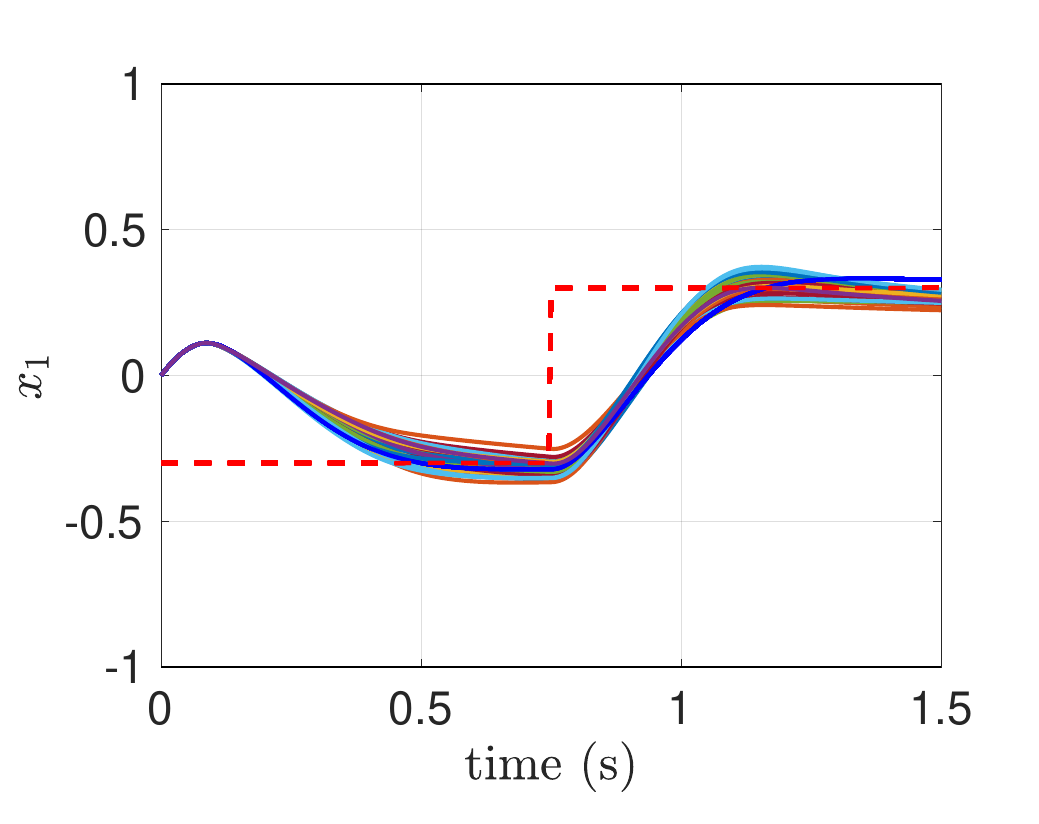}}
    \subfloat[]{\includegraphics[trim=2cm 7cm 2cm 7cm, clip=true, width=0.25\textwidth]{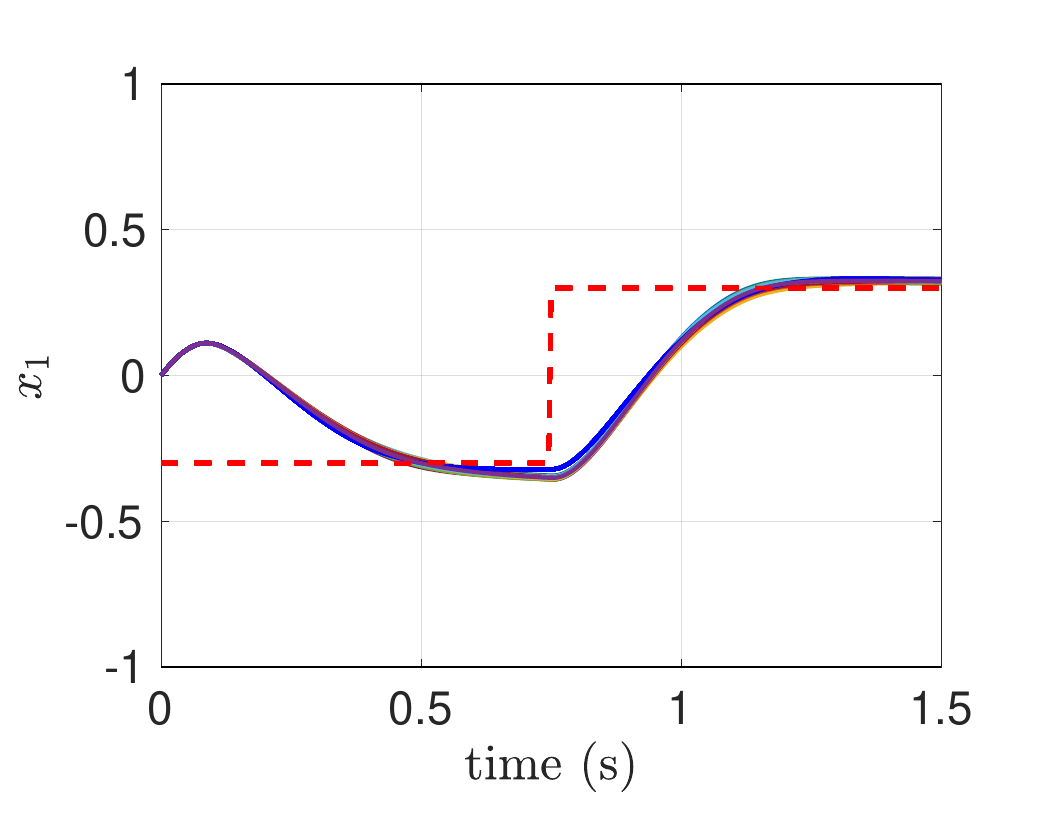}}
\subfloat[]{\includegraphics[trim=2cm 7cm 2cm 7cm, clip=true, width=0.25\textwidth]{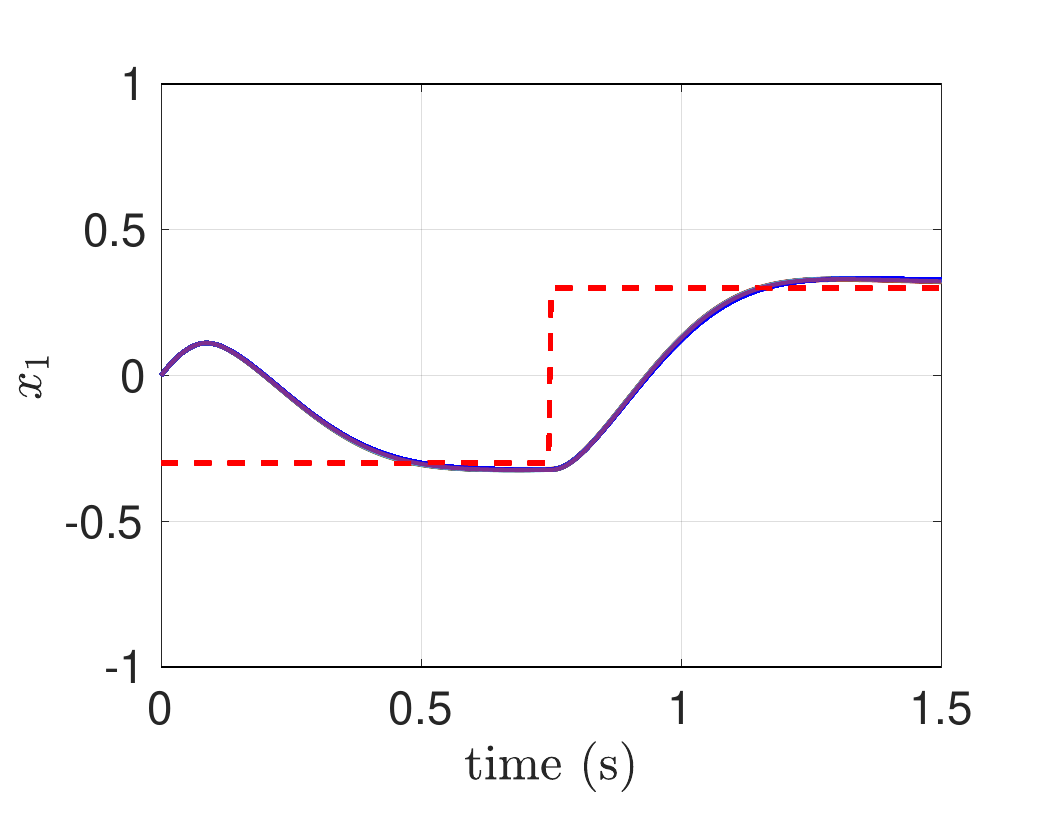}}
\subfloat[]{\includegraphics[trim=2cm 7cm 2cm 7cm, clip=true, width=0.25\textwidth]{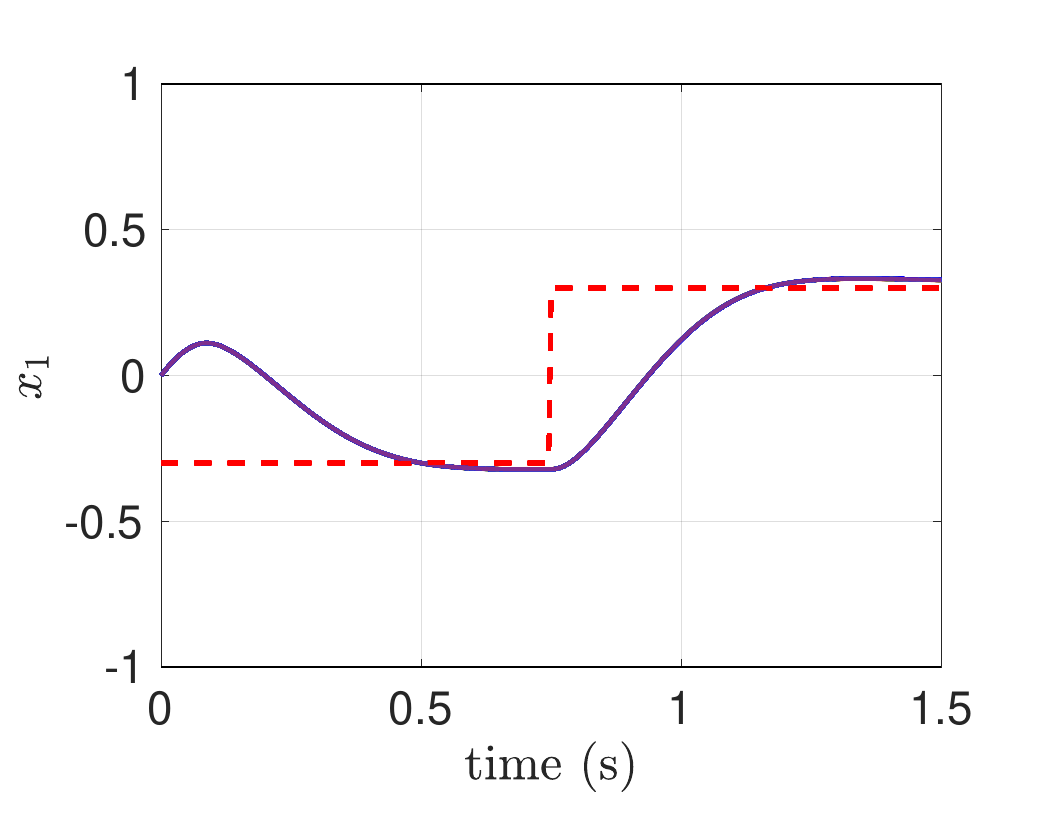}}
\caption{Error and prediction profile for Van der Pol oscillator \eqref{Eq: VdP}: (a) relative error in matrix $A$; (b) relative error in matrix $B$; (c) time-averaged relative prediction error; (d) optimal cost achieved by the LMPC \eqref{eq:LMPC}; (e)--(h) LMPC tracking performance (with model identified from data snapshots quantized by 50 independent dither signal realization) for word lengths $b=4,\ 6, \ 8,\ 10$ respectively; dashed red line is the reference signal for LMPC.}
\label{Fig: VanDerPol}
\end{figure*}

\begin{figure*}
\centering 
\subfloat[]{\includegraphics[trim=1.9cm 7cm 2cm 7cm, clip=true, width=0.25\textwidth]{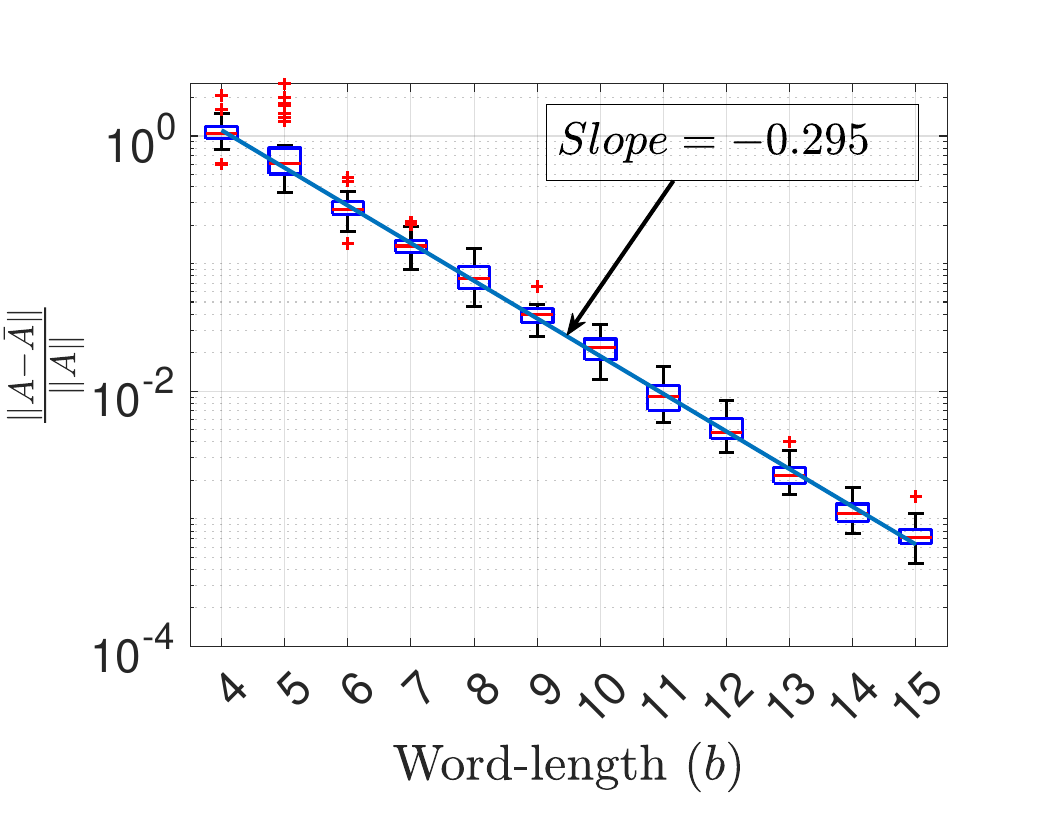}}
\subfloat[]{\includegraphics[trim=1.9cm 7cm 2cm 7cm, clip=true, width=0.25\textwidth]{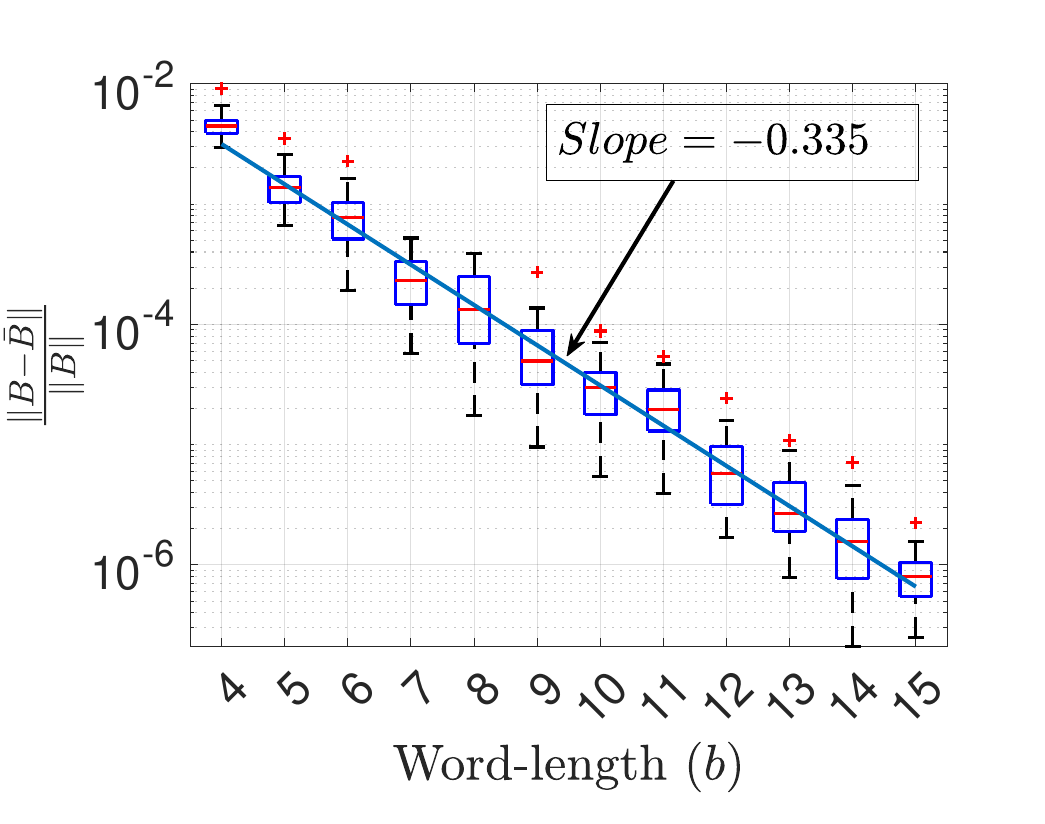}}
\subfloat[]{\includegraphics[trim=2cm 7cm 2cm 7cm, clip=true, width=0.25\textwidth]{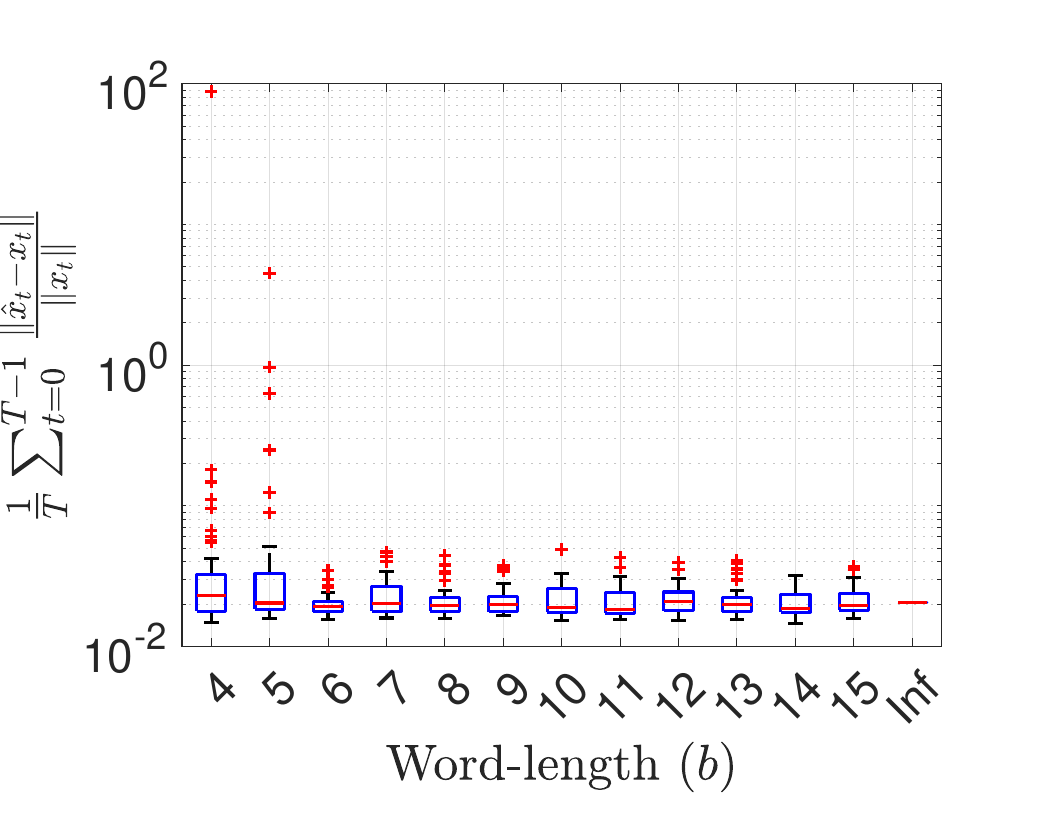}}
\subfloat[]{\includegraphics[trim=2cm 7cm 2cm 7cm, clip=true, width=0.25\textwidth]{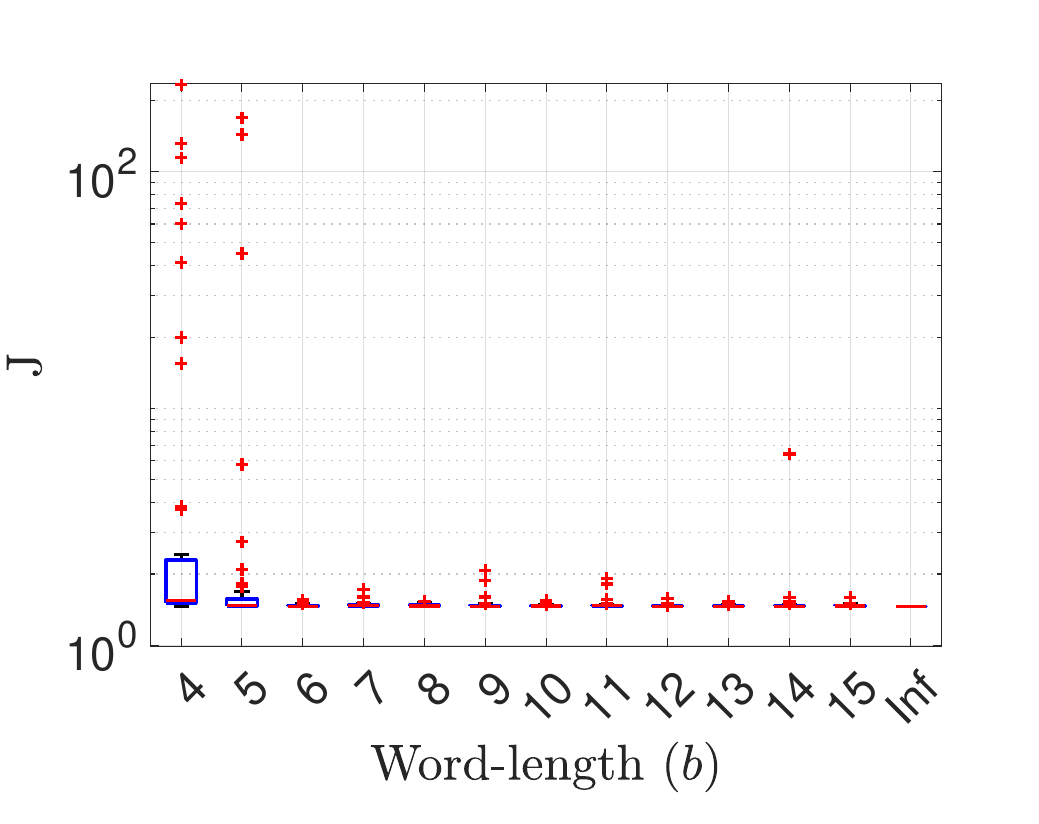}}\\
    \subfloat[]{\includegraphics[trim=2cm 7cm 2cm 7cm, clip=true, width=0.25\textwidth]{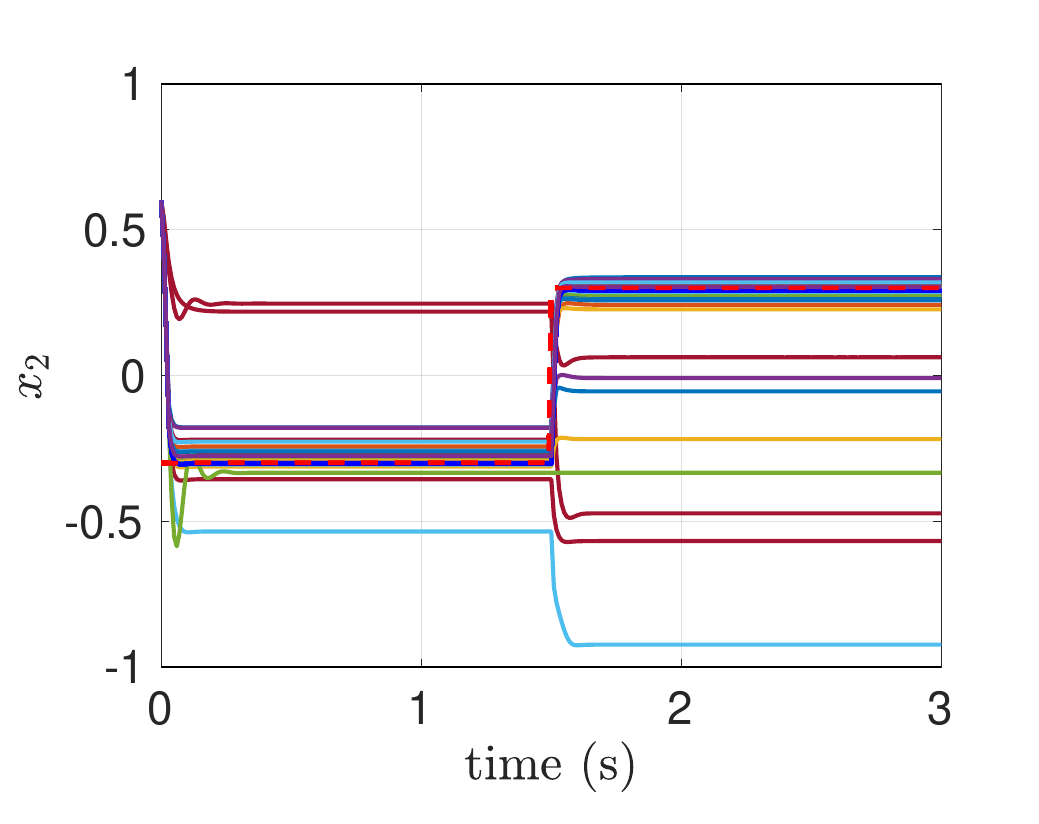}}
    \subfloat[]{\includegraphics[trim=2cm 7cm 2cm 7cm, clip=true, width=0.25\textwidth]{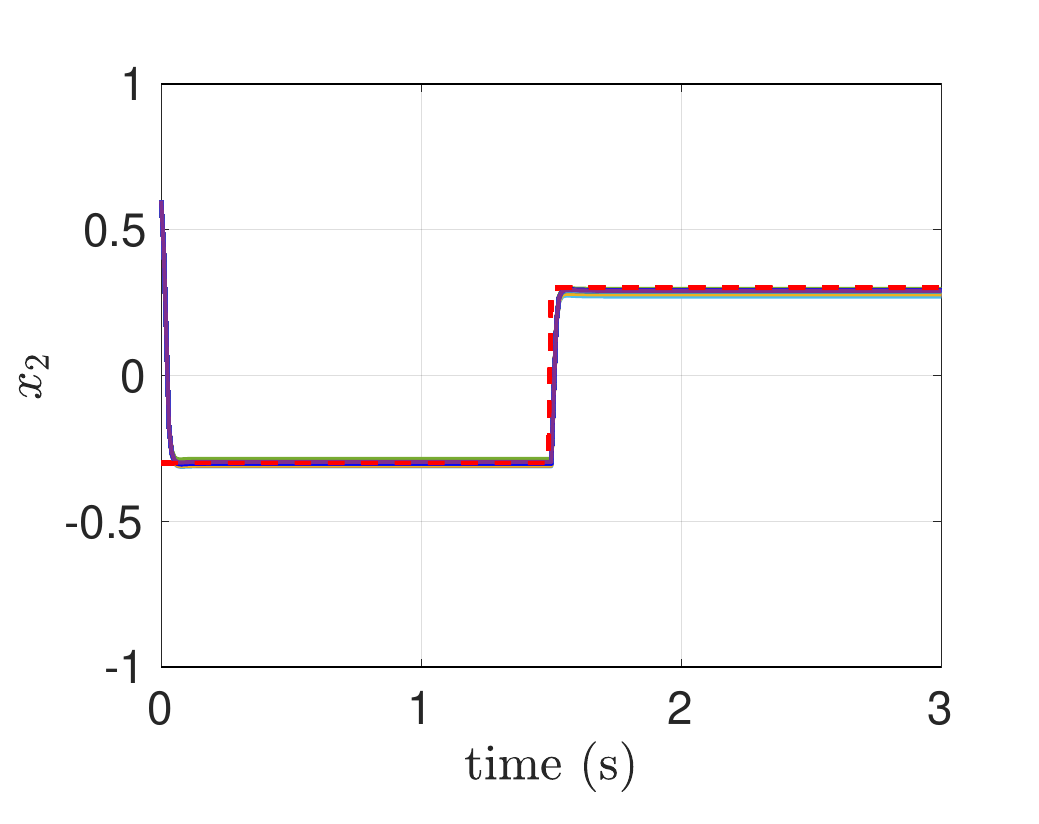}}
\subfloat[]{\includegraphics[trim=2cm 7cm 2cm 7cm, clip=true, width=0.25\textwidth]{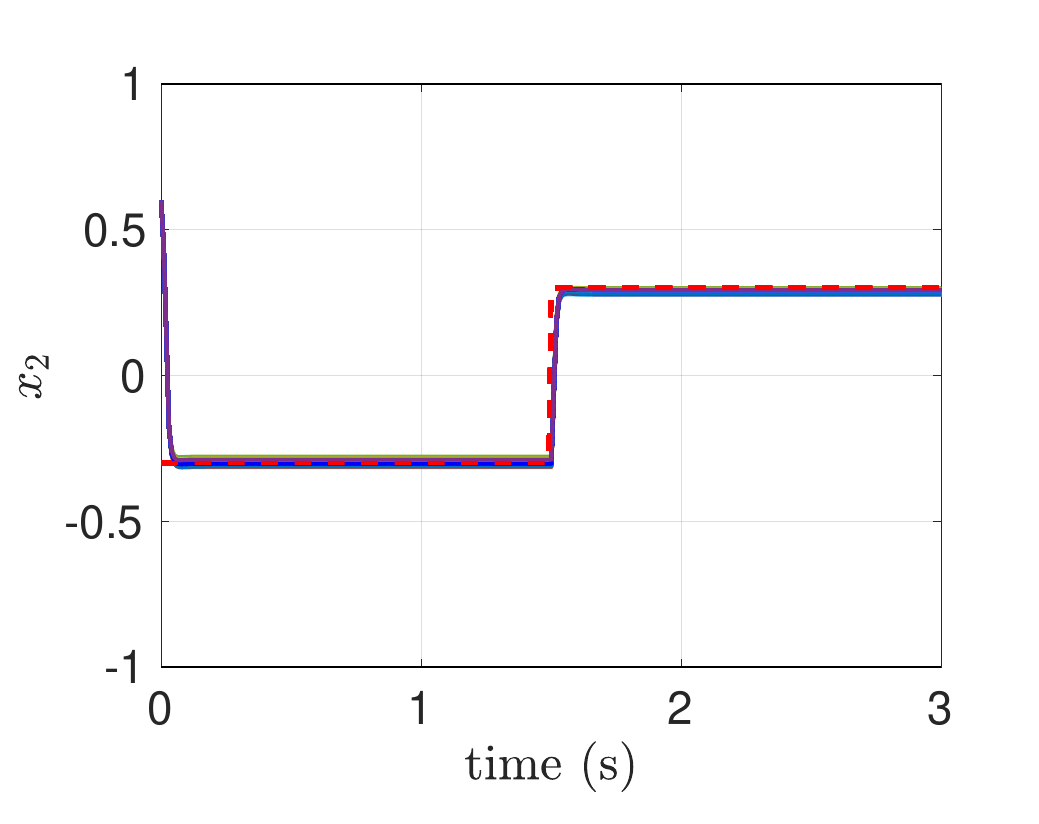}}
\subfloat[]{\includegraphics[trim=2cm 7cm 2cm 7cm, clip=true, width=0.25\textwidth]{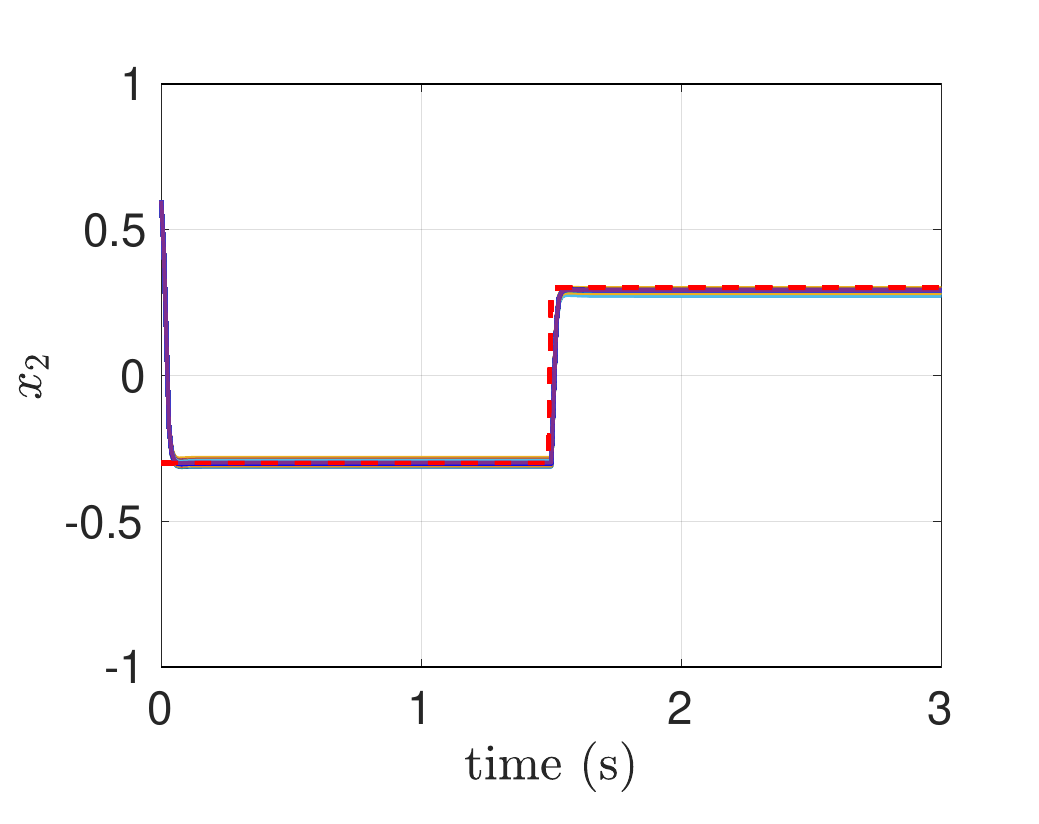}}
\caption{Error and prediction profile for motor \eqref{Eq: Motor}: (a) relative error in matrix $A$; (b) relative error in matrix $B$, (c) time-averaged relative prediction error; (d) optimal cost achieved by the LMPC \eqref{eq:LMPC}; (e)--(h) LMPC tracking performance (with model identified from data snapshots quantized by 50 independent dither signal realization) for word lengths $b=4,\ 6, \ 8,\ 10$ respectively; dashed red line is the reference signal for LMPC.}
\label{Fig: motor}
\end{figure*}

\subsection{Van der Pol oscillator}
The classical forced Van der Pol oscillator is examined for the second example. The following differential equations describe the dynamics of the system:
\begin{align}\label{Eq: VdP}
& \dot{x}_1=2 x_2 \nonumber \\
& \dot{x}_2=-0.8 x_1+2 x_2-10 x_1^2 x_2+u
\end{align}
Same sampling interval $\Delta t$, and training length $T$ are used to form the data matrices. The control input for each trajectory is a uniformly distributed random signal drawn from $[-1, 1]$. The simulation setting and lifting functions are same with the previous example of a negatively damped pendulum. The physical constraints on the control input and position $x_1$ are $u \in [-4, 4]$ and $x_1 \in [-1, 1]$ respectively. Fig.~\ref{Fig: VanDerPol} demonstrates the prediction and MPC performance with different word-length $b$. Fig.~\ref{Fig: VanDerPol}(a)-(c) shows similar trends for errors in linear predictor matrices $A$ and $B$, and the time-averaged prediction error. However, the model derived from data-snapshot quantized with a word-length $b=8$ outperforms the model derived from higher resolution. This effect might stem from the inherent regularization property of the quantization noise and should be investigated in future works. LMPC is used here with the identified linear predictor to track a reference position $x_{1, ref}$ with same $Q$, $R$, and $T_h$, as in the previous example. Fig.~\ref{Fig: VanDerPol}(d)--(h) also demonstrates a similar improvement in optimal cost and tracking performance of the LMPC with increasing word-length. Notably, LMPC performed better with a word-length $b=4$ than $b=5, 6$.

\subsection{Bilinear motor}
In this section, we apply the proposed approach to the control of a bilinear model of a DC motor. The model is derived from the nonlinear continuous-time dynamics of a thyristor-driven DC motor, as described in \cite{DANIELBERHE1998615}, and is represented by the following equations:
\begin{align} \label{Eq: Motor}
\begin{split}
& \dot{x}_1=-\left(R_a / L_a\right) x_1-\left(k_m / L_a\right) x_2 u+u_a / L_a \\
& \dot{x}_2=-(B / J) x_2+\left(k_m / J\right) x_1 u-\tau_l / J
\end{split}
\end{align}
where $x_1$ and $x_2$ are the armature current and angular velocity of the armature-shaft, $u$ is the input field current, $R_a, L_a, u_a$ are armature resistance, armature inductance and armature constant terminal voltage, and $B, J, \tau_l$ are viscous friction constant, moment of inertia and constant load torque. We use the following values of the motor parameters: $L_a = 0.314$, $R_a = 12.345$, $km = 0.253$, $J = 0.00441$, $B = 0.00732$, $\tau_l = 1.47$, and $u_a = 60$. The simulation and training settings remain the same as that of the pendulum and Van der Pol example. Note that, here, field current $u$ is our control input. The physical constraints on the control input and velocity $x_2$ are $u\in[-2,\ 2]$ and $x_2\in[-1,\ 1]$.
Fig.~\ref{Fig: motor}(a)-(c) shows similar trends for errors in linear predictor matrices $A$ and $B$, and the time-averaged prediction error.
LMPC is used with the identified linear predictor to track a reference angular velocity $x{2, ref}$ here with $Q = \operatorname{diag}([0\ 1])$, and same $R$ and $T_h$, as in the previous examples. Fig.~\ref{Fig: motor}(d)--(h) also demonstrates a similar improvement in optimal cost and tracking performance of the LMPC with increasing word-length.

\begin{figure*}
\centering 
\subfloat[]{\includegraphics[trim=1.9cm 7cm 2cm 7cm, clip=true, width=0.25\textwidth]{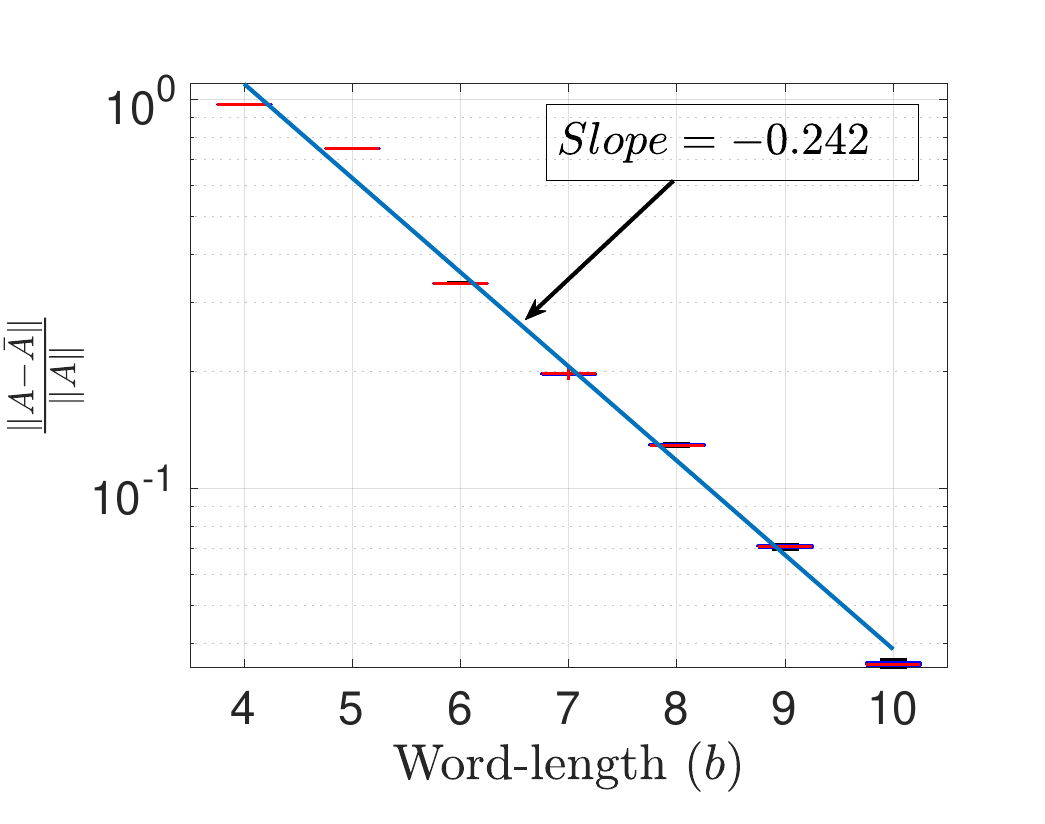}}
\subfloat[]{\includegraphics[trim=1.9cm 7cm 2cm 7cm, clip=true, width=0.25\textwidth]{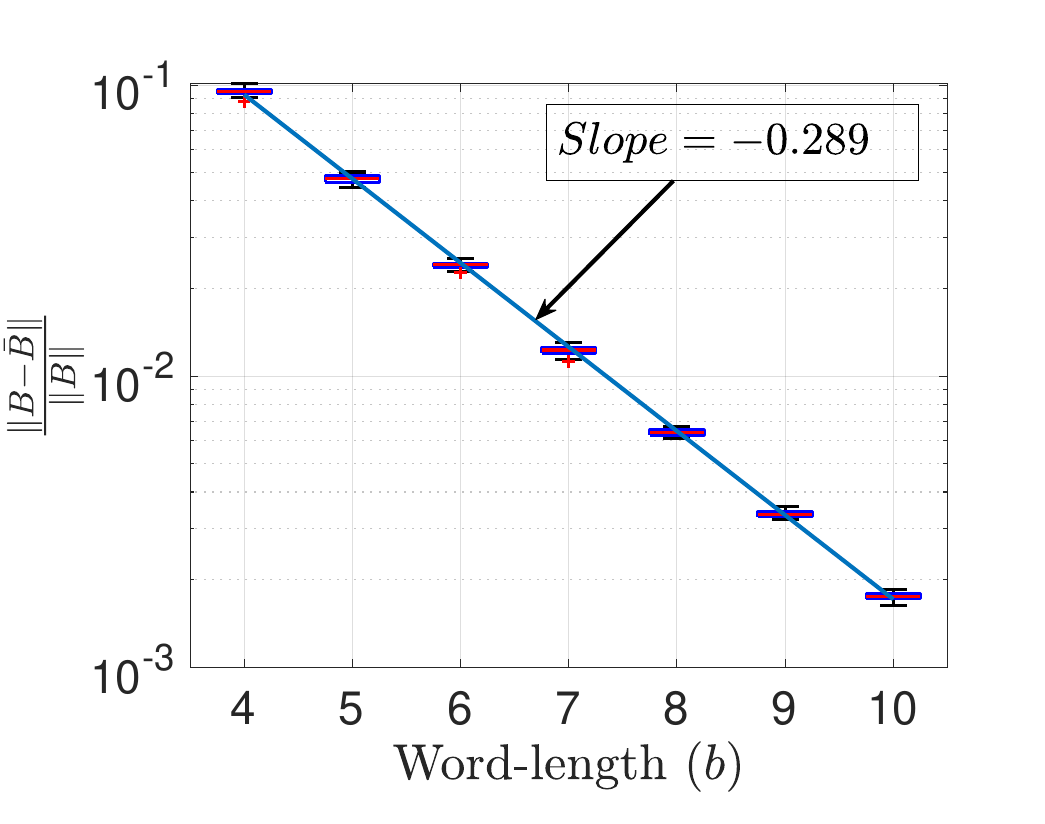}}
\subfloat[]{\includegraphics[trim=2cm 7cm 2cm 7cm, clip=true, width=0.25\textwidth]{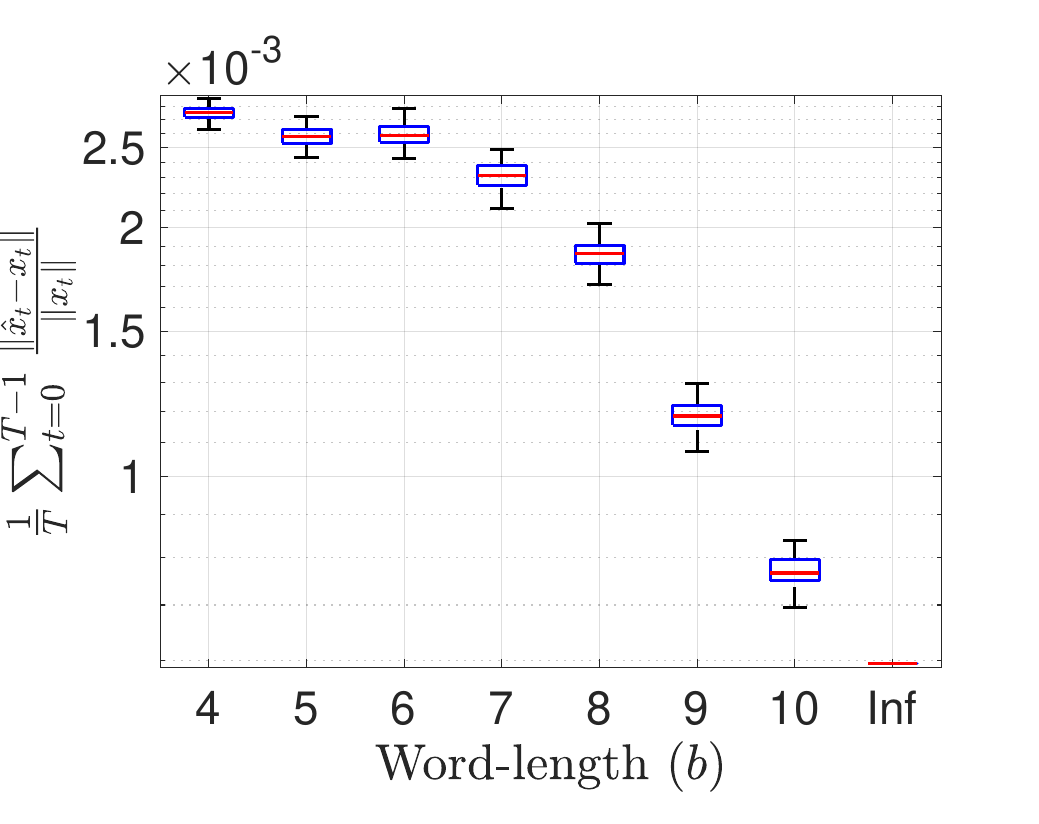}}
\subfloat[]{\includegraphics[trim=2cm 7cm 2cm 7cm, clip=true, width=0.25\textwidth]{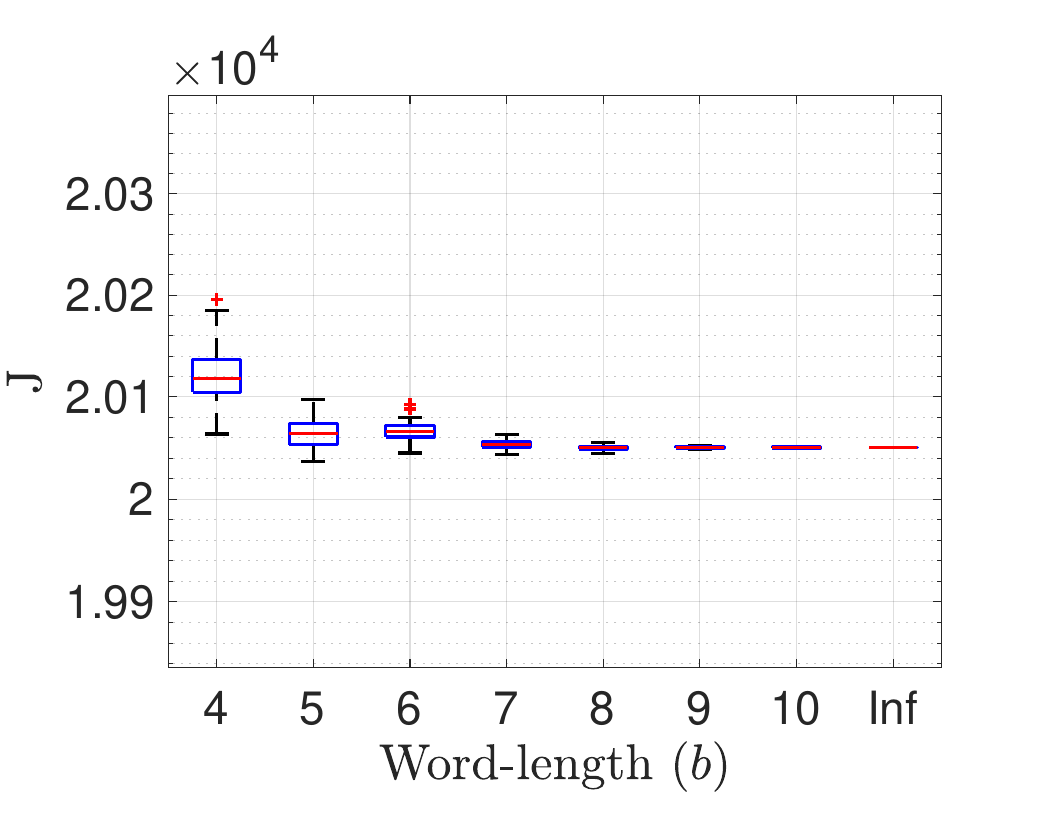}}\\
    \subfloat[]{\includegraphics[trim=1.9cm 7cm 2cm 7cm, clip=true, width=0.25\textwidth]{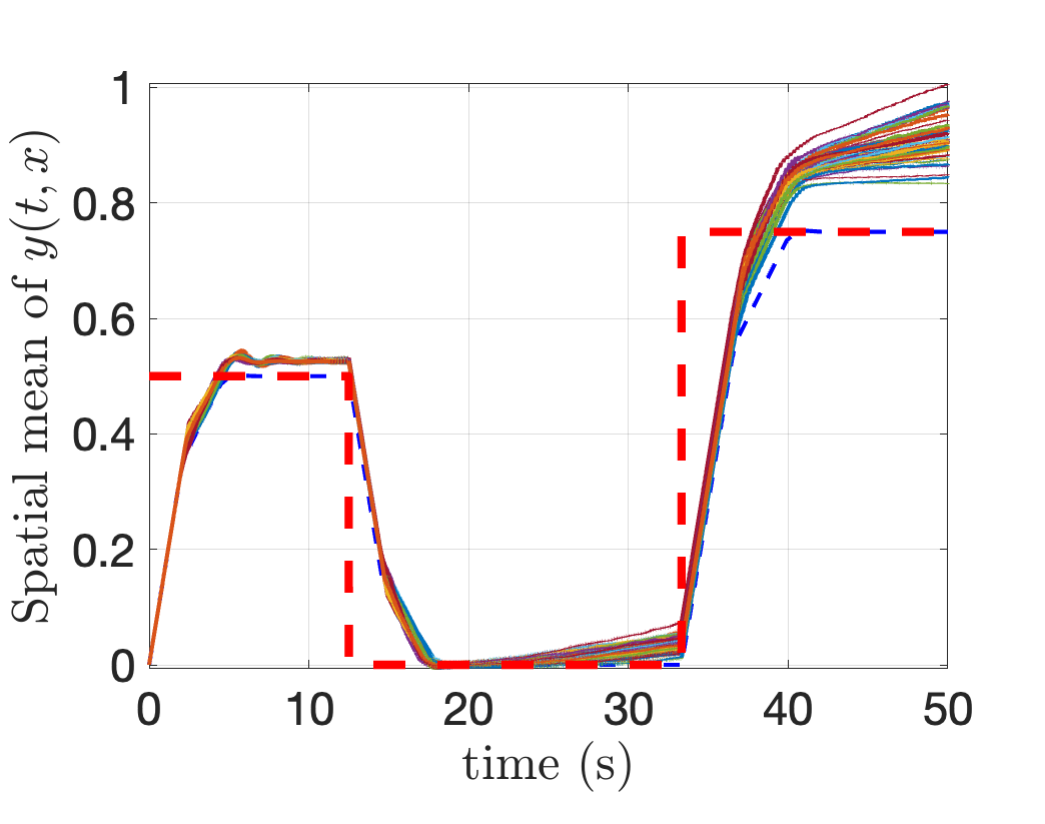}}
\subfloat[]{\includegraphics[trim=1.9cm 7cm 2cm 7cm, clip=true, width=0.25\textwidth]{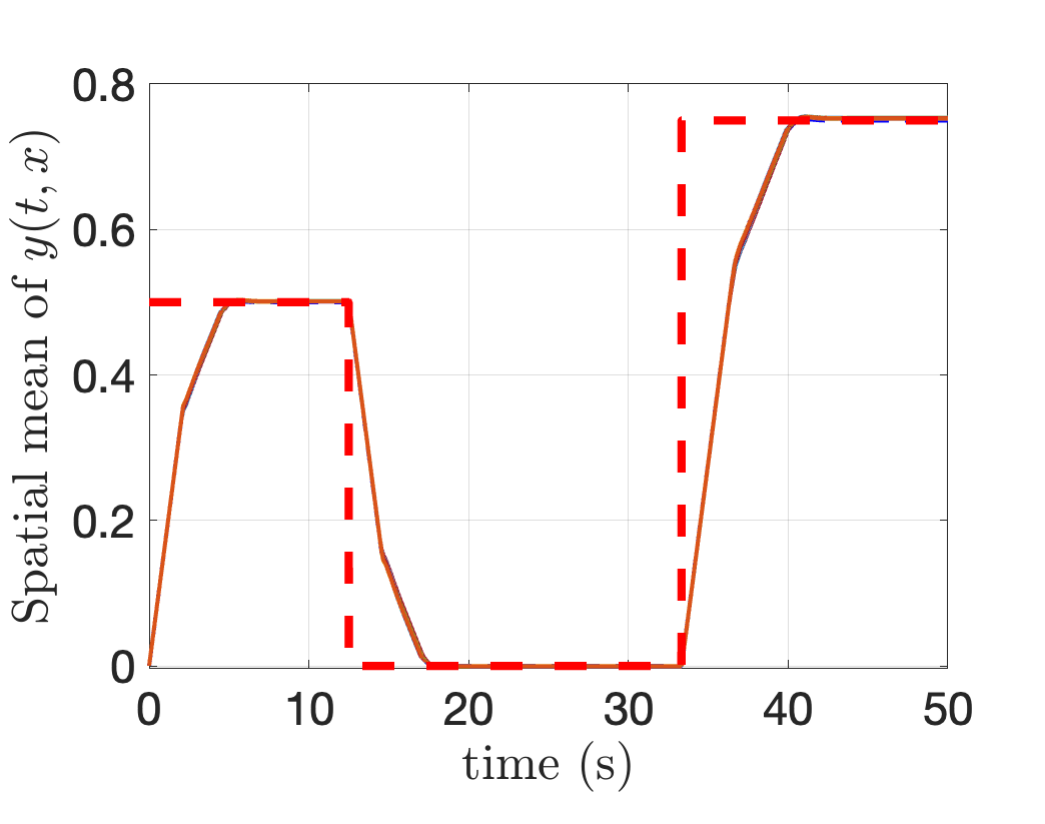}}
\subfloat[]{\includegraphics[trim=2cm 7cm 2cm 7cm, clip=true, width=0.25\textwidth]{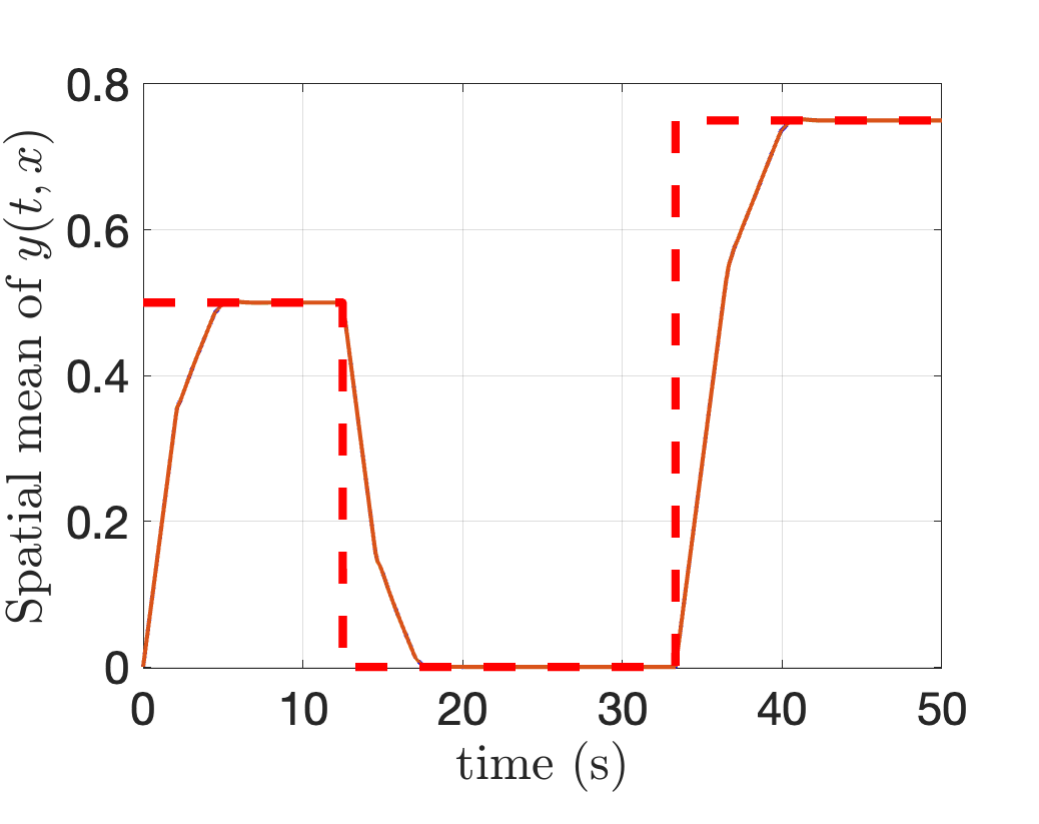}}
\subfloat[]{\includegraphics[trim=2cm 7cm 2cm 7cm, clip=true, width=0.25\textwidth]{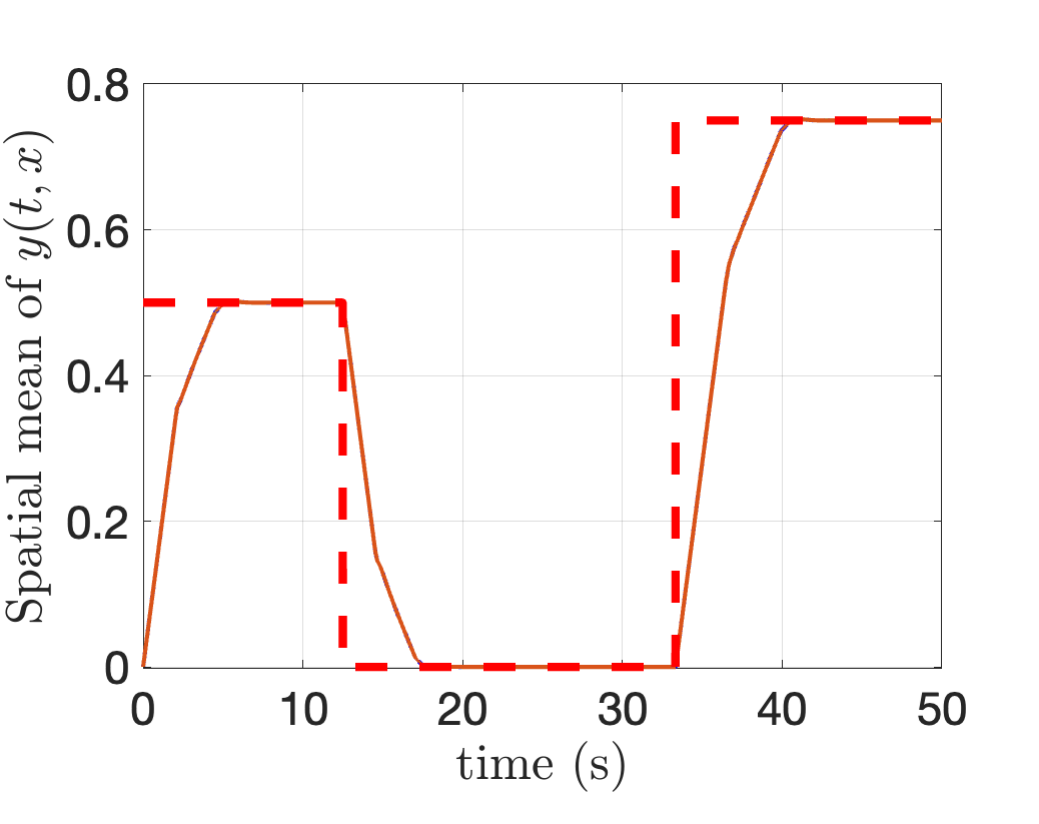}}\\
\subfloat[]{\includegraphics[trim=1.9cm 7cm 2cm 7cm, clip=true, width=0.25\textwidth]{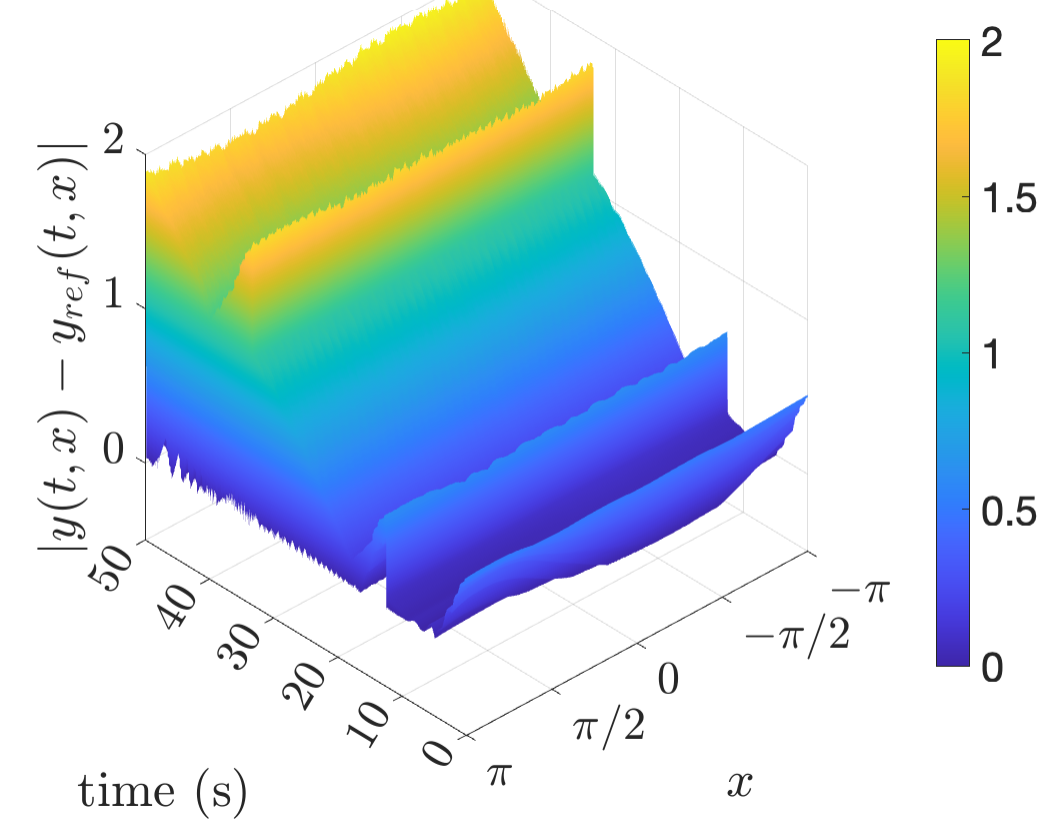}}
\subfloat[]{\includegraphics[trim=1.9cm 7cm 2cm 7cm, clip=true, width=0.25\textwidth]{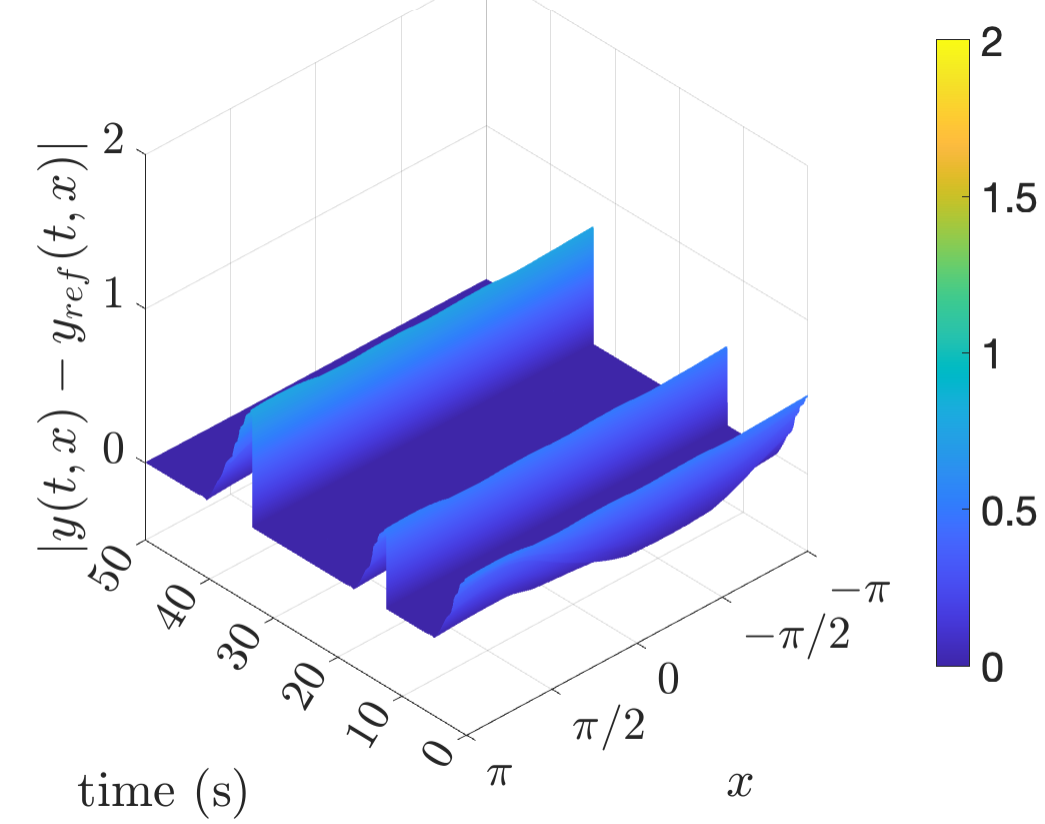}}
\subfloat[]{\includegraphics[trim=2cm 7cm 2cm 7cm, clip=true, width=0.25\textwidth]{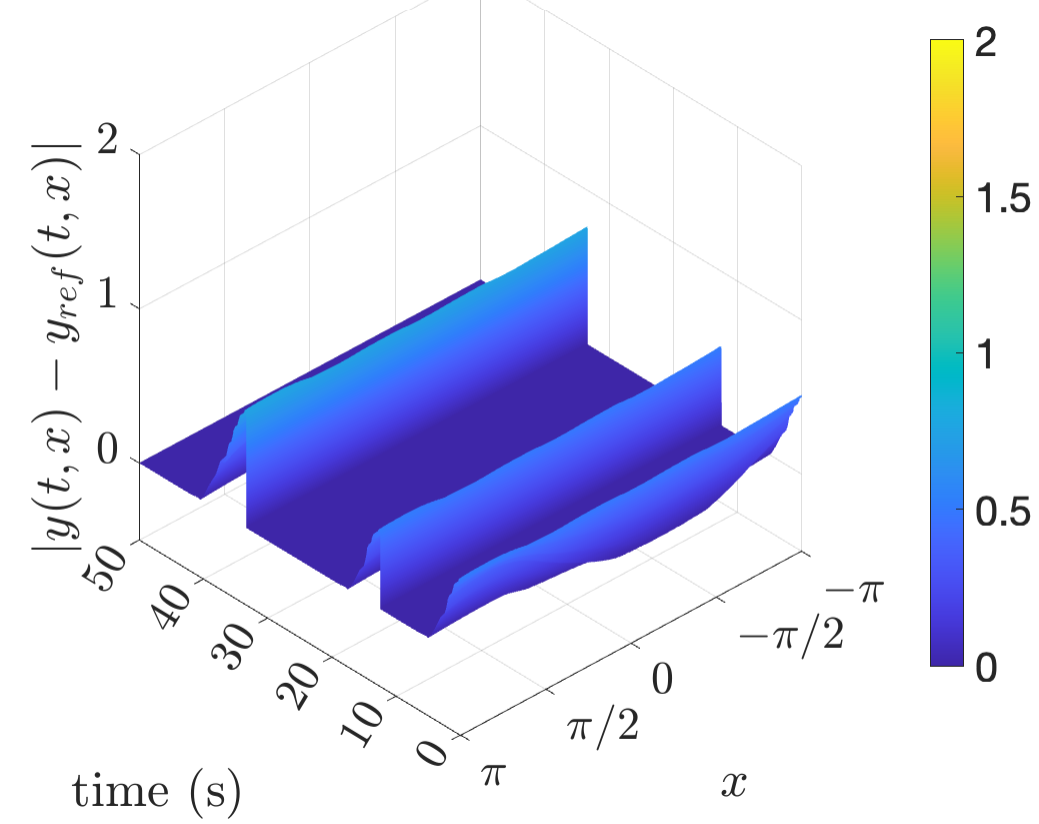}}
\subfloat[]{\includegraphics[trim=2cm 7cm 2cm 7cm, clip=true, width=0.25\textwidth]{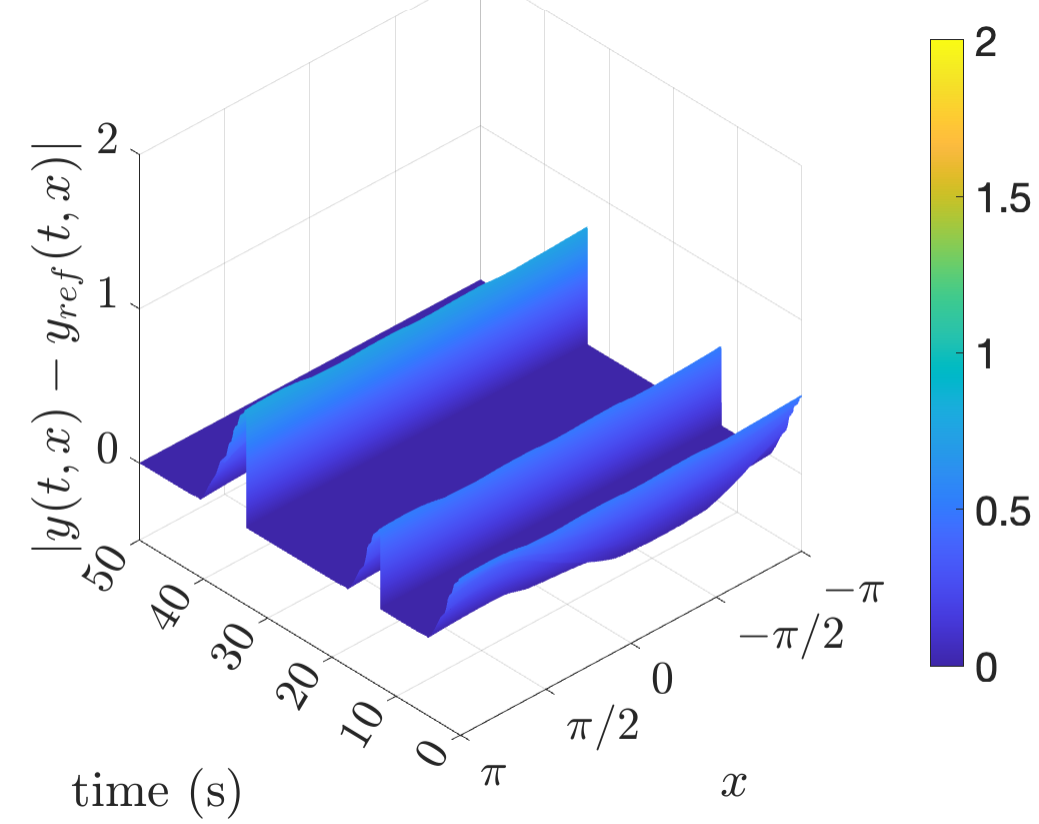}}

\caption{Error and prediction profile for KdV equation \eqref{eq:KDV_plant}: (a) relative error in matrix $A$, (b) relative error in matrix $B$, (c) time-averaged relative prediction error, (d) optimal cost achieved by the LMPC \eqref{eq:LMPC}, (e)--(h) LMPC tracking performance (with model identified from data snapshots quantized by 50 independent dither signal realization) for word lengths $b=4,\ 6, \ 8,\ 10$ respectively; dashed red line is the spatial mean of the reference signal for LMPC; (i)--(l) spatiotemporal LMPC tracking error for the same.}
\label{Fig: kdv}
\end{figure*}

\subsection{Korteweg-de Vries nonlinear PDE}
For the last example, the proposed method is applied to the
nonlinear Korteweg-deVries (KdV) equation:
\begin{align} \label{eq:KDV_plant}
    \frac{\partial y(t, x)}{\partial t}+y(t, x) \frac{\partial y(t, x)}{\partial x}+\frac{\partial^3 y(t, x)}{\partial x^3}=u(t, x)
\end{align}
where $y(t,x)$ and $u(t,x)$ the unknown function and control input, respectively. We assume periodic boundary conditions on the spatial domain $x \in[-\pi, \pi]$. The equation is discretized with time discretization of $\Delta t=0.01 \mathrm{~s}$ and spatial mesh of 128 points. The control input 
$u$ is structured as $u(t, x)=\sum_{i=1}^3 u_i(t) v_i(x)$ where  $v_i$ are predefined spatial profiles given by $v_i(x)=e^{-25\left(x-c_i\right)^2}$ with $c_1=-\pi / 2, c_2=0, c_3=\pi / 2$, and $u_i(t)$ are generated by control algorithm. The control inputs are constrained to $u_i(t) \in[-1,1]$. For EDMD, we simulate 200 trajectories over 1000 sampling periods. The initial conditions of these trajectories are random convex combinations of three fixed spatial profiles: $y_0^1=e^{-(x-\pi / 2)^2}, y_0^2=-\sin (x / 2)^2, y_0^3=e^{-(x+\pi / 2)^2}$; The control input for each trajectory is a random signal uniformly distributed on $[-1, 1]$. The observable functions $\varphi^i$ are chosen to be the state itself (i.e., $\varphi^i=x_i$), the element-wise square of the state (i.e., $\varphi^i=x_i^2$), the element-wise product of the state with its periodic shift (i.e., $\varphi^i=x_i\cdot x_{i+1}$) and the constant function, resulting in the dimension of the lifted state $N=3 \cdot 128+1=385$. 
Fig.~\ref{Fig: kdv}(a)-(c) shows the similar trends for errors in linear predictor matrices $A$ and $B$, and the time-averaged prediction error.
The control objective is to track a constant-in-space reference that varies in time in a piecewise constant manner. In order to do so, we utilize the LMPC \eqref{eq:LMPC} $Q = I,\ R = 0$, and the prediction horizon $T_h = 10$, i.e., $0.1$s.
Fig.~\ref{Fig: kdv}(d)--(l) demonstrates a similar improvement in optimal cost and tracking performance of the LMPC with increasing word-length.

\section{Special Case: Quantized Observables} \label{sec:SpecialCase}
While the previous analysis assumes the quantized data ($\tilde{x}_t$) is available, here we consider the case where the observable itself is quantized. 
That is, we consider the scenario where the $i$-th observable function becomes:
\begin{align}
    \tilde{\varphi}^i(x_t) = \Q(\varphi^i(x_t) + w^i_t ) - w^i_t. \label{eq:tildeObservablesspecialCase}
\end{align}
The quantization affects differently here compared to the previous case of $\bar{\varphi}_i(x_t)$ in \eqref{eq:tildeObservables}. 
Most surprisingly, some intrinsic properties of the observable functions (e.g., their gradients) which previously influenced the identification process (e.g., see \Cref{assm:BoundedPhi_derivative}, $\beta(\epsilon)$ and $\Gamma(\epsilon)$ in \Cref{thm:equivalence}), now do not have any such influence whatsoever.

\begin{thm}[Large data regime result]
\label{thm:equivalence:specialCase}
    As $T \to \infty$, $[\tilde{A}, \tilde{B}]$ converges almost surely to the solution of the following regularized least-square optimization   \vspace{ -1mm}
    \begin{align} \label{eq:equivalence_spl_case}
    \begin{split}
        &\underset{\substack{\mathcal{A} \in \R^{N \times N}\\ \mathcal{B}\in \R^{N\times m}}}{\min} \limsup_{T\to \infty}\frac{1}{T}  \| \Phi^{+} - \mathcal{A} \Phi - \mathcal{B}U\|^2 +  \frac{\epsilon^2}{12}\tr(\mathcal{G}^\top \mathcal{G}) \\
        &= \underset{\substack{\mathcal{A} \in \R^{N \times N}\\ \mathcal{B}\in \R^{N\times m}}}{\min} \limsup_{T\to \infty}\frac{1}{T}  \| \Phi^{+} - \mathcal{A} \Phi - \mathcal{B}U\|^2 +  \frac{\epsilon^2}{12}(\|\A\|^2 + \|\B\|^2)
        \end{split}\vspace{-2mm}
    \end{align} 
    where $\mathcal{G} = [\mathcal{A},\,\mathcal{B}]$.
\end{thm}
\begin{proof}
    The proof is presented in \Cref{AP:thm:equivalence:specialCase}.
\end{proof}

From \eqref{eq:equivalence_spl_case} we notice that neither the observable functions $\varphi_i$'s nor the data $\{x_t, u_t\}_{t\in \bN_0}$ influence the regularization term $\frac{\epsilon^2}{12}(\|\A\|^2 + \|\B\|^2)$, which is quite a contrast from the case studied under \Cref{thm:equivalence}. 
Furthermore, from \eqref{eq:equivalence_spl_case} we obtain 
\begin{align*}
    [\tilde{A}, \tilde{B}] = \lim_{T\to \infty}\frac{\Phi^+ \Psi}{T} \left( \frac{\Psi \Psi^\top}{T} + \frac{\epsilon^2}{12}I  \right)^{-1},
\end{align*}
where $\Psi = [\Phi^\top,~ U^\top]^\top \in \R^{(N+m)\times T}$.
Without any loss of generality, we assume that $\Psi$ is of full (row) rank for large enough $T$, and consequently, by application of Woodbury matrix inverse equality, we may write 
\begin{align*}
     [\tilde{A}, \tilde{B}] = & \underbrace{\lim_{T\to \infty}\frac{\Phi^+ \Psi^\top}{T} \left( \frac{\Psi \Psi^\top}{T}  \right)^{-1}}_{[A,B]} \\
     &~~ - 
     \lim_{T\to \infty}\frac{\Phi^+ \Psi^\top}{T}\left( \frac{\Psi \Psi^\top}{T}  \right)^{-1} \left( \left( \frac{\Psi \Psi^\top}{T}   \right)^{-1} + \frac{12}{\epsilon^2}I \right)^{-1}\left( \frac{\Psi \Psi^\top}{T}  \right)^{-1}
\end{align*}
If $G= [A, B]$ is the identified matrix under unquantized data, then, the last equation yields 
\begin{align*}
    [\tilde{A}, \tilde{B}] & = [A, B] - [A, B]\lim_{T\to \infty}\left( \left( \frac{\Psi \Psi^\top}{T}   \right)^{-1} + \frac{12}{\epsilon^2}I \right)^{-1}\left( \frac{\Psi \Psi^\top}{T}  \right)^{-1} \\
    & = [A, B] - [A, B]\lim_{T\to \infty} \frac{\epsilon^2}{12} \left( \frac{\Psi \Psi^\top}{T} + \frac{\epsilon^2}{12} I \right)^{-1}
\end{align*}
Consequently, we may quantify the normalized error in the system identification as:
\begin{align} \label{eq:modelMismatch}
    \frac{\|[\tilde{A}, \tilde{B}] - [A, B]\|}{\|[A, B]\|} \le    \frac{\epsilon^2}{12} \left\|\left( \lim_{T\to \infty}\frac{\Psi \Psi^\top}{T} + \frac{\epsilon^2}{12} I \right)^{-1}\right\|.
\end{align}
Equation \eqref{eq:modelMismatch} provides an upper bound on the model identification error as a function of both the quantization resolution $\epsilon$ and the data $\Psi$. 

It is noteworthy that the r.h.s. of \eqref{eq:modelMismatch} can be zero even when $\epsilon > 0$. 
To be stated precisely, if $\lim_{T\to \infty} \frac{\Psi \Psi^\top}{T}$ diverges, then  $\left\|\left( \lim_{T\to \infty} \frac{\Psi \Psi^\top}{T} + \frac{\epsilon^2}{12} I \right)^{-1}\right\|$ approaches to zero for any nonzero $\epsilon$. 
More formally, we have the following lemma. 

\begin{lm} \label{lem:infiniteData-finitResolution}
    Let the data matrices $\Phi^+$ and $\Psi$ satisfy 
    \begin{align} \label{eq:cond1}
        \lim_{T\to \infty} \frac{\Psi \Psi^\top}{T} \to \infty,
    \end{align}
   and $\Psi$ has full (row) rank. Let there exist $\sigma > 1$ such that 
    \begin{align} \label{eq:cond2}
        \lim_{T\to \infty} \frac{\Psi \Psi^\top}{T^\sigma} < +\infty, \qquad 
        \lim_{T\to \infty} \frac{\Phi^+ \Psi^\top}{T^\sigma} < +\infty.
    \end{align}
    Then, for any $\epsilon$, 
    \begin{align} \label{eq:zero_mismatch}
         \frac{\|[\tilde{A}, \tilde{B}] - [A, B]\|}{\|[A, B]\|} \to 0
    \end{align}
    almost surely as $T\to \infty$.
\end{lm}
\begin{proof}
    The proof is presented in \Cref{AP:lem:infiniteData-finitResolution}.
\end{proof}

\Cref{lem:infiniteData-finitResolution} essentially reveals that a sufficiently large amount of data can mitigate the effect of quantization regardless of how coarse the quantization error is. 
Existing work \cite{Arbabi2017} has already shown that, under certain cases, enough unquantized data ensures the mismatch between the finite dimensional representation of the Koopman operator and the true infinite dimensional one converges to zero. 
Now, with the help of \Cref{lem:infiniteData-finitResolution}, we can claim that the stated difference between the finite dimensional representation of the Koopman operator and the true infinite dimensional one converges to zero even under quantized data, provided the observables are directly quantized (instead of quantizing the state $x_t$) and the conditions of \Cref{lem:infiniteData-finitResolution} are satisfied. 

It is also worth noting that such a fundamental result holds due to use of \textit{dither quantization} which enables the analysis of this work. Other form of quantizations may not necessarily yield the same results. 
 
\section{Conclusions} \label{sec:conclusions}
In this paper, we present a least-square optimization method that estimates a Koopman-based linear predictor from dither quantized state and control data. 
We theoretically analyze the connection between the estimates obtained from the quantized data and that from unquantized data. 
The effect of quantization is analyzed and quantified for both finite and large data regimes. 
Our analysis shows the quantization resolution $\epsilon$ affects the estimates as $O(\epsilon)$ in finite data regime and $O(\epsilon^2)$ in large data regime. 
Our analysis is validated via repeated trials of experiments on multiple problems for both prediction and control. These insights pave the path to implement Koopman-based robust controllers with linear predictive model with bounded errors due to quantization.

\appendix \label{sec:appendix}
\subsection{Some useful results on quantization error} \label{sec:prelim_results}
This section provides some technical results that are used in the proof of Theorem~\ref{thm:equivalence}. 
Although similar results may be derived from textbook knowledge, we provide these proofs for completeness. 

\begin{lm}\label{lem:cross_error}
    Let $e^{x,i}_t = \tilde x^i_t - x^i_t$ be the quantization error of the $i$-th state at time step $t$. 
    Then, 
    \begin{align} \label{eq:errorCorrelation}
        \E[e^{x,i}_t e^{x,j}_s] = \begin{cases}
            \frac{\epsilon^2}{12}, \qquad i= j  ~~\mathrm{and}~~ t =s,\\
            0, \qquad~~ \mathrm{otherwise}.
        \end{cases}
    \end{align}
    Similarly, let $e^{u,i}_t = \tilde u^i_t - u^i_t$ be the quantization error of the $i$-th input at time step $t$.
    Then, 
    \begin{align}
        \E[e^{u,i}_t e^{u,j}_s] = \begin{cases}
            \frac{\epsilon^2}{12}, \qquad i= j  ~~\mathrm{and}~~ t =s,\\
            0, \qquad~~ \mathrm{otherwise}.
        \end{cases}
    \end{align}

\end{lm}
\begin{proof}
    Due to the dither quantization scheme and the dither noise being uniform in $\left[ -\frac{\epsilon}{2}, \frac{\epsilon}{2} \right]$, each $e^{x,i}_t$ is independent and also uniformly distributed between $\left[ -\frac{\epsilon}{2}, \frac{\epsilon}{2} \right]$. 
    Consequently, when $i\ne j$ or $t\ne s$, we have $\E[e^{x,i}_t e^{x,j}_s] = \E[e^{x,i}_t] \E[e^{x,j}_s] = 0$. 
    On the other hand, $\E[(e^{x,i}_t)^2] = \frac{1}{\epsilon}\int_{-\epsilon/2}^{\epsilon/2} e^2 \mathrm{d}e = \frac{\epsilon^2}{12}.$ Similar steps show the results for $e^{u,i}_i$. 
\end{proof}

\begin{corr} \label{corr:pqij}
    For any fixed $i,j \in \{1,\ldots, N\}$, let $p^{ij}_t = e^{x,i}_t e^{x,j}_t$, and $q^{ij}_t = e^{u,i}_t e^{u,j}_t$ where $i\ne j$. 
    Then,
    \begin{align}
         \lim_{T\to \infty}\frac{1}{T}\sum_{t=0}^{T-1} p^{ij}_t = 0,\
         \lim_{T\to \infty}\frac{1}{T}\sum_{t=0}^{T-1} q^{ij}_t = 0.
    \end{align}
\end{corr}

\begin{proof}
    One may notice that $\{p^{ij}_t\}_{t\ge 0}$ is an i.i.d. sequence of random variables with $\E[p^{ij}_t] = 0$ (due to Lemma~\ref{lem:cross_error}) and $\E[(p^{ij}_t)^4] < \infty$. 
    Therefore, from strong law of large numbers, $\frac{1}{T}\sum_{t=0}^{T-1} p^{ij}_t \to 0$ almost surely. Similarly, $\frac{1}{T}\sum_{t=0}^{T-1} q^{ij}_t \to 0$ almost surely as well.
\end{proof}

\begin{corr} \label{corr:yzij}
    For any fixed $i,j \in \{1,\ldots, N\}$, let $y^{ij}_t = e^{x,i}_t e^{x,j}_{t+1}$, $z^{ij}_t = e^{u,i}_t e^{u,j}_{t+1}$, and $s^{ij}_t = e^{u,i}_t e^{x,j}_{t+1}$ where $i\ne j$. 
    Then,
    \begin{align}
         \lim_{T\to \infty}\frac{1}{T}\sum_{t=0}^{T-1} y^{ij}_t = 0,~~
         \lim_{T\to \infty}\frac{1}{T}\sum_{t=0}^{T-1} z^{ij}_t = 0,~~
         \lim_{T\to \infty}\frac{1}{T}\sum_{t=0}^{T-1} s^{ij}_t = 0.
    \end{align}
\end{corr}

\begin{proof}
    The proof follows the same steps as in the proof of \Cref{corr:pqij}.
\end{proof}

\begin{corr} \label{corr:zi}
    Let $y^i_t = e^{x,i}_t e^{x,i}_{t+1}$ and $z^i_t = e^{u,i}_t e^{u,i}_{t+1}$. Then, 
    \begin{align}
         \lim_{T\to \infty}\frac{1}{T}\sum_{t=0}^{T-1} y^{i}_t = 0,~~
         \lim_{T\to \infty}\frac{1}{T}\sum_{t=0}^{T-1} z^{i}_t = 0.
    \end{align}
\end{corr}
\begin{proof}
    Although $\{y^i_t\}_{t\ge 0}$ is an identically distributed sequence, it is not independent. 
    Therefore, the standard strong law of large numbers does not apply readily. 
    
    To proceed with the proof, let us first note that, for all $\tau \ge 1$,
    \begin{align} \label{eq:uncorrelated}
        \E[y^i_t y^i_{t+\tau}] &= \E[e^{x,i}_t e^{x,i}_{t+1}e^{x,i}_{t+\tau}e^{x,i}_{t+\tau+1}] \nonumber \\
         & = \E[e^{x,i}_t] \E[e^{x,i}_{t+1}e^{x,i}_{t+\tau}e^{x,i}_{t+\tau+1}] = 0,
    \end{align}
    where the second inequality follow from the fact that $e^{x,i}_t$ is independent of $e^{x,i}_{t+1}e^{x,i}_{t+\tau}e^{x,i}_{t+\tau+1}$ for all $\tau \ge 1$. 
    Therefore, \eqref{eq:uncorrelated} proves pairwise uncorrelation of the sequence $\{y^i_t\}_{t\ge 0}$. 

    Now let us define the random variable $\vartheta_T = \sum_{t=0}^{T-1} y^i_t$. 
    We note that, 
    \begin{align*}
        \E[(\vartheta_T)^4] &= \E\left[ \!\bigg( \sum_{t=0}^{T-1} y^i_t \bigg)^{\!\! 4} \right] \\
        & = T \E[(y^i_0)^{4}] + 3T(T-1) \E[(y^i_0 y^i_1)^2],
    \end{align*}
    where we have used $\E[y^i_k (y^i_\ell)^3] = 0$ since $y^i_k$ and $y^i_\ell$ are uncorrelated for all $k\ne \ell$ and $\E[y^i_k] = 0$ for all $k$.
    Given that $e^{i}_t$ is uniformly distributed in $\left[ -\frac{\epsilon}{2}, \frac{\epsilon}{2} \right]$, there exists a $K < \infty$ such that 
    \begin{align*}
        K = 4 \max\{ \E[(y^i_0)^4],  \E[(y^i_0 y^i_1)^2]\},
    \end{align*}
    which then implies 
    \begin{align} \label{eq: KBound}
        \E[(\vartheta_T)^4] \le K T^2,
    \end{align}
    for all $T \ge 1$. 
    Next, we use this bound to show that $\frac{1}{T}\sum_{t=0}^{T-1} y^{i}_t  \to 0$ almost surely. 
    To that end, let us start with
    \begin{align*}
        \E \left[ \sum_{T \ge 1} \left( \frac{\vartheta_T}{T} \right)^4\right] = \sum_{T \ge 1} \E \left[ \left( \frac{\vartheta_T}{T} \right)^4 \right] \le \sum_{T \ge 1} \frac{K} {T^{2}} < \infty, 
    \end{align*}
    where the first equality follows from the Fubini-Tonelli Theorem, and the first inequality follows from \eqref{eq: KBound}.
    Having proven that $\E \left[ \sum_{T \ge 1} \left( \frac{\vartheta_T}{T} \right)^4\right] < \infty$, we may conclude that $\sum_{T \ge 1} \left( \frac{\vartheta_T}{T} \right)^4 < \infty$ almost surely.
    Therefore, since the series converges, the underlying sequence must converge to zero, which implies 
    \begin{align*}
        \left( \frac{\vartheta_T}{T} \right)^4 \to 0 \qquad \text{almost surely}. 
    \end{align*}
    Consequently, we may conclude that 
    \begin{align*}
        \frac{1}{T}\sum_{t=0}^{T-1} y^{i}_t = \frac{\vartheta_T}{T} \to 0 \qquad \text{almost surely}. 
    \end{align*}
    This concludes the proof of $\lim_{T\to \infty}\frac{1}{T}\sum_{t=0}^{T-1} y^{i}_t = 0$.
    The proof of $\lim_{T\to \infty}\frac{1}{T}\sum_{t=0}^{T-1} z^{i}_t = 0$ follows the same steps.
    
\end{proof}

\begin{lm} \label{lm:errorfunctions}
    Let $\{g_i(x_t)\}_{i=1:N}$ be a collection of scalar-valued function of $x_t$, and define the dither quantization error of these functions as 
    \begin{align*}
        e^{g,i}_t = \Q(g_i(x,t) + w^{g,i}_t) - w^{g,i}_t - g_i(x),
    \end{align*}
    where $w^{g,i}_t \sim \U(\big[-\frac{\epsilon}{2}, \frac{\epsilon}{2}\big]) $ is a uniformly distributed dither noise for all $i$ and $t$, where $\epsilon$ is the quantization resolution of $\Q$ (see \eqref{eq:quantizationResolution}). 
    Furthermore, $w^{g,i}_t$ and $w^{g,j}_s$ are independent for all $i\ne j$ or $t\ne s$. 
    Then, $e^{g,i}_t \sim \U(\big[-\frac{\epsilon}{2}, \frac{\epsilon}{2}\big]) $ for all $i$ and $t$, and $e^{g,i}_t$ and $e^{g,j}_s$ are independent for $i\ne j$ or $t\ne s$. 
    Consequently, 
    \begin{align} \label{eq:errorCorrelationFunctions}
        \E[e^{g,i}_t e^{g,j}_s] = \begin{cases}
            \frac{\epsilon^2}{12}, \qquad i= j  ~~\mathrm{and}~~ t =s,\\
            0, \qquad~~ \mathrm{otherwise}.
        \end{cases}
    \end{align}
   Moreover, for all $i,j \in \{1,\ldots, N\}$, we have 
   \begin{align} \label{eq:egijt}
        \lim_{T\to \infty}\frac{1}{T}\sum_{t=0}^{T-1} e^{g,i}_t e^{g,j}_{t+1} = 0,
   \end{align}
   and for all $i\ne j$ 
   \begin{align}
        \lim_{T\to \infty}\frac{1}{T}\sum_{t=0}^{T-1} e^{g,i}_t e^{g,j}_t = 0.
   \end{align}
\end{lm}
\begin{proof}
    The proof follows the same arguments as presented in \Cref{lem:cross_error}, \Cref{corr:pqij}, \Cref{corr:yzij}, and \Cref{corr:zi}. 
    We leave it as an exercise to the readers. 
\end{proof}

\begin{lm}[Kolmogorov's Strong law] \label{lm:kolmogorov}
    Let $v_1,v_2,\ldots$ be a sequence of independent random variables that are not necessarily identically distributed. 
    Furthermore, each $v_i$ has finite second moment and $ \lim_{T\to \infty}\sum_{t=1}^T \frac{1}{t^2} \text{Var}(v_t)  < +\infty$. 
    Then, 
    \begin{align*}
      \frac{1}{T}  \sum_{t=1}^T v_t  \longrightarrow \frac{1}{T}  \sum_{t=1}^T \E[ v_t ]\quad \quad \text{almost surely.}
    \end{align*}
\end{lm}

\begin{proof}
    See Theorem 2.3.10 in \cite{Sen1993}.
\end{proof}

\begin{corr} \label{corr:atet}
    Let $\{a_t\}_{t\in \bN}$ be a sequence of real numbers such that $|a_t| \le \bar{a}$ for all $t$. 
    Then, 
    \begin{align}
        \frac{1}{T}  \sum_{t=1}^T a_t e^{\kappa, i}_t \longrightarrow 0 \qquad \text{almost surely,}
    \end{align}
    where $\kappa = \{x, u, f\}$. 
\end{corr}
\begin{proof}
    The proof follows from \Cref{lm:kolmogorov}. 
\end{proof}

\subsection{Proof of \Cref{thm:equivalence}}  \label{AP:thm:equivalence}
We augment the state and input at time-step $t$ to create a new variable
\begin{align*}
    \xi_{t} = \left[
    \begin{array}{c}
         x_t  \\
         u_t 
    \end{array}
    \right]
\end{align*}
Define a lifting function on this augmented variable
\begin{align*}
    \psi(\xi_{t}) = \left[
    \begin{array}{c}
        \varphi \left(x_t\right)  \\
         u_t 
    \end{array}
    \right],
\end{align*}
where $\varphi(\cdot):\mc{M}\rightarrow \R^n$ are the observables as defined in \eqref{eq:observables}.
The linear predictor obtained by solving the least-squares problem \eqref{Eq: optimization} from unquantized data is
    \begin{align}     \begin{split}
        [A,\ B] = G  &=  \underset{\mc{G} \in \R^{N\times \left(N+m\right)}}{\argmin} \frac{1}{T}  \| \Phi^{+} - \mc{G} \Psi\|^2\\\nonumber
        &= \underset{\mc{G} \in \R^{N\times \left(N+m\right)}}{\argmin} \frac{1}{T}  \| \Phi^{+} - \mc{A} \Phi - \mc{B}U\|^2,
        \end{split}
    \end{align}
where
\begin{align*}
     \Psi &= \begin{bmatrix}
         \psi(\xi_0) ~  \psi(\xi_1) ~ \hdots ~  \psi(\xi_{T-1}) 
    \end{bmatrix},  \\
     \Phi^{+} &= \begin{bmatrix}
         \varphi(x_1) ~  \varphi(x_2) ~ \hdots ~  \varphi(x_{T}) 
    \end{bmatrix} ,\\
    \mc{G} &=\begin{bmatrix}
        \mc{A},~\mc{B}
    \end{bmatrix}
\end{align*}
Using the definition of residuals $r(x_{t+1},x_t, u_t)\triangleq \|\varphi(x_{t+1})- \mc{A}\varphi(x_t) -\mc{B}u_t\|^2 = \left\|\varphi\left(x_{t+1}\right)- \mc{G} \psi \left(\xi_t\right)\right\|^2$, 
 \begin{align}
     \| \Phi^{+} - \mc{G} \Psi\|^2 =\sum_{t=0}^{T-1} r\left(x_{t+1}, \xi_t\right)
 \end{align}
where we used $\xi_t = (x_t,\ u_t)$ in order to express $r(x_{t+1},x_t, u_t) = r(x_{t+1}, \xi_t)$.
For the quantized state and input data snapshots, we have:
\begin{align}  \label{eq:phi_sum}
\begin{split}
     \| \bar \Phi^{+} - \mc{G} \bar \Psi\|^2 
     &= \sum_{t=0}^{T-1}\| \bar \varphi(x_{t+1}) - \mc{G} \bar \psi(\xi_t)\|^2 \\
     &= \sum_{t=0}^{T-1}\|  \varphi(\tilde x_{t+1}) - \mc{G} \psi(\tilde\xi_t)\|^2 \\
     &=\sum_{t=0}^{T-1}\ \left\|\varphi\left(x_{t+1} + e^x_{t+1}\right)- \mc{G} \psi \left(\xi_t + \eta_t\right)\right\|^2\\
     &=\sum_{t=0}^{T-1} r\left( x_{t+1}+ e^x_{t+1},  \xi_t + \eta_{t}\right)
\end{split}
 \end{align}
where 
\begin{align*}
    \eta_t =\left[ \begin{array}{c}
          \tilde x_t - x_t   \\
          \tilde u_t - u_t
    \end{array}\right] = \left[
    \begin{array}{c}
         e^x_t  \\
         e^u_t  
    \end{array}\right]
\end{align*}
Expanding $r\left( x_{t+1}+ e^x_{t+1},  \xi_t + \eta_t\right)$ via Taylor series we get:
\begin{align} \label{eq:residual_large_data}
    r\left( x_{t+1}+ e^x_{t+1},  \xi_t + \eta_t\right) = r\left( x_{t+1},  \xi_t\right) + \lim_{n\rightarrow\infty}\sum\nolimits_{k=1}^n h_k(e^x_{t+1},\eta_t)
\end{align}
where, the $k$-th term $h_k(\cdot,\cdot)$ involves the $k$-th order derivative. For instance,
\begin{align*}
    h_1(e^x_{t+1},\eta_t) &= \nabla_{x_{t+1}}r(x_{t+1},\xi_t)^\top e^x_{t+1} + \nabla_{\xi_{t}}r(x_{t+1},\xi_t)^\top \eta_t, \\
    h_2(e^x_{t+1}, \eta_t) &= \dfrac{1}{2} (e^x_{t+1})^\top \nabla^2_{x_{t+1}}r(x_{t+1},\xi_t) e^x_{t+1} + \dfrac{1}{2} \eta_t^\top \nabla^2_{\xi_{t}}r(x_{t+1},\xi_t) \eta_t \\
     & + (e^x_{t+1})^\top \nabla_{x_{t+1}}\nabla_{\xi_{t}}r(x_{t+1},x_t) \eta_t.
\end{align*}

\noindent
For $T\rightarrow\infty$, we may write
\begin{align} \label{eq:residual_large_data_expanded}
    &\lim_{T\rightarrow\infty}\dfrac{1}{T}\sum\nolimits_{t=0}^{T-1} r\left( x_{t+1}+ e^x_{t+1},  \xi_t + \eta_t\right) =\\\nonumber
    &\lim_{T\rightarrow\infty}\dfrac{1}{T}\sum\nolimits_{t=0}^{T-1}r\left( x_{t+1},  \xi_t\right) + \lim_{n\rightarrow\infty}\sum\nolimits_{k=1}^n \lim_{T\rightarrow\infty}\dfrac{1}{T}\sum\nolimits_{t=0}^{T-1} h_k(e^x_{t+1},\eta_t),
\end{align}
where we have used  Assumption~3 and invoked Fubini's theorem to interchange the order of summation in the last term. Next, we will simplify each term  $\lim_{T\rightarrow\infty} \dfrac{1}{T}\sum\nolimits_{t=0}^{T-1} h_k(e^x_{t+1},\eta_t)$ using Kolmogorov's strong law of large numbers. 
For $k=1$:
\begin{align*}
    \lim_{T\rightarrow\infty} \dfrac{1}{T}\sum\nolimits_{t=0}^{T-1} h_1(e^x_{t+1},\eta_t) & = \lim_{T\rightarrow\infty} \dfrac{1}{T}\sum\nolimits_{t=0}^{T-1} \nabla_{x_{t+1}}r(x_{t+1},\xi_t)^\top e^x_{t+1} \\
    &\qquad + \lim_{T\rightarrow\infty} \dfrac{1}{T}\sum\nolimits_{t=0}^{T-1} \nabla_{\xi_{t}}r(x_{t+1},\xi_t)^\top \eta_t \\
   & \overset{\text{almost surely}}{\longrightarrow} 0, 
\end{align*}
where we have used Kolmogorov's strong law of large numbers: $\lim_{T\rightarrow\infty} \dfrac{1}{T}\sum\nolimits_{t=0}^{T-1} \nabla_{x_{t+1}}r(x_{t+1},\xi_t)^\top e^x_{t+1} \to \lim_{T\rightarrow\infty}  \nabla_{x_{t+1}}r(x_{t+1}, \xi_t)^\top \E[e^x_{t+1}]  = 0$ almost surely.\footnote{Applying law of large number requires $\nabla_{x_{t+1}}r(x_{t+1},\xi_t)^\top e^x_{t+1}$ to have a finite second moment and $\sum_{t=1}^\infty \frac{1}{t^2} \text{Var}(\nabla_{x_{t+1}}r(x_{t+1},\xi_t)^\top e^x_{t+1})$ to be finite.
Both of these conditions are satisfied due to Assumptions~2--4. } 
Similarly, we simplify the $k=2$ term: 
\begin{align*}
    & \lim_{T\rightarrow\infty} \dfrac{1}{T}\!\!\sum\nolimits_{t=0}^{T-1}h_2(e^x_{t+1}, \eta_t)  \\ 
    & =  \lim_{T\rightarrow\infty} \dfrac{1}{T}\!\!\sum\nolimits_{t=0}^{T-1} \dfrac{1}{2} (e^x_{t+1})^\top \nabla^2_{x_{t+1}}r(x_{t+1},\xi_t) e^x_{t+1}  \\
    & + \lim_{T\rightarrow\infty} \dfrac{1}{T}\!\!\sum\nolimits_{t=0}^{T-1}\dfrac{1}{2} \eta_t^\top \nabla^2_{\xi_{t}}r(x_{t+1},\xi_t) \eta_t  \\
     & + \lim_{T\rightarrow\infty} \dfrac{1}{T}\!\!\sum\nolimits_{t=0}^{T-1}(e^x_{t+1})^\top \nabla_{x_{t+1}}\nabla_{\xi_{t}}r(x_{t+1},\xi_t) \eta_t. 
\end{align*}
 Using Kolmogorov's law of large numbers, we have:
 \begin{align*}
     \dfrac{1}{T}\sum\limits_{t=0}^{T-1}  \dfrac{1}{2} (e^x_{t+1})^\top \nabla^2_{x_{t+1}}r(x_{t+1},\xi_t) e^x_{t+1} \to \frac{\epsilon^2}{24} \dfrac{1}{T}\sum\limits_{t=0}^{T-1}  \tr(\nabla^2_{x_{t+1}}r(x_{t+1},\xi_t)),
 \end{align*}
 where we have used $\E[e^x_t(e^x_t)^\top] = \frac{\epsilon^2}{12}I$ for all $t$ as proved in Lemma \ref{lem:cross_error} and Corollary \ref{corr:pqij}.
 Similarly, 
  \begin{align*}
     &\dfrac{1}{T}\sum\nolimits_{t=0}^{T-1} \dfrac{1}{2} \eta_t^\top \nabla^2_{\xi_{t}}r(x_{t+1},\xi_t) \eta_t \to \frac{\epsilon^2}{24} \dfrac{1}{T}\sum\nolimits_{t=0}^{T-1}  \tr(\nabla^2_{\xi_{t}}r(x_{t+1},\xi_t)), \\
     & \dfrac{1}{T}\sum\nolimits_{t=0}^{T-1} (e^x_{t+1})^\top \nabla_{x_{t+1}}\nabla_{\xi_{t}}r(x_{t+1},\xi_t) \eta_t \to 0,
 \end{align*}
 where the last result is obtained by using $\E[e^x_{t+1}\eta_t^\top] = 0$ from Corollary \ref{corr:yzij} and \ref{corr:zi}.
Now, notice that 
 \begin{align*}
     \nabla^2_{x_{t+1}}r(x_{t+1},\xi_t) &= 2(\nabla_{x_{t+1}}\varphi(x_{t+1}))(\nabla_{x_{t+1}}\varphi(x_{t+1}))^\top \\
      &+ 2\sum_{i=1}^N  [\varphi(x_{t+1}) - \mc{G}\psi(\xi_t)]_i \nabla^2_{x_{t+1}} \varphi^i(x_{t+1}),
 \end{align*}
 where $[\varphi(x_{t+1}) - \mc{G}\psi(\xi_t)]_i$ is the $i$-th component of the vector. Therefore,
 \begin{align*}
    \tr(\nabla^2_{x_{t+1}}r(x_{t+1},\xi_t)) &= 2\|\nabla_{x_{t+1}}\varphi(x_{t+1})\|^2 \\
    & + 2(\varphi(x_{t+1}) - \mc{G} \psi(\xi_t))^\top g(x_{t+1}),
\end{align*}
where {\small$g(x_{t+1})\in \R^N$} with {\small$\tr(\nabla^2_{x_{t+1}}\varphi^i(x_{t+1}))$} being its $i$-th component. Similarly, we obtain
\begin{align*}
    \tr(\nabla^2_{\xi_{t}}r(x_{t+1},\xi_t)) &= 2\|\mc{G}\nabla_{\xi_{t}}\psi(\xi_{t})\|^2  \\
    & - 2(\varphi(x_{t+1}) - \mc{G} \psi(\xi_t))^\top \mc{G} h(\xi_t).
\end{align*}
where {\small$h(\xi_{t})\in \R^N$} with {\small$\tr(\nabla^2_{\xi_{t}}\psi^i(\xi_{t}))$} being its $i$-th component.

\noindent
Therefore,
\begin{align*}
   \dfrac{1}{T}\sum\nolimits_{t=0}^{T-1} h_1(e^x_{t+1},\eta_t) \to \epsilon^2\left(\alpha_2 + \tr(\mc{G}\beta_2) + \tr(\mc{G}^\top \mc{G}\Gamma_2)  \right),
\end{align*}
where \vspace{-0.3cm}
\begin{align*}
    \alpha_2 &= \lim_{T\to \infty}\dfrac{1}{12T}\sum\nolimits_{t=0}^{T-1} \|\nabla_{x_{t+1}}\varphi(x_{t+1})\|^2 + \varphi(x_{t+1})^\top g(x_{t+1}) \\
    \beta_2 &= - \lim_{T\to \infty}\dfrac{1}{12T}\sum\nolimits_{t=0}^{T-1} \psi(\xi_t)g(x_{t+1})^\top + h(\xi_t)\varphi(x_{t+1})^\top,\\
    \Gamma_2 &= \lim_{T\to \infty}\dfrac{1}{12T}\sum\nolimits_{t=0}^{T-1} \nabla_{\xi_{t}}\psi(\xi_{t}) \nabla_{\xi_{t}}\psi(\xi_{t})^\top + h(\xi_t) \psi(\xi_t)^\top.
\end{align*}
Similarly, for any $k$ we have
\begin{align*}
     \dfrac{1}{T}\!\!\sum\limits_{t=0}^{T-1}\! h_k(e^x_{t+1},\eta_t) \overset{\text{almost}}{\underset{\text{surely}}{\to}}  \begin{cases}
         \epsilon^k\left(\alpha_k + \tr(\mc{G}\beta_k) + \tr(\mc{G}^\top \mc{G}\Gamma_k)  \right), ~ k = \text{even} \\
         0,\qquad \qquad  \text{otherwise}.
     \end{cases}
\end{align*}
Adding all the residuals, We obtain
\begin{align}\label{eq:residuals}
    \lim_{T\rightarrow\infty}\dfrac{1}{T}\!\!\sum\nolimits_{t=0}^{T-1} r(x_{t+1} & + e^x_{t+1}, \xi_t+\eta_t) \to \lim_{T\rightarrow\infty}\dfrac{1}{T}\!\!\sum\nolimits_{t=0}^{T-1} r(x_{t+1}, \xi_t) \nonumber \\
    &+ \alpha(\epsilon) + \tr(\mc{G} \beta(\epsilon)) + \tr(\mc{G}^\top \mc{G} \Gamma(\epsilon))  ,    
\end{align}
where 
\begin{align*}
   \alpha(\epsilon) = \sum_{k=1}^\infty \epsilon^{2k} \alpha_{2k}, \ \beta(\epsilon) = \sum_{k=1}^\infty \epsilon^{2k} \beta_{2k}, \ \Gamma(\epsilon) = \sum_{k=1}^\infty \epsilon^{2k} \Gamma_{2k}.
\end{align*}
Consequently, from \eqref{eq:phi_sum} and \eqref{eq:residuals}
\begin{align} 
\begin{split}
   \bar G &= \left[\bar A \; \bar B \right] \\
   &= \argmin_{\mc{G} \in \R^{N\times \left(N+m \right)}} \frac{1}{T} \| \bar \Phi^{+} - \mc{G} \bar \Psi\|^2  \\ 
    &\underset{T\rightarrow\infty}{\overset{\operatorname{a.s.}}{\longrightarrow}} \argmin_{\mc{G} \in \R^{N\times \left(N+m \right)}} \limsup_{T\rightarrow\infty} \frac{1}{T}  \| \Phi^{+} - \mc{G} \Psi\|^2 \\
    &\qquad + \tr(\mc{G}\beta(\epsilon)) + \tr(\mc{G}^\top \mc{G} \Gamma(\epsilon)).
    \end{split}
\end{align}

This completes the proof of \Cref{thm:equivalence}. \hfill $\blacksquare $

\subsection{Proof of \Cref{thm:equivalence:specialCase}} \label{AP:thm:equivalence:specialCase} 

Let us define
\begin{align*}
    \tilde{\Psi} &= \begin{bmatrix}
         \begin{bmatrix}\tilde\varphi(x_0)\\\tilde u_0\end{bmatrix} ~  \begin{bmatrix}\tilde\varphi(x_1)\\\tilde u_1\end{bmatrix} ~ \hdots ~ \begin{bmatrix}\tilde\varphi(x_{T-1})\\\tilde u_{T-1}\end{bmatrix}
    \end{bmatrix},  \\
    \tilde \Phi &= \begin{bmatrix}
        \tilde \varphi(x_0) ~ \tilde \varphi(x_1) ~ \hdots ~ \tilde \varphi(x_{T-1}) 
    \end{bmatrix},~\\
    \tilde \Phi^{+} &= \begin{bmatrix}
        \tilde \varphi(x_1) ~ \tilde \varphi(x_2) ~ \hdots ~ \tilde \varphi(x_{T}) 
    \end{bmatrix},~\\
    \bar U &= \begin{bmatrix}
        \tilde u_1 ~ \tilde u_2 ~ \hdots ~ \tilde u_{T-1} 
    \end{bmatrix},
\end{align*}
and where $\tilde{\varphi}(\cdot)$ is defined in \eqref{eq:tildeObservablesspecialCase}.
Now, we may write
\begin{align*}
     \| \tilde \Phi^+ - \mathcal{G} \tilde \Psi\|^2 =  \sum_{t=0}^{T-1}\| \tilde \varphi(x_{t+1}) - \A \tilde \varphi(x_t) - \B\tilde{u}_t\|^2.
\end{align*}
Let us further define $e^{\varphi,i}_t \triangleq \tilde\varphi^i(x_t) - \varphi^i(x_t)$ to be the quantization error on the $i$-th observable at time $t$, which is now a zero-mean i.i.d. process due to the dither quantization. 
This implies $\E[e^{\varphi, i}_t e^{\varphi, j}_s] = 0$ when $i\ne j$ or $t\ne s$. 
Let us further define $e^\varphi_t = [e^{\varphi, 1}_t,\ldots, e^{\varphi, N}_t]^\top$.
We follow the same definition  $e^{u,i}_t = \tilde{u}^i_t - u^i_t$ as in  \Cref{lem:cross_error} and $e^u_t = [e^{u, 1}_t,\ldots, e^{u, m}_t]^\top$.

Expanding $\| \tilde \varphi(x_{t+1}) - \A \tilde \varphi(x_t) - \B\tilde{u}_t\|^2$ yields:
\begin{align} \label{eq:tilde_expansion_specialCase}
    \| \tilde \varphi(x_{t+1}) & - \A \tilde \varphi(x_t) - \B\tilde{u}_t\|^2 \\
      = &~  \|  \varphi(x_{t+1}) - \A  \varphi(x_t) - \B u_t + e^\varphi_{t+1} - \A e^\varphi_t - \B e^u_t \|^2 \nonumber \\ \nonumber
     = &~ \|  \varphi(x_{t+1}) - \A  \varphi(x_t) -\B u_t\|^2 + \| e^\varphi_{t+1} \|^2 + \|\A e^\varphi_t\|^2  + \| \B e^u_t \|^2 \\ \nonumber
     &~~  - 2(e^\varphi_{t+1})^\top \A e^\varphi_t + 2 (e^\varphi_{t+1})^\top (\varphi(x_{t+1}) - \A  \varphi(x_t) - \B u_t) \\ \nonumber
     &~~ - 2(e^\varphi_{t+1})^\top \B u_t - 2 (e^\varphi_t)^\top \A^\top (\varphi(x_{t+1}) - \A  \varphi(x_t) - \B u_t) \\
     &~~ - 2 (e^\varphi_t)^\top \A^\top \B u_t - 2(e^u_t)^\top \B^\top (\varphi(x_{t+1}) - \A  \varphi(x_t) -\B u_t). \nonumber
\end{align}
Now recall that $\{e^{\varphi,i}_t, \}_{t=  0:T}^{i=1:N}$ and $\{e^{u,j}_t, \}_{t=  0:T}^{j=1:m}$ are i.i.d sequences, therefore, using the \textit{law of large numbers}, we may write 
\begin{align}
    \lim_{T\to \infty}\frac{1}{T}\sum_{t=0}^{T-1} e^{\varphi, i}_t = 0, \qquad \lim_{T\to \infty}\frac{1}{T}\sum_{t=0}^{T-1} e^{u, j}_t = 0
\end{align}
for all $i = 1, \ldots, N$ and $j=1,\ldots, m$, with almost sure probability. 
Similarly, we may also write 
\begin{align*}
    \lim_{T\to \infty} \frac{1}{T} \sum_{t=0}^{T-1}\|e^\varphi_{t+1}\|^2 = \E[\|e^\varphi_0\|^2] = \sum_{i=1}^N \E[(e^i_0)^2] \overset{(\dagger)}{=} N \frac{\epsilon^2}{12}
\end{align*}
almost surely, where $(\dagger)$ follows from \Cref{lm:errorfunctions}. 
Similarly, 
\begin{align*}
    \lim_{T\to \infty} \frac{1}{T} \sum_{t=0}^{T-1} (e^\varphi_{t+1})^\top \mc{A} e^\varphi_t = 0, 
\end{align*}
almost surely due to \eqref{eq:egijt} in \Cref{lm:errorfunctions}.
Using \Cref{corr:atet} one may verify that 
\begin{align*}
    &\lim_{T\to \infty} \frac{1}{T} \sum_{t=0}^{T-1} (e^\varphi_{t+1})^\top (\varphi(x_{t+1}) - \A  \varphi(x_t) - \B u_t) \longrightarrow 0, \\ 
     & \lim_{T\to \infty} \frac{1}{T} \sum_{t=0}^{T-1} (e^\varphi_{t+1})^\top \B u_t \longrightarrow 0, \\
     & \lim_{T\to \infty} \frac{1}{T} \sum_{t=0}^{T-1} (e^\varphi_t)^\top \A^\top (\varphi(x_{t+1}) - \A  \varphi(x_t) - \B u_t) \longrightarrow 0, \\
     & \lim_{T\to \infty} \frac{1}{T} \sum_{t=0}^{T-1} (e^\varphi_t)^\top \A^\top \B u_t \longrightarrow 0, \\
    & \lim_{T\to \infty} \frac{1}{T} \sum_{t=0}^{T-1} (e^u_t)^\top \B^\top (\varphi(x_{t+1}) - \A  \varphi(x_t) -\B u_t) \longrightarrow 0.
\end{align*}
Therefore, for a large enough $T$, we may write
\begin{align} \label{eq:mainEquation}
\begin{split}
    \frac{1}{T}\sum_{t=0}^{T-1} \| \tilde \varphi(x_{t+1}) & - \A \tilde \varphi(x_t) - \B u_t\|^2 \\
    \overset{\text{a.s.}}{\underset{T\rightarrow \infty}{\longrightarrow}} ~~~ & \frac{1}{T} \sum_{t=0}^{T-1} \| \varphi(x_{t+1}) - \A  \varphi(x_t) - \B u_t\|^2  \\
    &+ N \frac{\epsilon^2}{12} + \frac{\epsilon^2}{12} \| \A\|^2 + \frac{\epsilon^2}{12} \| \B\|^2. 
    \end{split}
\end{align}
Thus, 
\begin{align*}
    &[\tilde{A},\,\tilde{B}] = \underset{{\mathcal{A},\mathcal{B}}}{\argmin} \ \| \tilde \Phi^+ - \mathcal{A} \tilde \Phi - \mathcal{B} \bar U\|^2 \\
    &= \underset{{\mathcal{A},\mathcal{B}}}{\argmin}\ \frac{1}{T}\sum_{t = 0}^{T-1} \| \tilde \varphi(x_{t+1})  - \mathcal{A} \tilde \varphi(x_t) - \mathcal{B} \tilde{u}_t\|^2 \\
    & \overset{\text{a.s.}}{\underset{T\rightarrow \infty}{\longrightarrow}} ~ \underset{{\mathcal{A},\mathcal{B}}}{\argmin} \limsup_{T\to \infty} \frac{1}{T}\! \sum_{t = 0}^{T-1} \| \varphi(x_{t+1}) - \mathcal{A}  \varphi(x_t) - \mathcal{B} u_t\|^2 \\
    & + \frac{\epsilon^2}{12} \left(\|\A\|^2 + \|\B\|^2\right). 
\end{align*}
This concludes the proof of \Cref{thm:equivalence:specialCase}. 
\hfill $\blacksquare $ 
\subsection{Proof of \Cref{lem:infiniteData-finitResolution}} \label{AP:lem:infiniteData-finitResolution}

Let us redefine the EDMD problem \eqref{Eq: optimization} under quantized data as 
\begin{align} \label{eq:modified_EDMD}
    [A, B] = \underset{\mc{A} {\in \R^{N \times N}},\,\mc{B} \in {\R^{N \times m}}}{\argmin} \dfrac{1}{T^\sigma}\|\Phi^{+} - \mc{A} {\Phi} - \mc{B} {U}\|^2 
\end{align}
for some $\sigma > 1$ that satisfies \eqref{eq:cond2} in \Cref{lem:infiniteData-finitResolution}.
Notice that the minimizer to \eqref{eq:modified_EDMD} does not depend on $\sigma$ for any finite $T$. 
Using asymptotics of minimizers of convex processes \cite{hjort2011asymptotics}, we may claim that the minimizer to $\dfrac{1}{T^\sigma}\|\Phi^{+} - \mc{A} {\Phi} - \mc{B} {U}\|^2$ coincides with the minimizer to $\|\Phi^{+} - \mc{A} {\Phi} - \mc{B} {U}\|^2$ even when $T\to \infty$.  
In other words, we have 
\begin{align*}
    [A, B] = \lim_{T\to \infty}\frac{\Phi^+ \Psi^\top}{T^\sigma} \left( \frac{\Psi \Psi^\top}{T^\sigma}  \right)^{-1}
\end{align*}
to be a well-defined quantity. 
Now, the r.h.s. of \eqref{eq:modelMismatch} goes to $0$ regardless of the value of $\epsilon$ under \eqref{eq:cond1} of this lemma. 
Consequently, we have \eqref{eq:zero_mismatch}.
\hfill $\blacksquare$

\bibliographystyle{ieeetr}
\bibliography{references, ref1}

\begin{IEEEbiography}[{\includegraphics[width=1in,height=1in,clip,keepaspectratio]{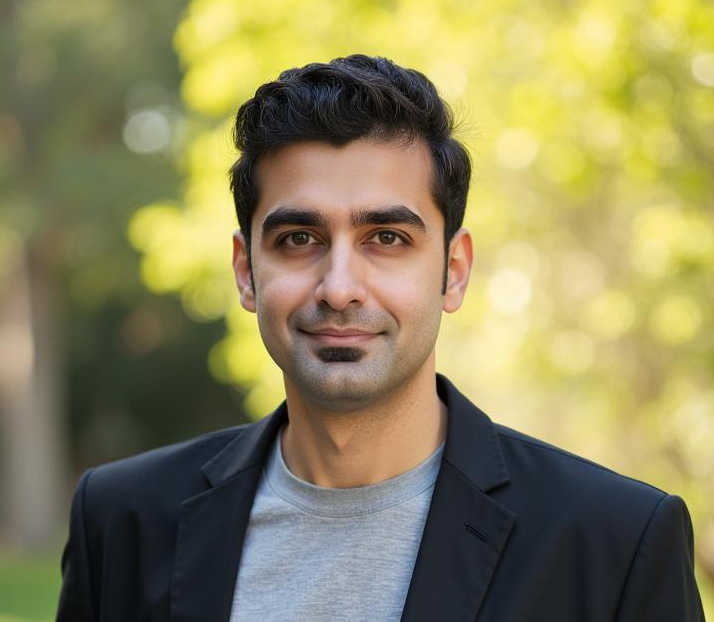}}]{Shahab Ataei} is a Ph.D. student in Electrical Engineering at The Ohio State University. He received his B.Sc. and M.Sc. degrees in Electrical Engineering from Amirkabir University of Technology in 2017 and 2021, respectively. His research interests include safe controller design for networked control systems and cyber-physical systems.

\end{IEEEbiography}

\begin{IEEEbiography}[{\includegraphics[width=1in,height=1 in, keepaspectratio]{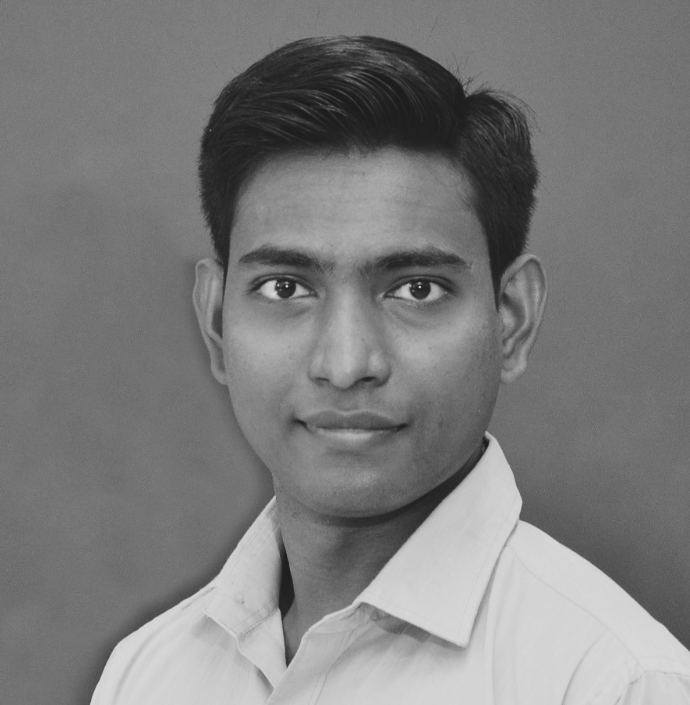}}]
{Dipankar Maity} (Senior Member, IEEE) received the B.E. degree in electronics and telecommunication engineering from Jadavpur University, India, in 2013, and the Ph.D. degree in electrical and computer engineering from the University of Maryland, College Park, MD, USA in 2018. He is an Assistant Professor with the Department of Electrical and Computer Engineering, University of North Carolina at Charlotte, Charlotte, NC, USA. He was a Postdoctoral Fellow with the Georgia Institute of Technology, Atlanta, GA, USA. During his Ph.D., he was a Visiting Scholar at the Technical University of Munich (TUM) and at the KTH Royal Institute of Technology, Stockholm, Sweden. His research interests include temporal logic-based controller synthesis, control under communication constraints, intermittent-feedback control, stochastic games, and integration of these ideas in the context of cyber-physical systems.
\end{IEEEbiography}

\begin{IEEEbiography}[{\includegraphics[width=1in,height=1.25in,clip,keepaspectratio]{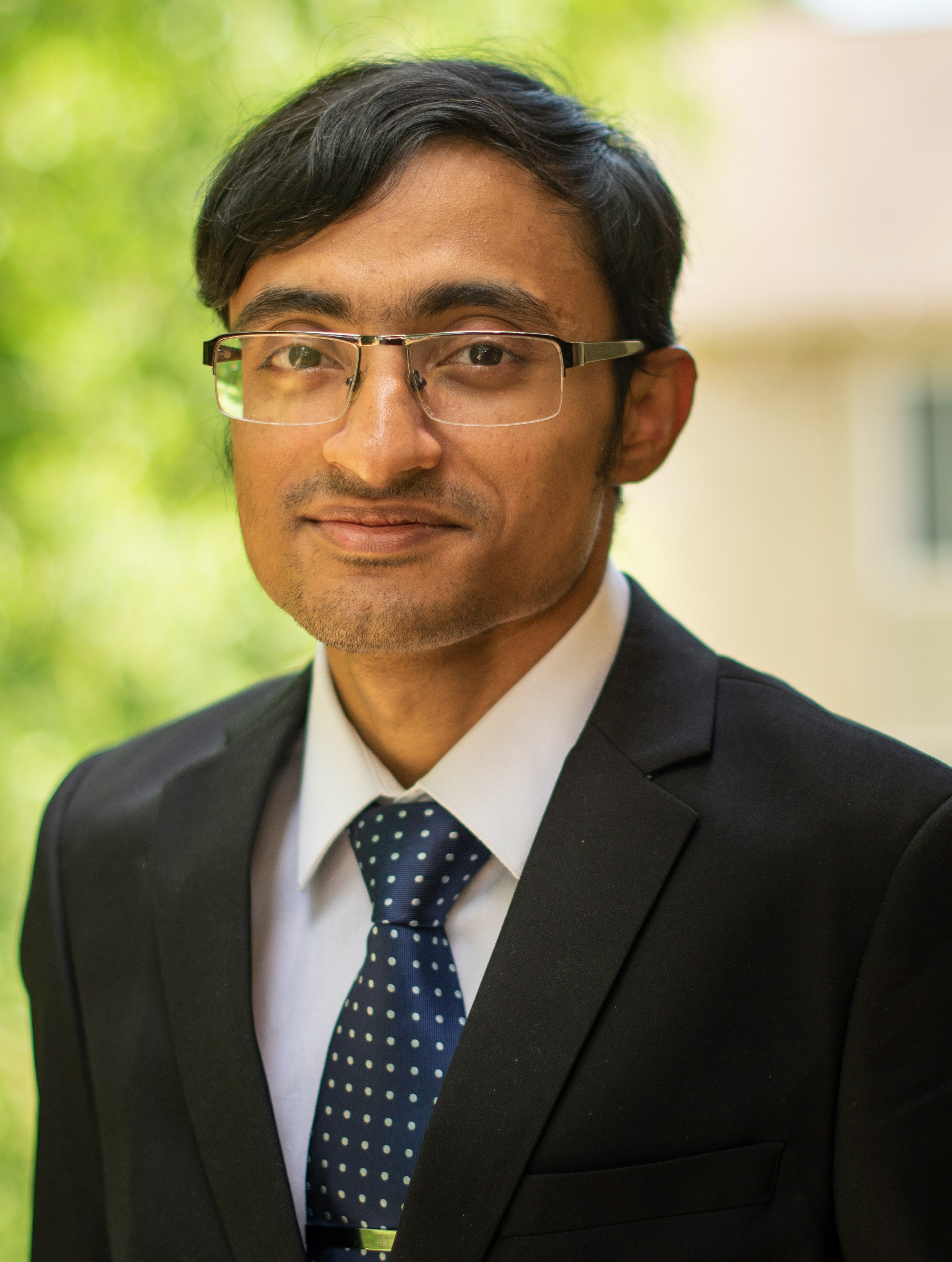}}]{Debdipta Goswami}(Member, IEEE)
is an assistant professor in the Department of Mechanical and Aerospace Engineering and Department of Electrical and Computer Engineering at The Ohio State University. He received the Ph.D. degree in Electrical and Computer Engineering from the University of Maryland, USA in 2020 and the B.E. degree in Electronics and Telecommunication Engineering from Jadavpur University, India in 2015. He was a Postdoctoral Research Associate at Princeton University. His current research interests include machine learning and data-driven methods for control systems, mission planning of unmanned aerial vehicles, and operator theoretic approach to dynamical systems.
\end{IEEEbiography}

\balance
\end{document}